\documentclass[nhess]{copernicus}
\usepackage{graphicx, amsthm, parskip}
\usepackage{amsmath, amssymb, textcomp, latexsym,subfig,float,natbib}
\usepackage{verbatim}

\begin{document}

\title{Statistical emulation of a tsunami model for sensitivity analysis and uncertainty quantification}

\author[1]{A. Sarri}
\author[2,1]{S. Guillas}
\author[3]{F. Dias}

\affil[1]{Institute of Risk and Disaster Reduction, University College London, UK}
\affil[2]{Department of Statistical Science, University College London, UK}
\affil[3]{School of Mathematical Sciences, University College Dublin, Ireland}

%% The [] brackets identify the author to the corresponding affiliation, 1, 2, 3, etc. should be inserted.

\runningtitle{Statistical Emulation of a landslide-generated tsunami model}

\runningauthor{A. Sarri, S. Guillas, F. Dias}

\correspondence{A. Sarri,\\ (andria.sarri.10@ucl.ac.uk)}

\received{}
\pubdiscuss{} %% only important for two-stage journals
\revised{}
\accepted{}
\published{}

%% These dates will be inserted by the Publication Production Office during the typesetting process.

\firstpage{1}

\maketitle

\begin{abstract}
Due to the catastrophic consequences of tsunamis, early warnings need to be issued quickly in order to mitigate the hazard. Additionally, there is a need to represent the uncertainty in the predictions of tsunami characteristics corresponding to the uncertain trigger features (e.g. either position, shape and speed of a landslide, or sea floor deformation associated with an earthquake). Unfortunately, computer models are expensive to run. This leads to significant delays in predictions and makes the uncertainty quantification impractical. Statistical emulators run almost instantaneously and may represent well the outputs of the computer model. In this paper, we use the Outer Product Emulator to build a fast statistical surrogate of a landslide-generated tsunami computer model. This Bayesian framework enables us to build the emulator by combining prior knowledge of the computer model properties with a few carefully chosen model evaluations. The good performance of the emulator is validated using the Leave-One-Out method.
\end{abstract}

%% only used for copernicus2.cls
%\abstract{
 %TEXT
 %\keywords{TEXT}}

\introduction
A tsunami is a series of powerful water waves generated by earthquakes, volcanic eruptions, underwater landslides as well as local landslides along the coast. Their main characteristic is the high speed of propagation. As emphasized by the recent tragic events in March 2011 in Japan and in December 2004 in Indonesia, tsunamis may be extremely catastrophic: they are able to destroy buildings, roads and generally the infrastructure is seriously affected. But the most tragic part is that tsunamis can lead to the loss of human lives. A deep knowledge of tsunamis is required in order to predict the maximum runups and rundowns, and also to give early warning notices to the regions that may be affected.

Since the most common sources for tsunamis are earthquakes, earthquake-generated tsunamis have been extensively investigated. Landslide-generated tsunamis have been much less studied and the existing knowledge about them is more limited. They are characterised by relatively short periods, compared to the earthquake-generated ones, resulting to stronger viscous damping. Hence, they do not travel as long distances as the earthquake-generated tsunamis do. Therefore, one of their characteristics is that their whole life cycle takes place near the source. Nevertheless, they can reach high amplitudes and can also become extremely harmful \citep{synolakisetal2002,tintietal2008}. The more challenging part in landslide-generated tsunami modelling results from the fact that they are not instantaneously generated, as the earthquake-generated tsunamis are, and their generation depends strongly on how the shape of the sea floor changes with time \citep{bardetetal2003}. 

\cite{oldpaper} performed the first experiments for landslide-generated tsunamis, where a sliding mass was moved down an incline. More recently, it was observed by \cite{liuetal2005} that larger wave maximum elevations occur for subaerial compared to submerged slides. Also, \cite{forecasting_impulse_waves} showed that the maximum wave amplitude depends on both the duration of the underwater motion and the front shape of the landslide. Studies about tsunamis generated by a sliding mass on a plane beach have also performed by \cite{liu_lynett}. The authors have investigated the whole life cycle of the tsunami: initially there is a high amplitude near the source, then the wave motion is predominantly near the shore, followed by edge waves along the shoreline and no motion near the source.

\cite{renzi2008} made an important contribution by developing an analytical three-dimensional model for landslide-generated tsunamis based on the forced linear long-wave equation of motion, considering a plane beach with a constant slope. The inputs of the model are the initial position, speed and spread ratio of the landslide and the output is the sea free-surface elevation at specific times and locations. Additionally, by comparing to available experimental data, they showed that the model represents the overall behaviour of the wave with acceptable accuracy. However, the predicted water elevations appear to be over-estimated, which was attributed to neglecting energy dissipation and dispersive effects. \cite{renzi2012} extended the landslide-generated tsunami model of \cite{renzi2008} to consider arbitrary initial position, speed and spread ratio. Furthermore, landslides in their framework can have a shape other than Gaussian. They investigated how these physical parameters and the shape of the landslide affect the resulting wave elevation. \cite{renzi2012} also analyzed the effect of the continental platform on the wave elevation. 

This paper presents a proof-of-concept case study for the statistical analysis of a landlide-generated tsunami model, by employing the analytical model constructed by \cite{renzi2008}. The main strategy of the analysis is to build a statistical emulator that accurately represents the analytical model, which can be used for fast predictions, quantification of uncertainties and sensitivity analysis. In Section 2, a more detailed explanation of the statistical emulator is presented. Section 3 describes the concept of a special form of emulator, named the Outer Product Emulator. An analytic description for the appropriate parameter selections and calculations required to build it are also presented. Section 4 describes the concept of the experimental design and its implementation. Section 5 shows the application of the Outer Product Emulator and its validation for the \cite{renzi2008} analytical model. The resulting emulator is then used for extremely efficient sensitivity and uncertainty analyses in Section 6.

\section{Statistical emulator}
An emulator is a simple statistical model that approximates a simulator, where a simulator is a deterministic input-output computer model (analytical model, complex statistical -e.g. stochastic- model, or most commonly a numerical solver of a large system of equations such as PDEs). Given some inputs $\textbf{x}$, the simulator output is given by $\textbf{y}=f(\textbf{x})$ and the emulator is denoted by $\hat{f}(\textbf{x})$, which indicates that it is an approximation of the simulator. In most cases, running simulators is very time and resource consuming, so one can only afford a limited number of runs. The use of emulators comes as a solution to this problem, since emulators run almost instantaneously. However, due to the fact that they are approximations of the computer model, some error is introduced by using them. So, emulators are recommended to be used only in the case when the simulator is expensive to evaluate. The error amount can be estimated since they can make probabilistic predictions of the output that the simulator would produce if it was exercised over certain regions of the input space. Therefore, the main use of statistical emulators is for fast predictions of the simulator output.

Analyses such as uncertainty and sensitivity analyses, as well as calibration, require a large number of evaluations of the expensive simulator and this means that they can become impractical. An emulator can be built and used to make such demanding analyses more efficiently. The uncertainty analysis provides us with a knowledge of the distribution of the simulator output. The sensitivity analysis investigates how each of the inputs affect the output. Calibration consists of fitting a model to the available observations by adjusting its parameters (we are not considering calibration in this paper). 

The emulator is created by employing a number of simulator evaluations. The error in its predictions is inversely related to the number of simulator evaluations. Therefore, a significantly large number of evaluations can make this error negligible, but this is unusual due to the simulator computational complexity. Also, since the emulator represents a deterministic model, it is also a deterministic model where the simulator has been exercised: it predicts perfectly, with zero error, the output at points that have been used in the creation of the emulator. At new points, the emulator gives a distribution for $f(\textbf{x})$ with mean value $\hat{f}(\textbf{x})$ and standard deviation which represents the error in the prediction and hence how close it is likely to be to the true simulator output $f(\textbf{x})$. 

Bayesian statistical analysis, through the emulators, can be much more efficient than other methods to quantify uncertainties, e.g. the standard Monte Carlo method for which the simulator must be run repeatedly. In a Bayesian analysis we first build a representative emulator for the simulator and then use it for further analysis. \cite{OO_unc,Oakley02probabilisticsensitivity} and  \cite{OHag2006} focused on a Bayesian approach for uncertainty and sensitivity analysis. They concluded that a Bayesian approach is more efficient than the Monte Carlo method as it uses a significantly smaller number of model runs. One can take advantage of this by running the model at higher resolution. 

The form of the emulator used in this analysis is the Gaussian Process (GP). A GP is an extension of the familiar and popular Normal distribution, also called Gaussian. Nice mathematical properties of the Normal distribution carry over to the GP and therefore the GP is the principal tool for creating an emulator, together with prior knowledge about the simulator. It is worthy to say that the term ``prior knowledge" is used to indicate the initial beliefs about the simulator before the use of the available data. An unknown function $f(.)$ has a GP distribution if for any set of input points $\{x_{1},\ldots,x_{n}\}$, the set of outputs $\{f(x_{1}),\ldots,f(x_{n})\}$ follows a multivariate Normal distribution. The simulator is represented by a GP distribution with mean function $m_{0}(.)$ and covariance function $V_{0}(.,.)$, i.e.
\begin{equation}
f(.)|\beta,\sigma^2,B \thicksim GP(m_{0}(.),V_{0}(.,.))
\end{equation}
where the symbol $\thicksim$ stands for ``is distributed as". The mean function is described by 
\begin{equation}
m_{0}(x)=h(x)^{T}\beta,
\end{equation}
in which $h(.)$ is the set of regression functions and $\beta$ is the vector of the unknown coefficients. The functions $h(.)$ are chosen to represent the main form of the actual simulator $f(.)$. The covariance function, which generates some additional variations as well as uncertainty, is given by 
 \begin{equation}
 V_{0}(x,x')=\sigma^{2}C(x,x';B)
 \end{equation}
where $C(.,.;B)$ is a correlation function whose shape is known but with unknown correlation parameters $B$, also called hyperparameters. A common choice for $C(.,.;B)$ is
 \begin{equation}
 C(x,x';B) = \exp\{-(x-x')^{T}B(x-x')\}
 \end{equation}
where $B$ is a diagonal matrix of the so-called smoothing parameters $b_{ii}$.  The inverse square roots of these parameters, $1/\sqrt b_{ii}$, are known as the correlation length scales. The $b_{ii}$ (or the correlation length scales) describe how rapidly the output responds to changes in each input; the correlation lengths scales give an indication of the distance in the input space for which correlation between the simulator outputs is either significant or negligible. 

\section{Outer Product Emulator}
In the case where the simulator has multiple outputs, the creation of a surrogate model is more complicated. The simplest approach is to build separate independent emulators for each output.  However this method has a major drawback: it ignores the correlations between the outputs. \citet{Rougier2008} proposed an approximate multivariate emulator, named the Outer Product Emulator (OPE), that creates one emulator for all the outputs, simplifying the process by using separable functions in inputs and outputs.

Therefore, the main advantage of the OPE is that the building cost is significantly smaller compared to a general multivariate emulator. The construction and use of an OPE can be fast, even in the case where many simulator evaluations and a large number of outputs exist. This property of the OPE is very important for the case investigated in this work. Indeed, the wave shape is not oscillating periodically and hence the frequency of the oscillation is not constant, so a large number of simulator evaluations is necessary. We have to run the simulator at small time steps and hence a large number of evaluations are collected to describe the outputs. This is the primary reason why we decided to use the OPE for the analysis. 

\cite{rohtua} describe further this special form of statistical emulation. The OPE has the form:
\begin{equation}
 f_{i}(r) = \sum_{j=1}^{\nu} \beta_{j}g_{j}(r,s_{i}) + \epsilon(r,s_{i})
\end{equation}
where $f_{i}(r)$ is the $i^{th}$ simulator output at input $r$, $g_{j}$ is the set of regressors, $\beta_{j}$ are the unknown coefficients and $\epsilon$ is the residual. Additionally, $s_{i}$ represents the output domain - e.g. time, space - corresponding to the $i^{th}$ simulator run.

In order to build an emulator, appropriate distributions for $\beta$ and $\epsilon$ must be chosen. A convenient choice is the Normal Inverse Gamma distribution that enables the use of conjugacy (so posterior estimates can be computed explicitly without resorting to Markov Chain Monte Carlo as in more standard fully Bayesian emulators), described by
\begin{equation}
\beta|\tau,B \thicksim N(m,\tau V)
\end{equation}
\begin{equation}
\epsilon|\tau,B \thicksim GP(0,\tau\kappa_\lambda(.))
\end{equation}
\begin{equation}
\tau | B\thicksim IG(a,d)
\end{equation}
where $B= \{m,V,a,d,\kappa_\lambda(.)\}$ is the set of the hyperparameters and $\kappa_\lambda(.)$ is the covariance function of the residuals with correlation lengths $\lambda$. Also, $N$ and $IG$ denote the Normal and Inverse Gamma distribution, respectively. Summing up, 
\begin{equation} 
\{\beta,\epsilon\} \thicksim NIG(m,V,a,d)
\end{equation}
where the hyperparameters $a$ and $d$ denote the degrees of freedom and the scale respectively. 

Furthermore, a choice for the appropriate regression $g_{j}(.)$ and covariance functions of the residual $\kappa_\lambda(.)$ is needed. There are two main characteristics that distinguish the OPE from a standard multivariate emulator. The first is that the covariance function of the residuals is separated in inputs $r$ and outputs $s$. This property can be represented by the equation 
\begin{equation}
\kappa_\lambda(r,s,r',s') = \kappa_\lambda^r(r,r') \times \kappa_\lambda^s(s,s')
\end{equation} 
The second characteristic is that the set of the regressor functions, $G$, is the outer product of the set of regressors for inputs, $G^r \,{\buildrel \Delta \over=}\,\{g_{j_r}^r(r)\}_{j_r=1}^{\nu_r}$, with the set of regressors for outputs, $G^s\,{\buildrel \Delta \over=}\,\{g_{j_s}^{s}(s)\}_{j_s=1}^{\nu_s}$, where the expression $\alpha{\buildrel \Delta \over=} \beta$ indicates that the term $\alpha$ is equal by definition to the term $\beta$. Therefore, the functions $g_{j}$ are given by $g_{j}(r,s)=g_{j_r}^r(r) \otimes g_{j_s}^{s}(s)$, where $\otimes$ is the outer product symbol and $j=\{1,\ldots,\nu\}$, where $\nu=\nu_r\times \nu_s$.

\subsection{Maximizing the marginal likelihood}
In order to find the most accurate representation of the simulator, appropriate values for the correlation lengths and other unknown parameters can be estimated by maximising the corresponding marginal likelihood \citep{GPbook} before getting posterior ditributions of emulated simulator outputs. In the application described in Section 5, this technique is used to obtain representative values for the four correlation lengths, one for each of the three inputs and one for the output. Starting from the general equation of the emulator, that is 
\begin{eqnarray}
y = f(x) = h(x) + \epsilon(x) &=& g(x)^{T}\beta + \epsilon(x) \nonumber \\
	                                 &=& Q(x)\beta + \epsilon(x),
\end{eqnarray}
where 
\begin{equation}
\epsilon\thicksim GP(0,\tau\kappa_{\lambda}),
\end{equation}
\begin{equation}
\beta\thicksim N(0,\tau V),
\end{equation}
we assume that the mean value of the unknown coefficients is zero and also that $V$ can be defined as $V=\sigma^2I$, with the common multiplier parameter to be described by 
\begin{equation}
\tau \thicksim IG(a,d)
\end{equation}
Therefore, this reformulation of the prior distributions entails that the regression functions multiplied by the unknown coefficients $\beta$, i.e. the function $h(.)$, has a Normal prior distribution given by
\begin{equation}
h|B \thicksim N(0, \tau QVQ^{T})
\end{equation}
The likelihood function is described as follows:
\begin{equation}
y|h,B \thicksim N(h,\tau\kappa_{\lambda})
\end{equation}
The marginal likelihood ( can be obtained from the integral of the likelihood times the prior, i.e.
\begin{equation}
p(y|B ) =  \int p(y|h,B)p(h|B) dh
\end{equation}
Therefore the marginal likelihood has a Normal distribution described by
\begin{equation}
y|B   \thicksim N(0,\tau\kappa_{\lambda} + \tau QVQ^{T})
\end{equation}
Consequently, the log marginal likelihood function is
\begin{equation}
\label{marginal}
\Lambda = \log(p(y)) = - \frac{1}{2}f^{T}C^{-1}f  - \frac{1}{2}\log |C| + constant
\end{equation}
where $C=\tau \left(\kappa_{\lambda} + QVQ^{T}\right)$.
The derivative, with respect to the correlation lengths, of the log marginal likelihood is given by
\begin{equation}
\nabla{\Lambda} =  \frac{1}{2}f^{T}C^{-1}\frac{\partial{C}}{\partial{\lambda}} C^{-1} f  - \frac{1}{2}tr(C^{-1}\frac{\partial{C}}{\partial{\lambda}})
\end{equation}
In order to calculate $C^{-1}$, the Cholesky decomposition is used. Optimization methods are used to help us with the maximization of the marginal likelihood function in order to find correlation lengths.

\subsection{Hyperparameters selection}
The final step in the process of building the prior emulator for the simulator is the selection of the hyperparameters $\{m,V,a,d\}$. To determine adequate hyperparameters, the simple approximation method presented by \cite{rohtua} is used. The idea is to average the simulator output $f_i(r)$ over the inputs $r$ and output $i$, which means that $f_i(r)$ is replaced by $f(x)$, and also to assume that $x$ has a uniform distribution. Using the mean and variance of the simulator output, $f(x)$,  the hyperparameters are estimated. Completing the selection of the hyperparameters yields the prior emulator. 

The prior emulator is combined with a sample of simulator's evaluations, called the training sample, giving the posterior emulator. The resulting emulator gives a prediction distribution for each point in the evaluations' output domain. These predictions are Student-t distributed with parameters (mean, variance and degrees of freedom) that are calculated according to the procedure explained in \cite{Rougier2008}.

After building the emulator, the next step is to test how accurately it represents the simulator. This process is called validation, and it is recommended to be performed before making use of the emulator. We use the so-called ``leave-one-out" diagnostic (LOO): one evaluation is left out and predicted using an emulator constructed from the rest of the training data set. We repeat this for all the evaluations. Therefore, the ability of the emulator to represent the simulator can be quantified.  

\section{Experimental Design}
One of the most important steps in the analysis is the experimental design. This is the process of finding a space filling design that covers the input space sufficiently. Due to the fact that the input points are selected strategically, the amount of useful information passed to the emulator can be maximized. Hence, the required number of simulator runs for an accurate emulator can be reduced, resulting in a more efficient procedure. 

Many different experimental designs exist. The simplest one is the regular grid, where equally spaced points are selected for each parameter. However, even with the simplicity of this design, some drawbacks exist by using it. The most important one is its ``collapsing'' property, where multiple grid points have the same coordinate value when projected onto a parameter axis. This means that a limited information is obtained from these points. For example, for a three-dimensional input space, in order to obtain $n$ distinct evaluations for each of the three parameters, the total number of required simulator runs is $n^3$, which is highly inefficient.

The Latin Hypercube design (LHD) is an experimental design that is constructed to avoid the ``collapsing'' property of grids. The LH design selects $n$ different sample points from each of the $k$ variables $X_{1},\ldots,X_{k}$ using the following process. First of all, the range of each variable is divided into $n$ equal probability and non-overlapping intervals. Then, one value from each interval is selected randomly with respect to the probability density of the interval. The $n$ values obtained for $X_{1}$ are paired randomly with the $n$ values for $X_{2}$. These $n$ resulting pairs are then combined randomly with the $n$ values for $X_{3}$ resulting into $n$ triplets. The same process continues until $n$ $k-$tuplets are formed, which is the LH sample.  

However, only a subset of LH designs are space filling. To ensure a space filling input selection, we adopt the so-called ``maximin'' Latin Hypercube Design. The specific design follows the same process as the LHD to choose the sample points, although it has an additional constraint that is to maximise the minimum distance between the points. Therefore, a maximum coverage of the input space is achieved. 

\cite{comparisonLH_grid} made a comparison of the Latin Hypercube with the regular grid design for the multivariate emulation. They report that the emulators built using the LHD make significantly improved predictions relative to the emulators created using a regular grid training sample. Furthermore, they concluded that the LH emulators are more accurate compared to the regular grid emulators in sensitivity analysis of a single-parameter model. 

\section{Application to the SR tsunami model}

\subsection{Model description}
\begin{figure}[htb]
\vspace*{2mm}
\begin{center}
\includegraphics[trim=4.5cm 8.3cm 3.8cm 6.8cm, clip=true,height=0.13\textwidth]{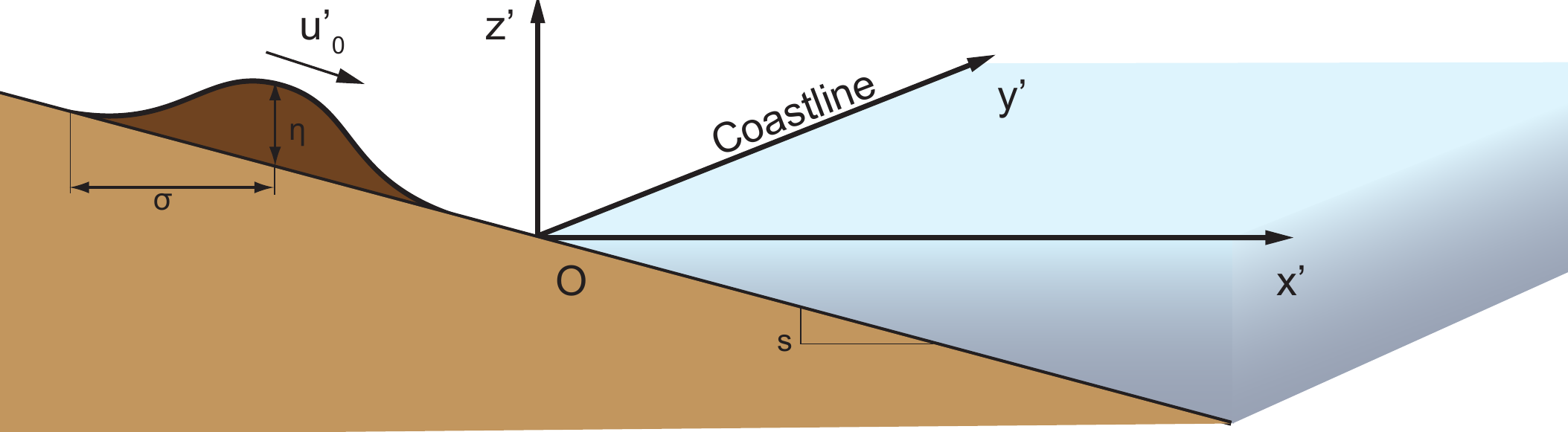}
\end{center}
\caption{Sketch illustrating the landslide's motion as considered in Sammarco and Renzi's analytical model. The $y'$-axis represents the shoreline, while the $x'$-axis is perpendicular to it.}
\label{landfig}
\end{figure}

In this section the methods described above are applied to find an accurate statistical representation of the landslide-generated tsunami analytical model of \cite{renzi2008}, abbreviated as the SR model. This model takes as inputs the initial position $x_0$, the speed $u_0$ and also the spread ratio or shape $c$ of the landslide, where the ``spread ratio" is defined as the ratio of the landslide's characteristic length over the characteristic width. Figure \ref{landfig} illustrates this specific analytical model set up. 

All the coordinates, functions and parameters used in the model are non-dimensional:

\begin{equation}
\begin{split}
x&=\frac{x'}{\sigma}, \quad  \ y=\frac{y'}{\sigma}, \quad \ t=\sqrt{\frac{gs}{\sigma}}t', \quad \zeta=\frac{\zeta'}{\eta}, \\ u_0 &= \frac{1}{\sqrt { \sigma gs}} u'_0, \quad \ c=\frac{\sigma}{\lambda}
\end{split}
\end{equation}
where the primes denote dimensional values, $\sigma$ is the landslide characteristic horizontal length, $s$ is the beach slope, $\eta$ denotes the landslide maximum vertical thickness, $\zeta$ is the the non-dimensional sea free-surface elevation, $\lambda$ is the landslide characteristic width, $t$ is the time and $g$ is the acceleration due to gravity. 

When the landslide starts moving from the origin, which is the position where the sea surface meets the sloping beach, $x_{0}$ is equal to zero. Also, negative values of $x_0$ indicate that the landslide initiates from a subaerial position, whereas positive values of $x_0$ indicate submerged slides. The output of this model is the sea free-surface elevation of the wave at given time and location. A plane beach with constant slope is considered and it is important to notice that the landslide continues to move even after it falls into the water. This causes the existence of high wave elevations even at large times.  

\begin{figure*}[t]
\vspace*{2mm}
\begin{center}
\subfloat[t=0.5]{\label{polar05}\includegraphics[trim=1.2cm 4.5cm 2.65cm 0.5cm, clip=true,height=0.185\textwidth]{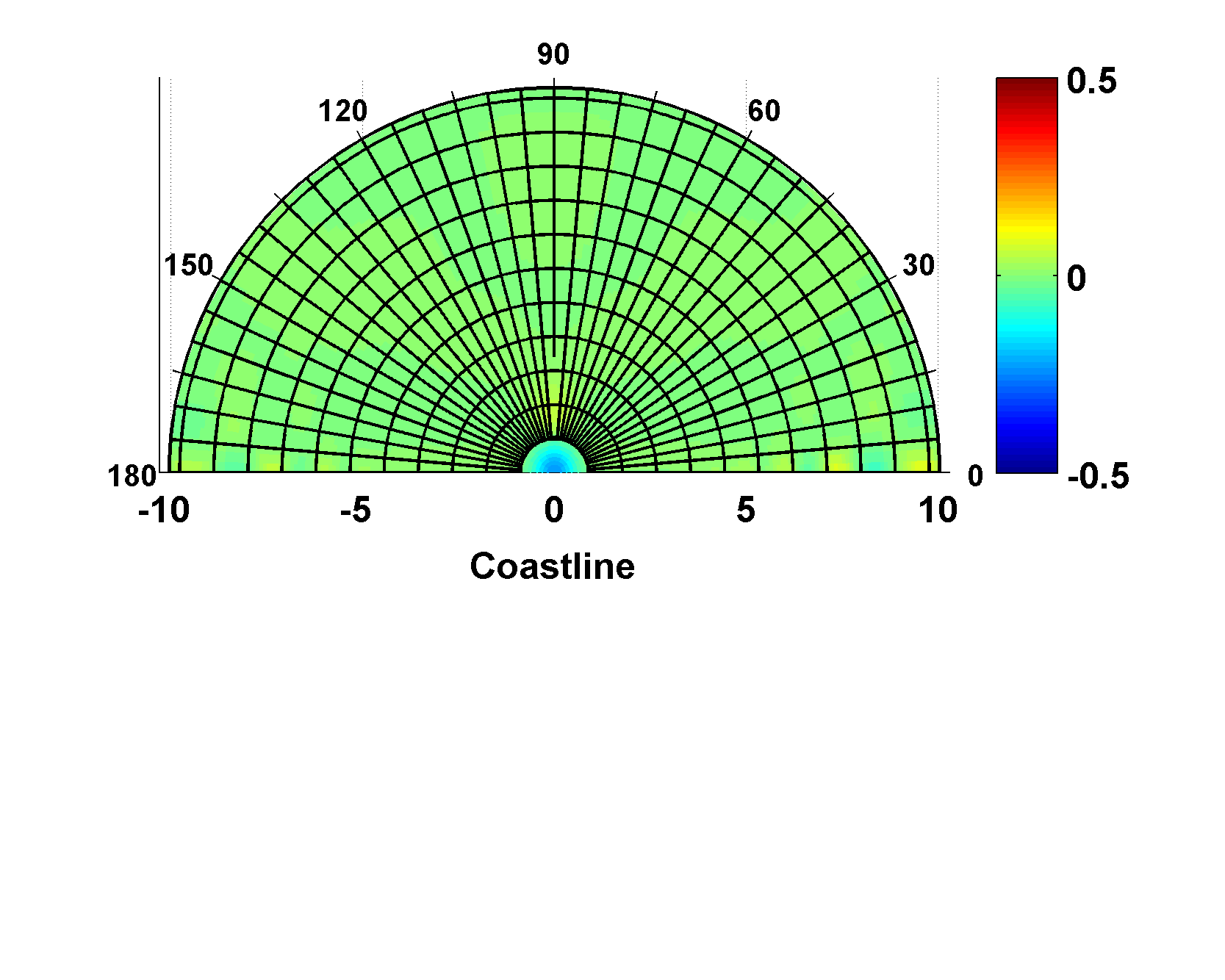}}
\subfloat[t=1]{\label{polar1}\includegraphics[trim=1.2cm 4.5cm 2.65cm 0.5cm, clip=true,height=0.185\textwidth]{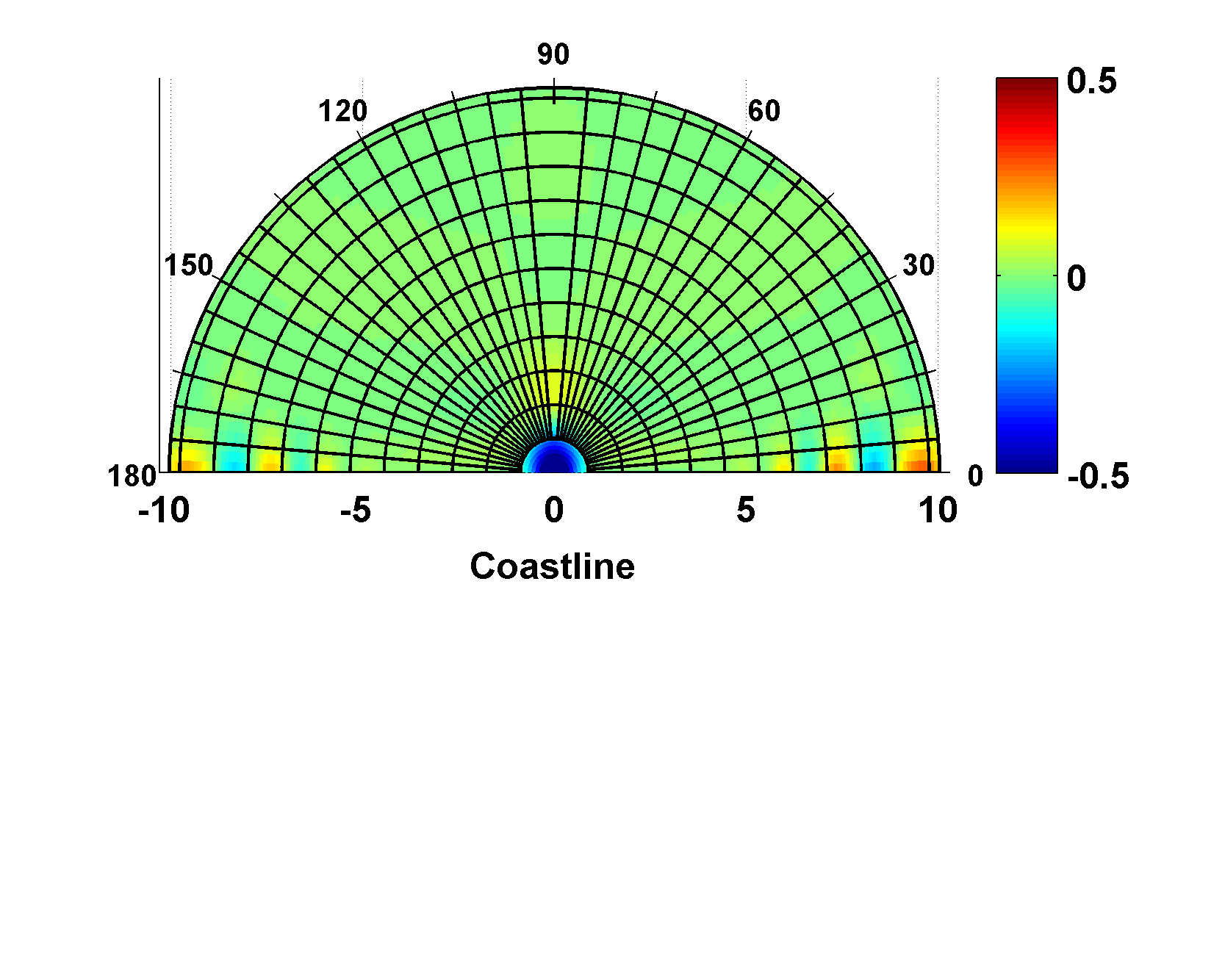}}
\subfloat[t=1.5]{\label{polar15}\includegraphics[trim=1.2cm 4.5cm 0cm 0.5cm, clip=true,height=0.183\textwidth]{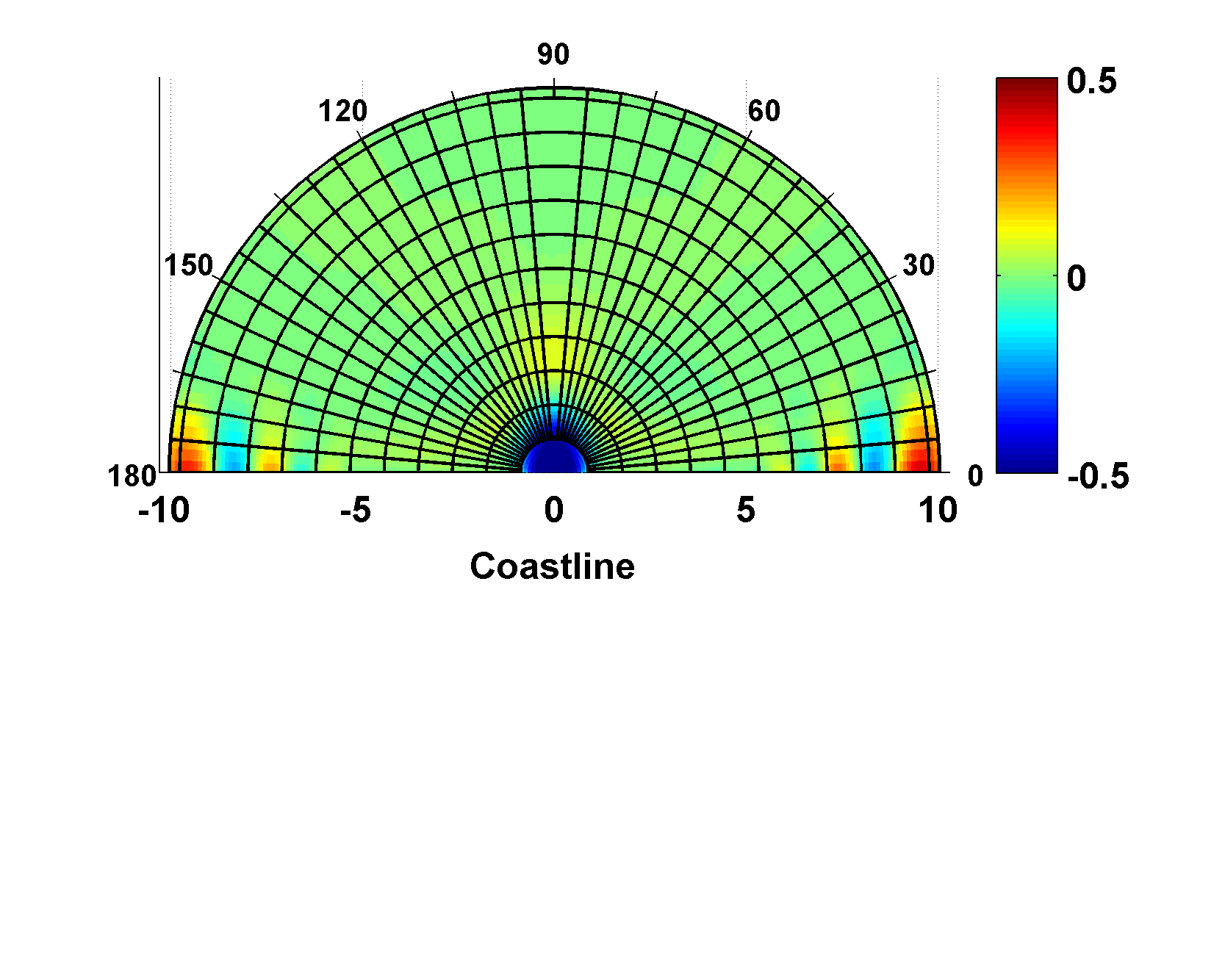}}\\
\subfloat[t=2]{\label{polar2}\includegraphics[trim=1.2cm 4.5cm 2.65cm 0.5cm, clip=true,height=0.185\textwidth]{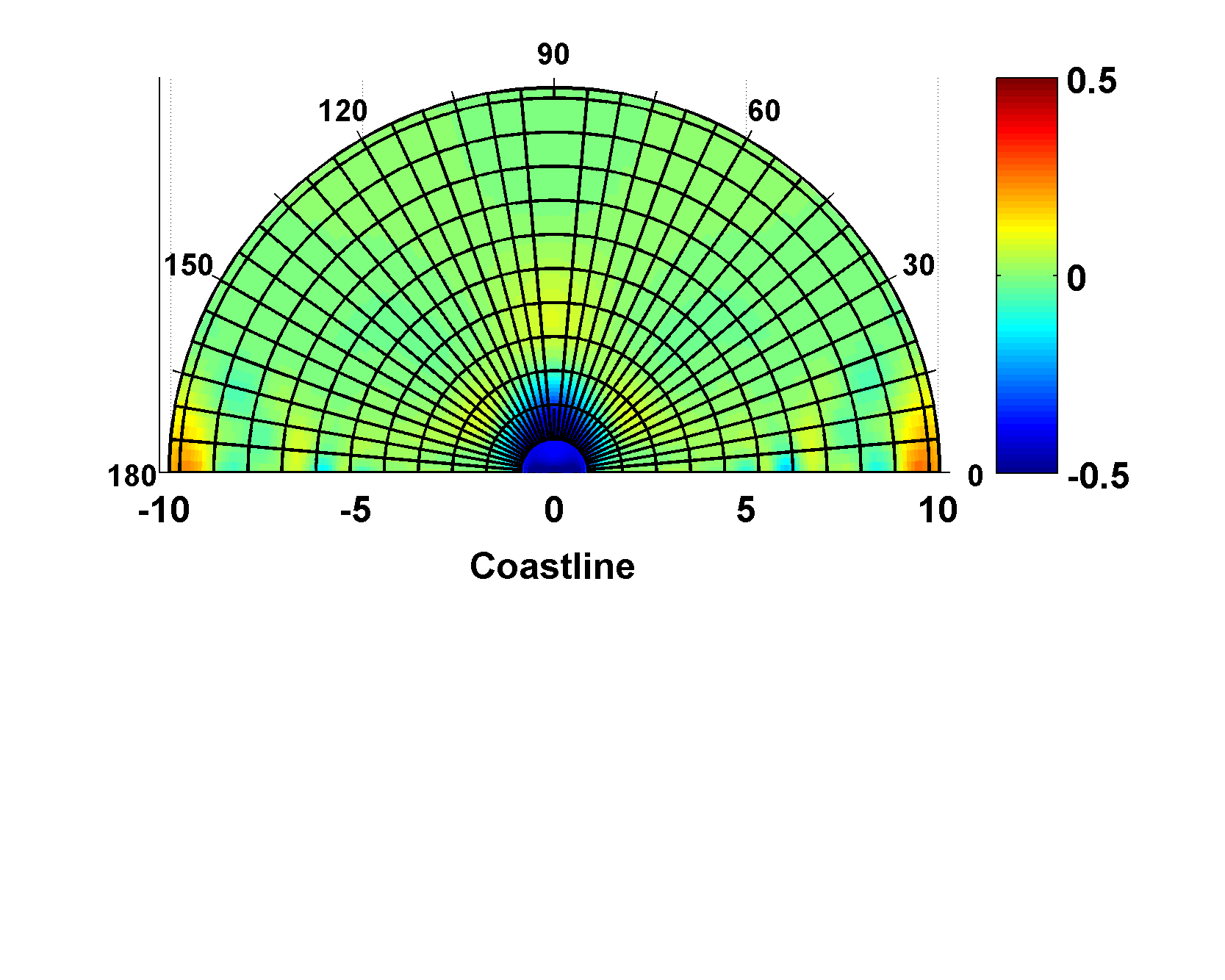}}
\subfloat[t=2.5]{\label{polar25}\includegraphics[trim=1.2cm 4.5cm 2.65cm 0.5cm, clip=true,height=0.185\textwidth]{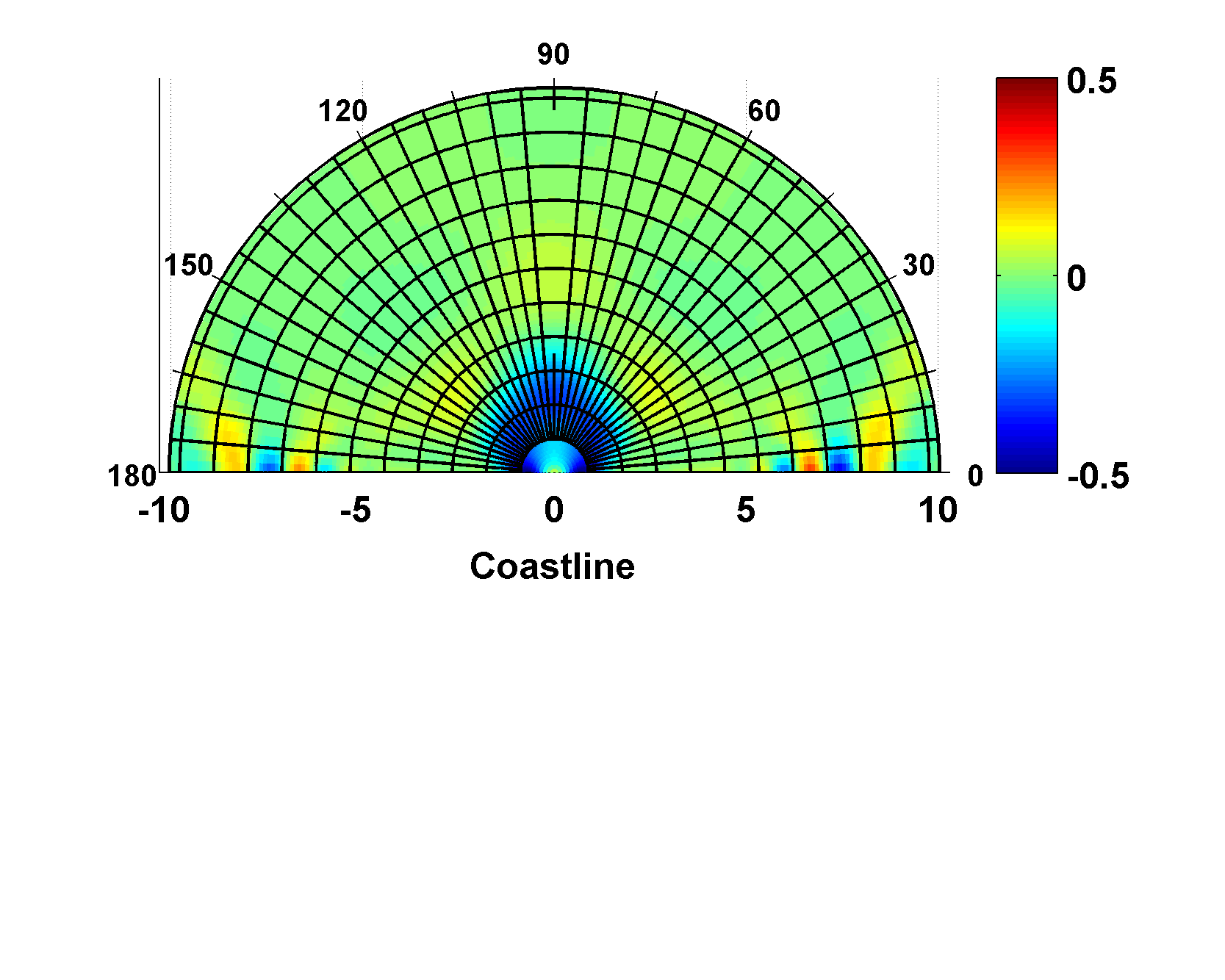}}
\subfloat[t=3]{\label{polar3}\includegraphics[trim=1.2cm 4.5cm 0cm 0.5cm, clip=true,height=0.183\textwidth]{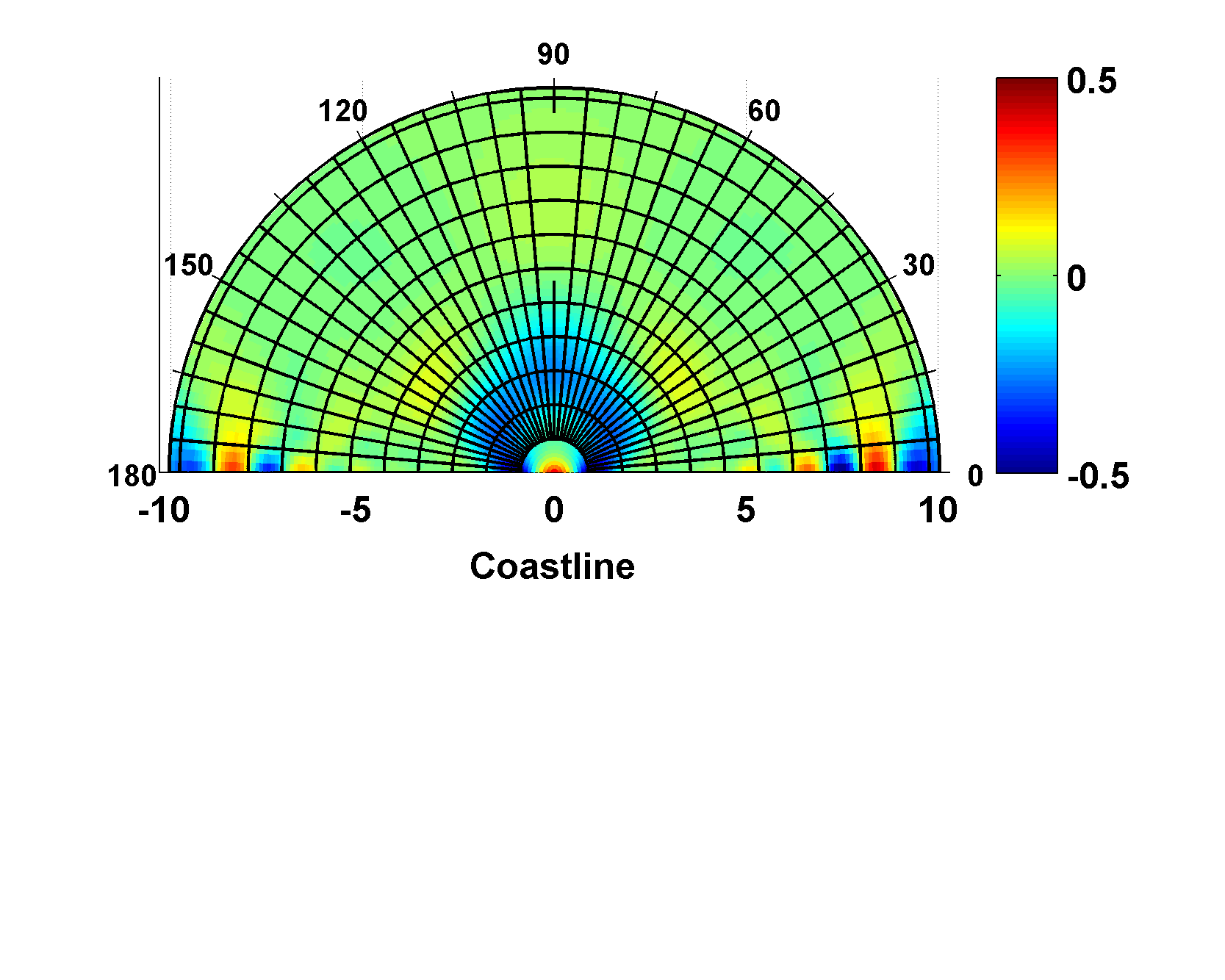}}\\
\subfloat[t=5]{\label{polar5}\includegraphics[trim=1.2cm 4.5cm 2.65cm 0.5cm, clip=true,height=0.185\textwidth]{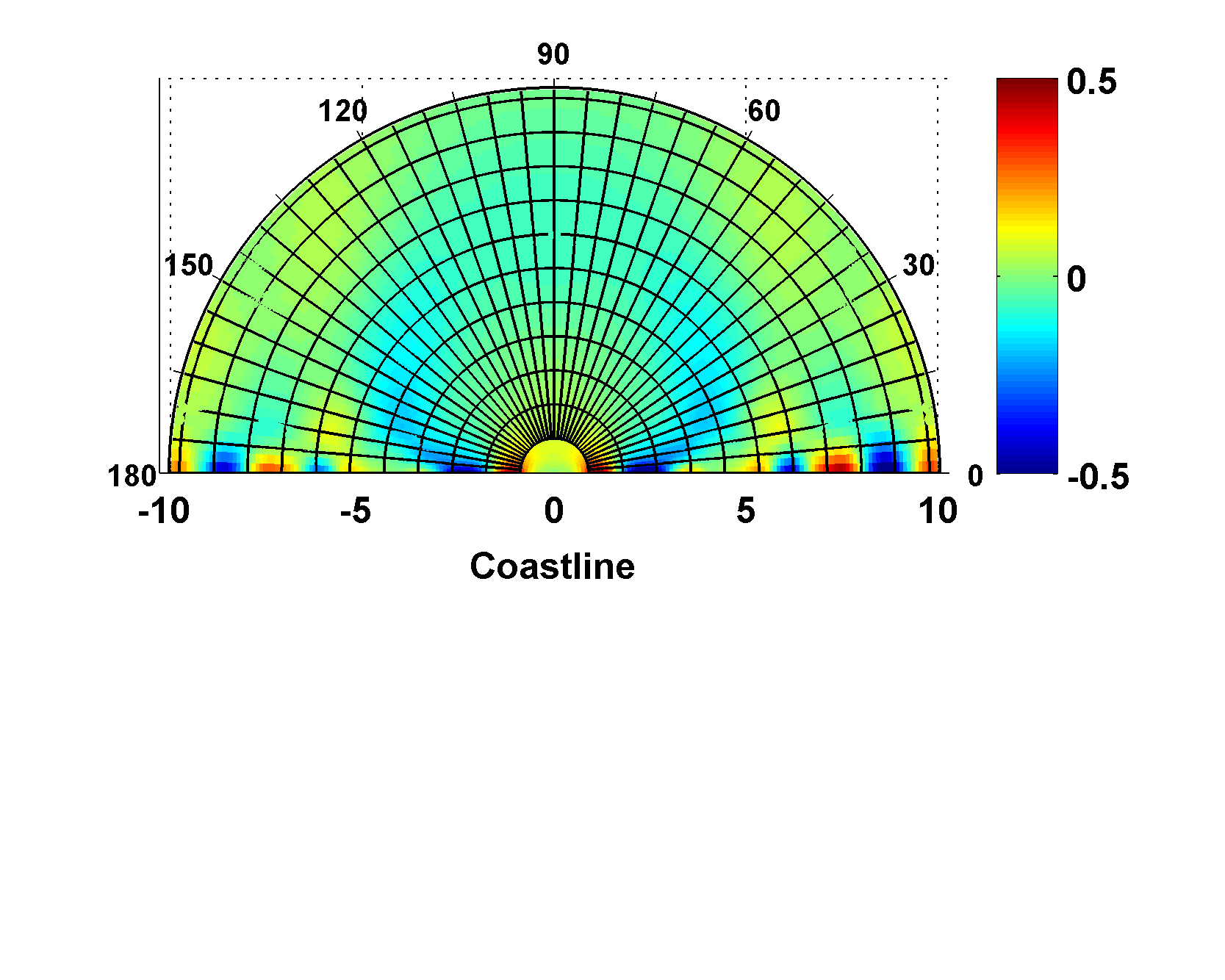}}
\subfloat[t=10]{\label{polar10}\includegraphics[trim=1.2cm 4.5cm 2.65cm 0.5cm, clip=true,height=0.185\textwidth]{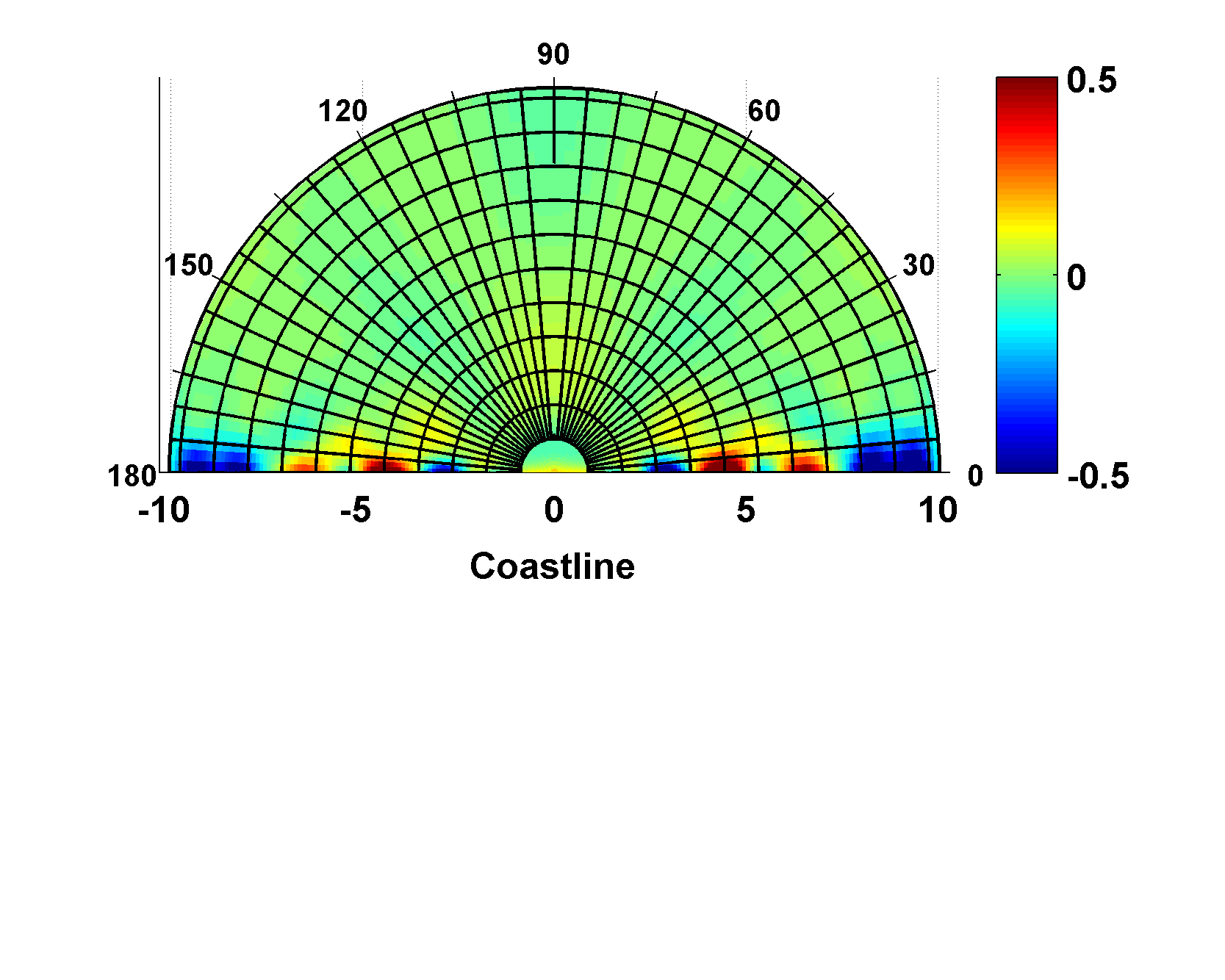}}
\subfloat[t=20]{\label{polar20}\includegraphics[trim=1.2cm 4.5cm 0cm 0.5cm, clip=true,height=0.183\textwidth]{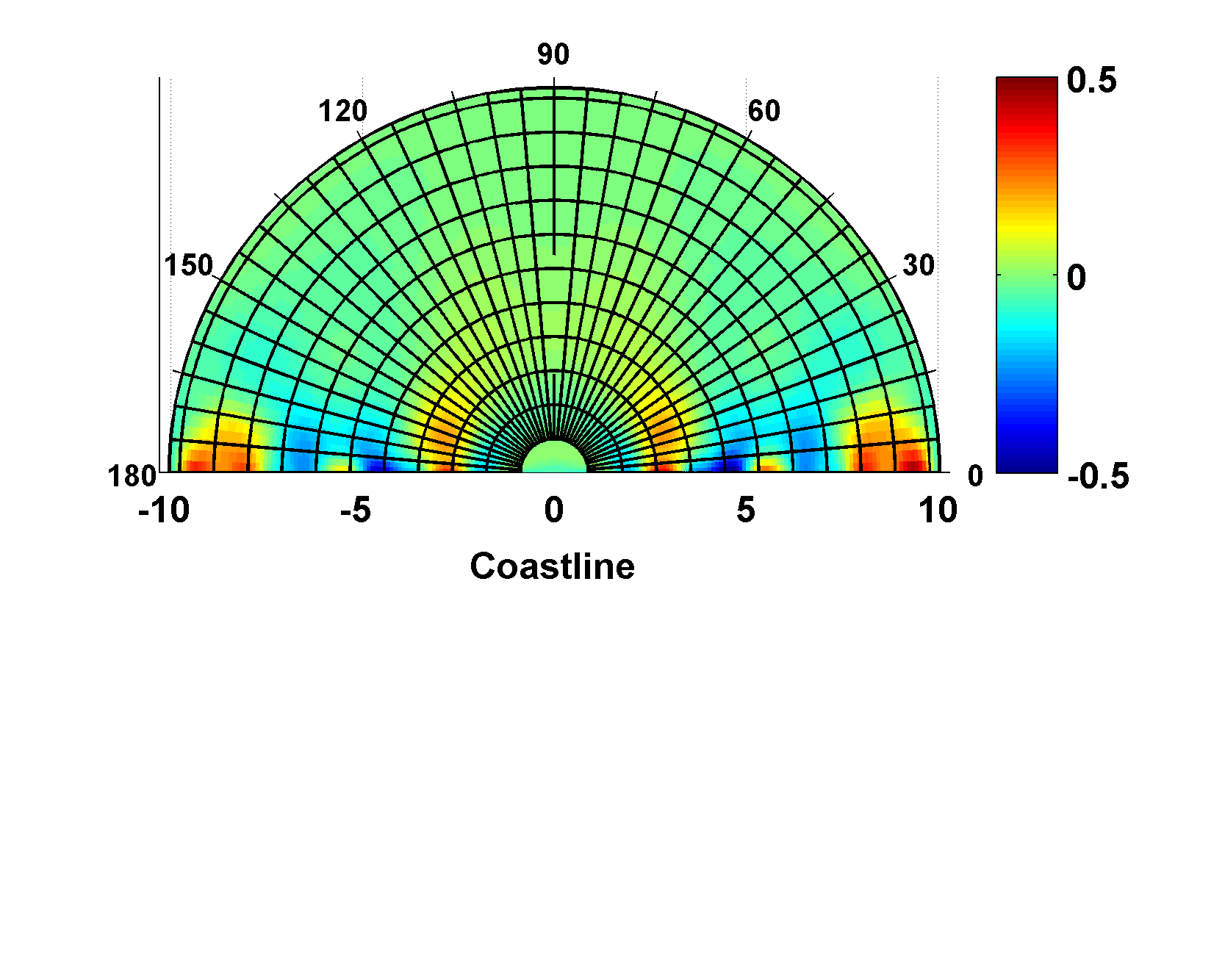}}
\end{center}
\caption{Sea free-surface elevation of the landslide generated tsunami observed at different times with non-dimensional inputs $(x_0,u_0,c)=(0,1,2)$. The horizontal axis represents the shoreline and the vertical axis points to the offshore direction.}
\label{polar}
\end{figure*}

By considering this model, \cite{renzi2008} came to the conclusion that the landslide generates a wave field that is composed by two components, oscillatory and evanescent. The life cycle of the wave can be visualized in Fig. \ref{polar}, where the sea free-surface elevation of the landslide-generated tsunami wave is shown in polar coordinates at times $t=0.5, 1, 1.5, 2, 2.5, 3, 5, 10, 20$. The initial position of the landslide is at the origin, the speed is equal to 1 and the spread ratio of the landslide is equal to 2, which means that the characteristic length is twice the size of the characteristic width.

When the landslide occurs, it displaces water forward and an elevation wave is generated, that propagates mostly in the offshore direction. Also a depression wave occurs near the origin (see Fig. \ref{polar05}). Later on, the elevation wave spreads along the shoreline, while the depression wave extends around the origin (see Figs \ref{polar1}, \ref{polar15}, \ref{polar2}). At larger times, a second elevation wave is generated at the origin and the depression wave spreads out (see Figs \ref{polar3}, \ref{polar5}). Finally, at even larger times, the wave motion is dominated by edge waves propagating along the shoreline, with no motion around the origin (see Fig. \ref{polar10}, \ref{polar20}). From this study, it is concluded that the first generated waves are not those with the larger amplitude. This indicates that in order to capture the maximum elevation, the model has to be evaluated up to a significantly large time $t$.

\subsection{Training sample}
In this work, a statistical emulator is constructed looking at specific locations; meaning that its output is only time-depended. Specifically, seven locations along the shoreline ($x=0$) at $y=2,4,6,7,8,8.38$ and $10$ have been investigated. The time domain is selected to be between 0 and 35. Small time steps are required in order to have sufficient information to capture the wave shape with sufficient detail: specifically $dt=0.2$ was chosen for the analysis. 

The first step of the analysis is the experimental design. Using the ``maximin" Latin Hypercube design method, as detailed in Section 4, forty points, $(x_{0},u_{0},c)$, are chosen to cover the three-dimensional parameter space. This is a compromise in order to have a significantly good coverage of the design space as well as a significantly small computation cost. The input domain is chosen to be $x_0\in[-3,1]$, $u_0\in[1,2]$ and $c\in[0.5,3]$.

\begin{figure}[htb]
\vspace*{2mm}
\begin{center}
\subfloat{\includegraphics[trim=4cm 8cm 2.5cm 9cm, clip=true,height=0.345\textwidth]{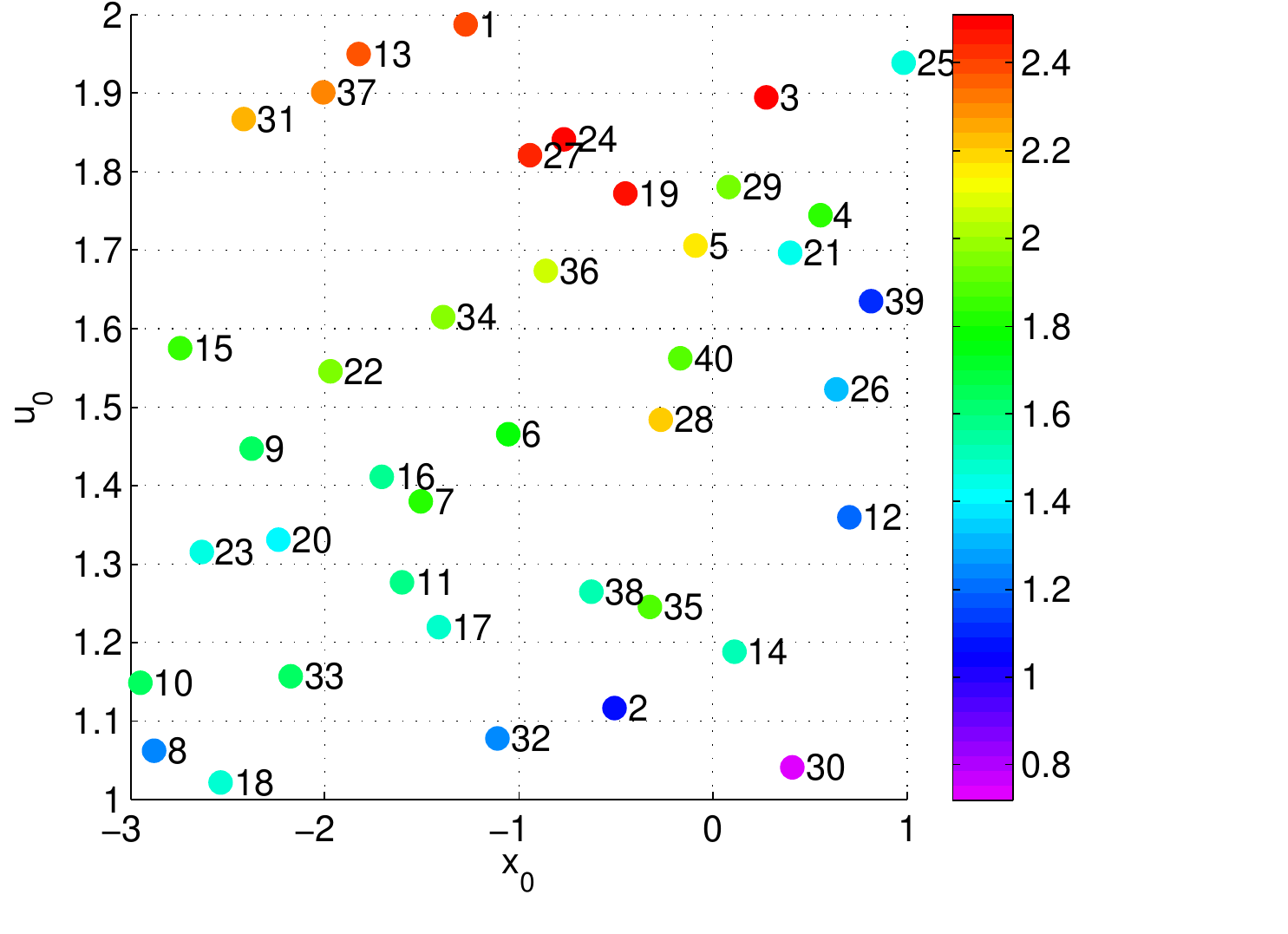}}\\
\subfloat{\includegraphics[trim=4cm 8cm 2.5cm 9cm, clip=true,height=0.345\textwidth]{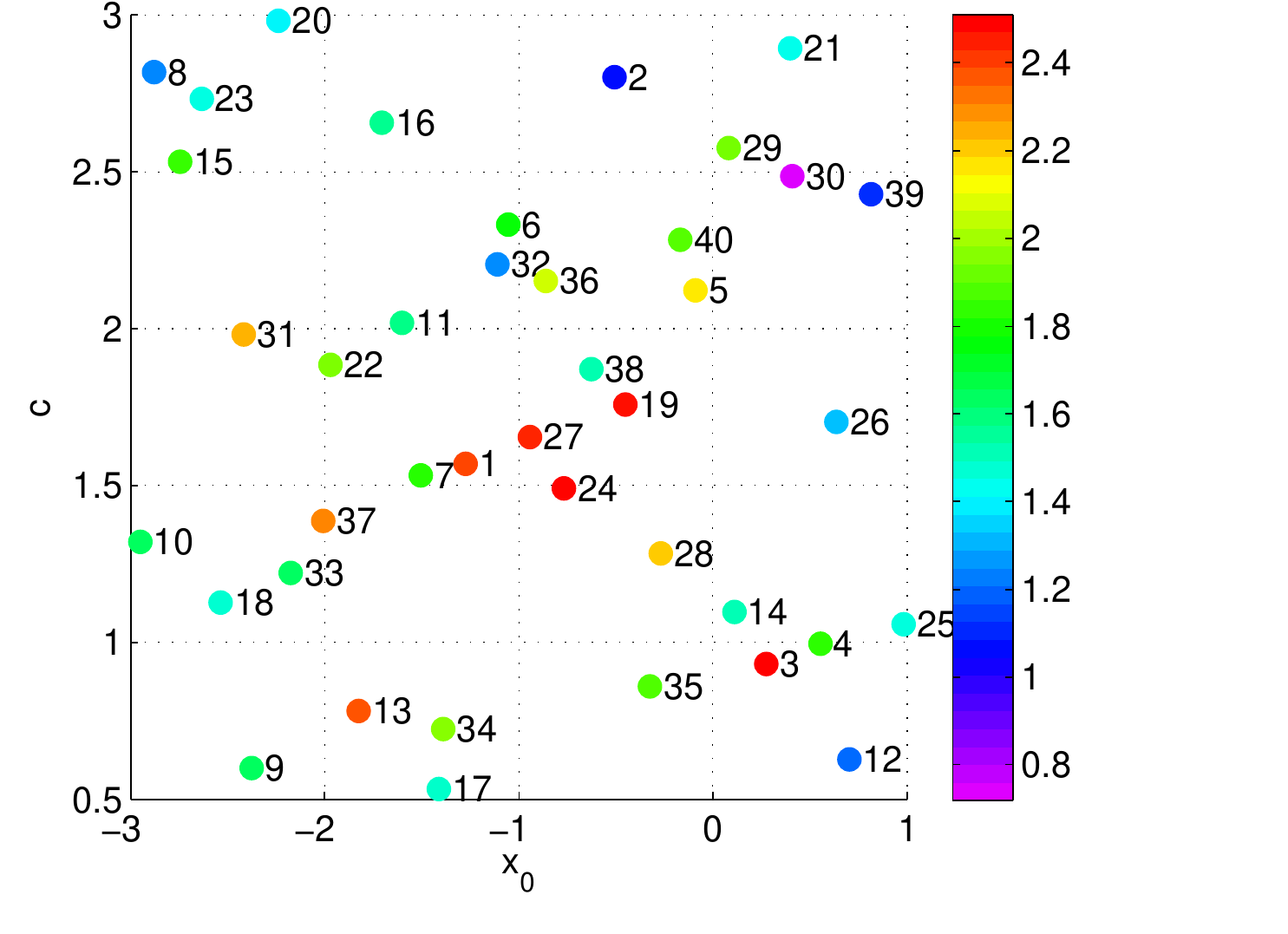}}\\
\subfloat{\includegraphics[trim=4cm 8cm 2.5cm 9cm, clip=true,height=0.345\textwidth]{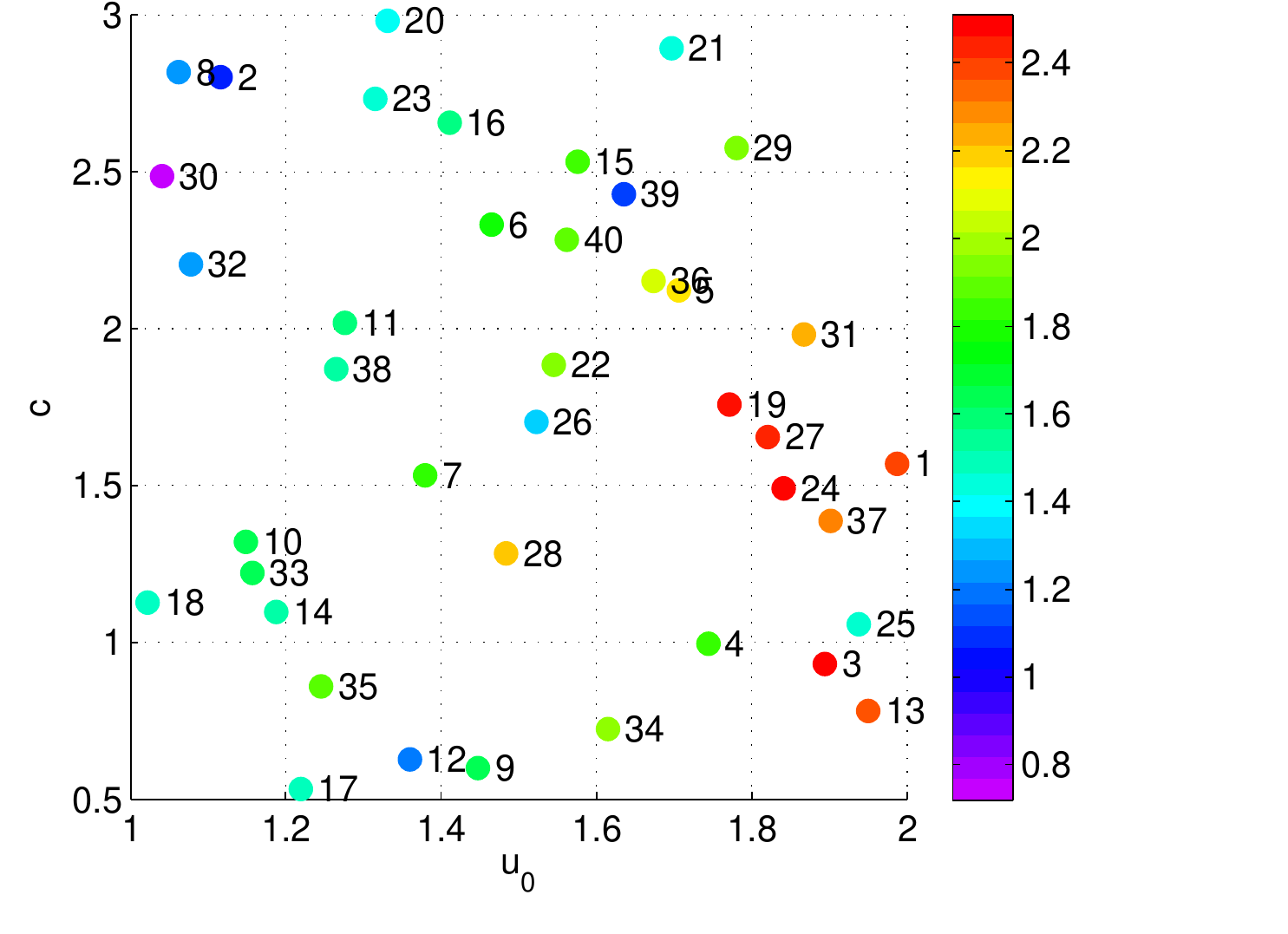}}
\end{center}
\caption{Maximum sea free-surface elevation at the location $(x,y) = (0,8.38)$ for time between 0 and 35 for each of the 40 design input points selected using the ``maximin" LHD method. Three quantities are varied: the landslide's speed, its initial location and its shape, that are given in non-dimensional form as in Eq. (21). }
\label{1}
\end{figure}

The positions of the forty inputs in the parameter space are shown in Fig. \ref{1}. The colour at each point indicates the maximum sea free-surface elevation, for the location $x=0$ and $y=8.38$, i.e. along the shoreline and far away from the source. The figure shows that the maximum wave elevation significantly depends on the landslide's speed: the larger the speed $u_0$, the larger the maximum elevation. Furthermore, it can be observed that the maximum wave elevation shows higher amplitudes when the landslide starts from a subaerial close to the origin position and also when the landslide spread ratio is less than $2$. However, the dependence of the maximum elevation on the initial position and spread ratio of the landslide is not as obvious as that on the speed.

For example consider points 13 and 25. They both represent landslides characterised by high speed and spread ratio close to one. However point 13 is a subaerial case while point 25 is a submerged one. This yields a significant difference in the maximum sea free-surface elevation, with the subaerial case being much higher. 

The simulator's evaluations for the other six locations along the shore yields similar conclusions about the dependency of the maximum sea free-surface elevation to the input parameters. 

\subsection{OPE prior choices}
The next step in the analysis involves the appropriate prior choices for the regression and residuals covariance functions for inputs $r$ and outputs $s$. In the case of the SR model, $r$ is equal to $(x_{0},u_{0},c)$ and $s$ is time $t$. The set of input regression functions, $G^{r}{\buildrel \Delta \over=}\, \{g_{1}^{r},\ldots,g_{\nu_{r}}^{r}\}$, where $\nu_{r}$ is the number of input regressors, consists of a combination of appropriate choices of polynomials for each of the three input parameters. For each input parameter, a linear and a quadratic polynomial, plus a constant term, are chosen, resulting to a total of seven input regressors. Since the simulator's output variation with respect to $r$ is smooth, the use of higher order polynomials is unnecessary, which would additionally increase the prior uncertainty. The chosen polynomials are shifted into the unit interval $[0,1]$ and their coefficients are selected so that the two functions for each input parameter are orthonormal with a uniform weighting function. Combining all the inputs' functions, the set of chosen input regressors is the following: 
\begin{eqnarray}
G^{r} &=& \{1, \sqrt3\frac{(x_0+3)}{4}, -3\sqrt5\frac{(x_0+3)}{4} + 4\sqrt5(\frac{x_0+3}{4})^2,  \nonumber\\
	 &  & \sqrt3(u_0-1), -3\sqrt5(u_0-1) + 4\sqrt5(u_0-1)^2,  \nonumber\\
	 &  & \sqrt3\frac{(c-0.5)}{2.5}, -3\sqrt5\frac{(c-0.5)}{2.5} + 4\sqrt5(\frac{c-0.5}{2.5})^2\}
\end{eqnarray}

After choosing the regression functions for the inputs, we need to make an appropriate choice for the regression functions for the output, $G^{s}{\buildrel \Delta \over=}\, \{g_{1}^{s},\ldots,g_{\nu_{s}}^{s}\}$, where $\nu_{s}$ is the number of output regressors. Fourier terms are chosen of the form $\sin(\frac{2\pi t}{T})$ and  $\cos(\frac{2\pi t}{T})$, where $T$ is the period of the oscillation, in addition to a constant term. However, since the sea free-surface elevation waves do not oscillate with constant period, this selection is challenging. To make this selection, we consider the range of oscillating frequencies present in the wave and using the LOO diagnostic method (explained in more detail in Section 5.4), we choose the smallest set of frequencies that give the most accurate predictions, since as for the case of input regressors an unnecessary large number of regressors is not desirable. The selected set of frequencies is the following: $\{\frac{1}{6},\frac{1}{5},\frac{1}{4},\frac{1}{3},\frac{1}{2}\}$. Therefore, the set of output regression functions is given by
\begin{eqnarray}
G^{s} &=& \{1,\sin(\pi t/3),\cos(\pi t/3),\sin(2\pi t/5),\cos(2\pi t/5), \nonumber\\
	 &  & \sin(\pi t/2),\cos(\pi t/2),\sin(2\pi t/3),\cos(2\pi t/3), \nonumber\\
	 & & \sin(\pi t),\cos(\pi t)\}
\end{eqnarray}

Power exponential functions are chosen for input and output residuals covariance functions, $\kappa^{r}$ and $\kappa^{s}$:
\begin{eqnarray}
\kappa^{r} &=& \exp{(-(\frac{|x_0 - x_{0}'| }{ \lambda_{x}})^{3/2})}\times\exp{(-(\frac{| u_0 - u_{0}'| }{ \lambda_{u}})^{3/2})}  \nonumber\\
	&  & \times \exp{(-(\frac{| c - c'| }{ \lambda_{c}})^{3/2})}
\end{eqnarray}
and
\begin{equation}
\kappa^{s} = \exp{(-(\frac{| t_1 - t_2| }{ \lambda_{t}})^{3/2})}
\end{equation}
respectively, where $\lambda_x$,  $\lambda_u$, $\lambda_c$ represent the correlation lengths for inputs and $\lambda_t$ denotes the output (i.e. time) correlation length. The values of the correlation lengths can be varied in order to adjust the fit of the emulator. The correlation lengths are chosen by maximizing the marginal likelihood. Since $\tau$ appears in the equation of the marginal likelihood \eqref{marginal}, in order the process of maximizing the marginal likelihood to be feasible, $\tau$ has been treated as a constant and estimated by the process simultaneously with the correlation lengths. The estimated value for $\tau$ is not used further in the analysis since $\tau$ was considered as constant only for practical purposes for this process and it is everywhere else considered as a scalar variable that is described by an Inverse Gamma distribution. Furthermore, note that the $3/2$ exponent is chosen so that the covariance is smooth enough, but not too much as the usual choice of square power is infinitely smooth and hence may not be realistic for such a complex simulator.

The last step for the creation of the prior emulator for the SR model is to make a choice for the values of the hyperparameters $\{m,V,a,d\}$. To do so we follow the method described by \citet{rohtua}. We have already assumed $m=0$. The hyperparameter $a$, which is equal to the number of degrees of freedom, takes the value 3 in the case of the SR model. Also, after the simple calculations recommended by \citet{rohtua}, it is concluded that $\sigma^2=0.257$ and $d=0.208$. Hence, $V$ can be easily obtained from $V=\sigma^{2}I$. By fixing these parameters, the creation of the prior emulator is completed. Using the evaluations of the 40 selected design points, the prior emulator is updated to obtain the posterior, which is the statistical emulator. Evaluating the statistical emulator at a given input point, $(x_0,u_0,c)$, results in predictions of the output's distribution for all the points in the time domain, in this case from 0 to 35, every 0.2 time step, i.e. 176 prediction distributions.

\subsection{Emulator's validation}
After the creation of the emulator, the LOO validation method is applied, resulting in 40 LOO diagnostic plots. These diagnostics give information about the predictive power, capabilities and shortcomings of the emulator, since we can estimate the amount of the error induced by using the emulator instead of the simulator. Some of the diagnostic plots for the location $(x,y) = (0,8.38)$ are shown in Fig. \ref{diagnostics}. Similar diagnostic plots are created for all the other locations investigated. In general, the LOO diagnostics allow us to conclude that in most of the cases the emulator predicts very well the simulator evaluations, capturing both shape and the maximum wave elevations (peaks). Additionally almost always the simulator's evaluation line is within the 95\% prediction credible interval (ideally it should be within this interval 95\% of the time).

\begin{figure*}[htb]
\vspace*{2mm}
\begin{center}
\subfloat[$(x_0,u_0,c)=(-2.88,1.06,2.82)$ -- point number 8]{\label{LOO8}\includegraphics[trim=0cm 0cm 0cm 2cm, clip=true,height=0.335\textwidth]{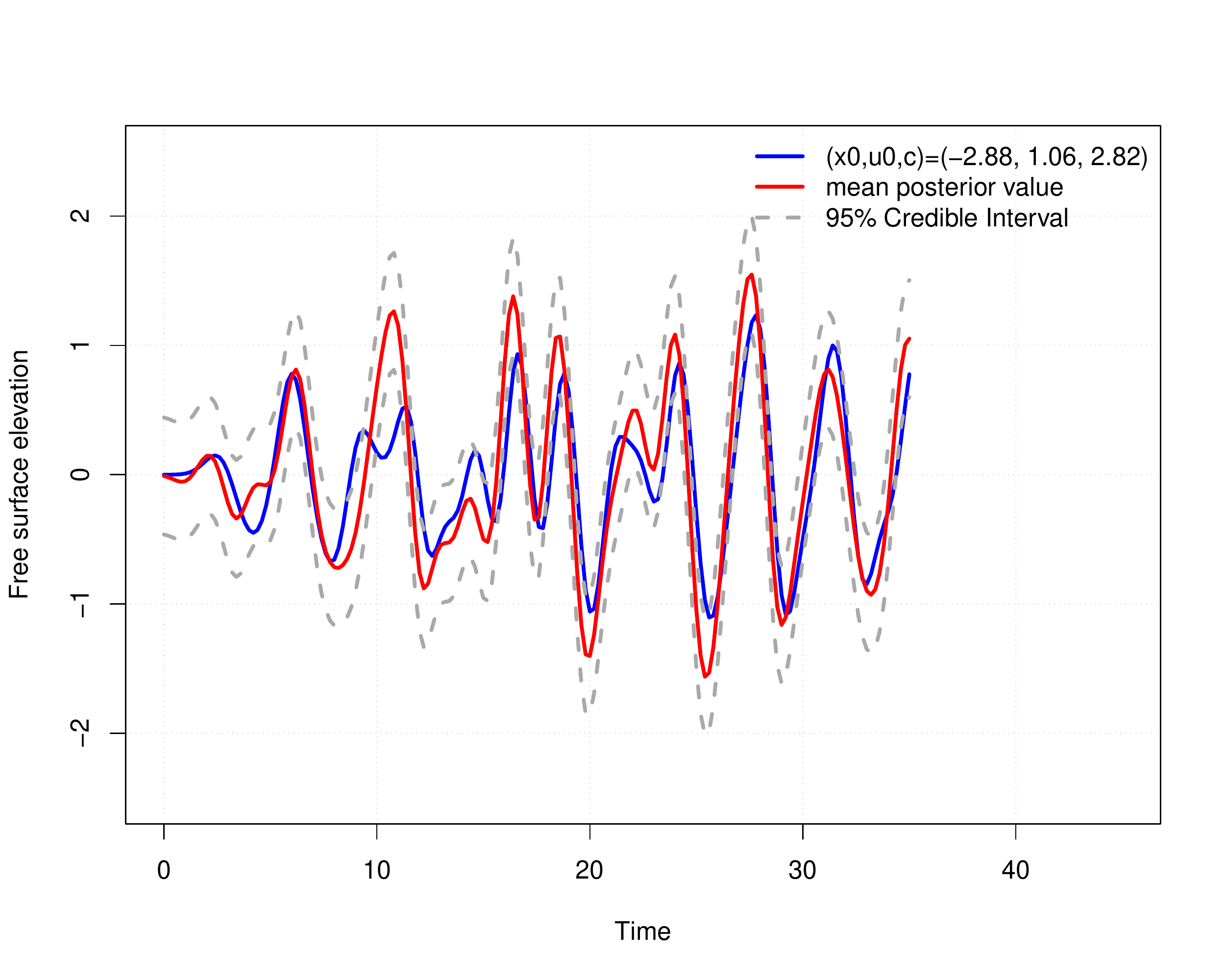}}
\subfloat[$(x_0,u_0,c)=(-2.96,1.15,1.32)$ -- point number 10]{\label{LOO10}\includegraphics[trim=0cm 0cm 0cm 2cm, clip=true,height=0.335\textwidth]{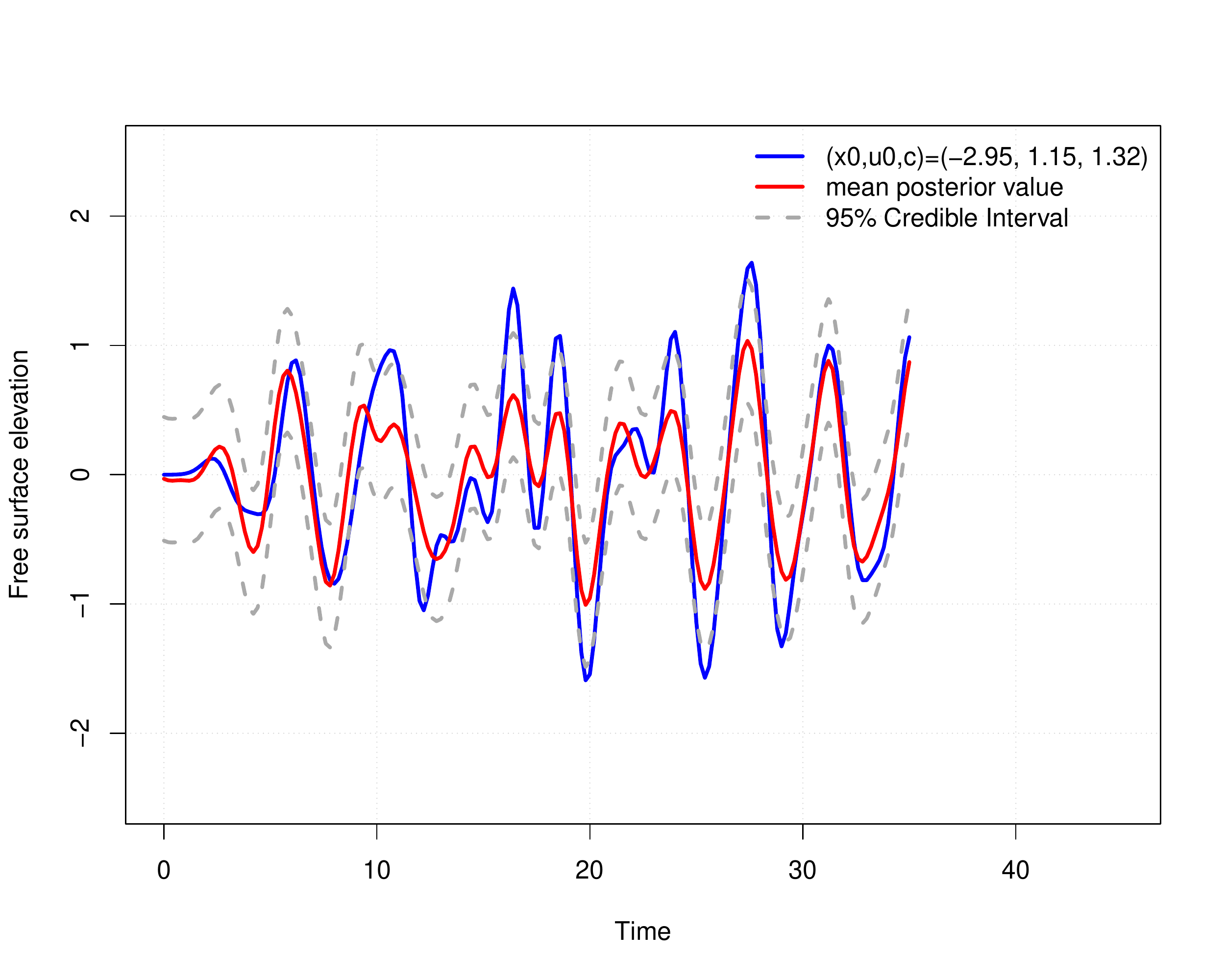}}\\
\subfloat[$(x_0,u_0,c)=(0.70,1.36,0.63)$ -- point number 12]{\label{LOO12}\includegraphics[trim=0cm 0cm 0cm 2cm, clip=true,height=0.335\textwidth]{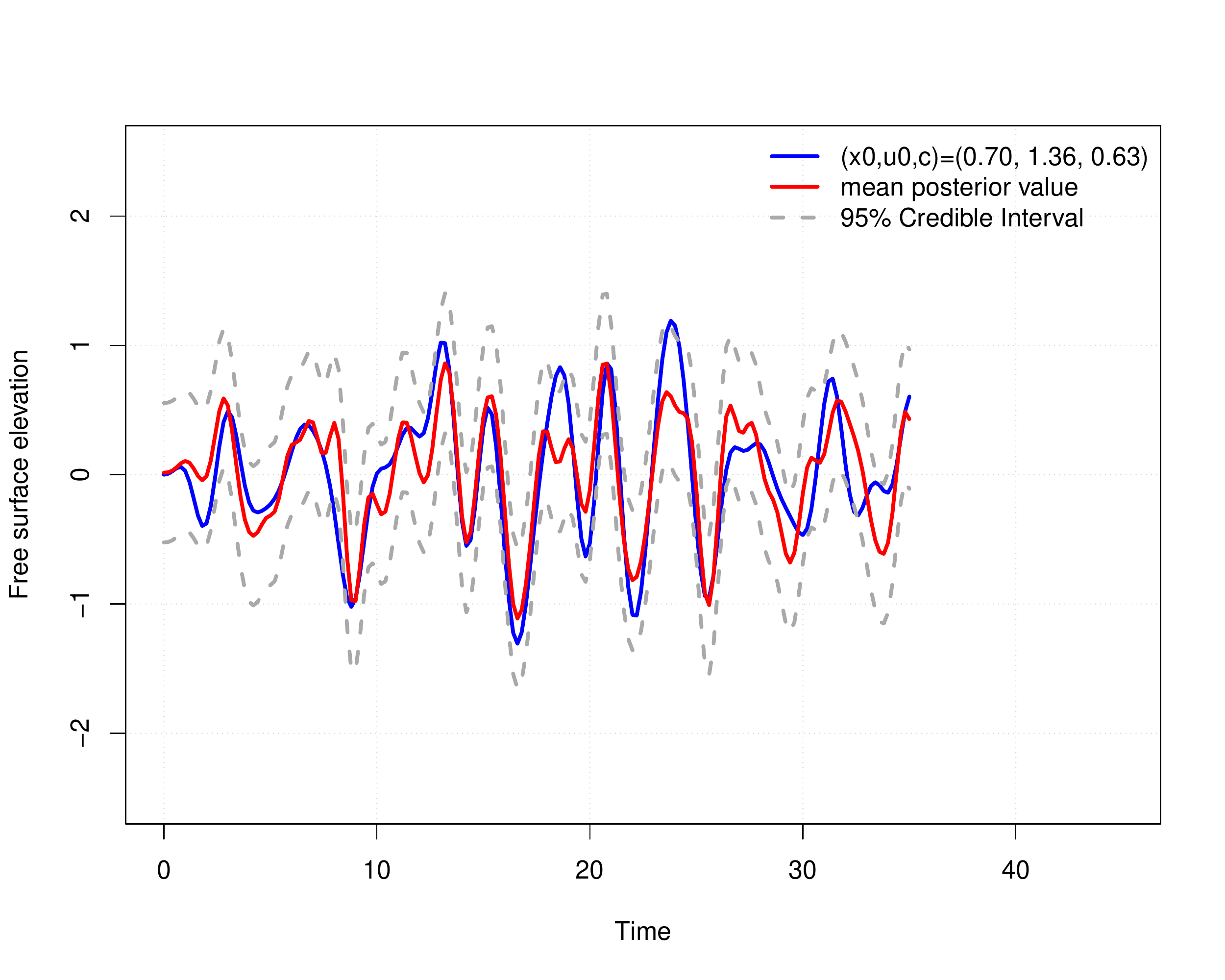}}
\subfloat[$(x_0,u_0,c)=(-0.45,1.77,1.76)$ -- point number 19]{\label{LOO19}\includegraphics[trim=0cm 0cm 0cm 2cm, clip=true,height=0.335\textwidth]{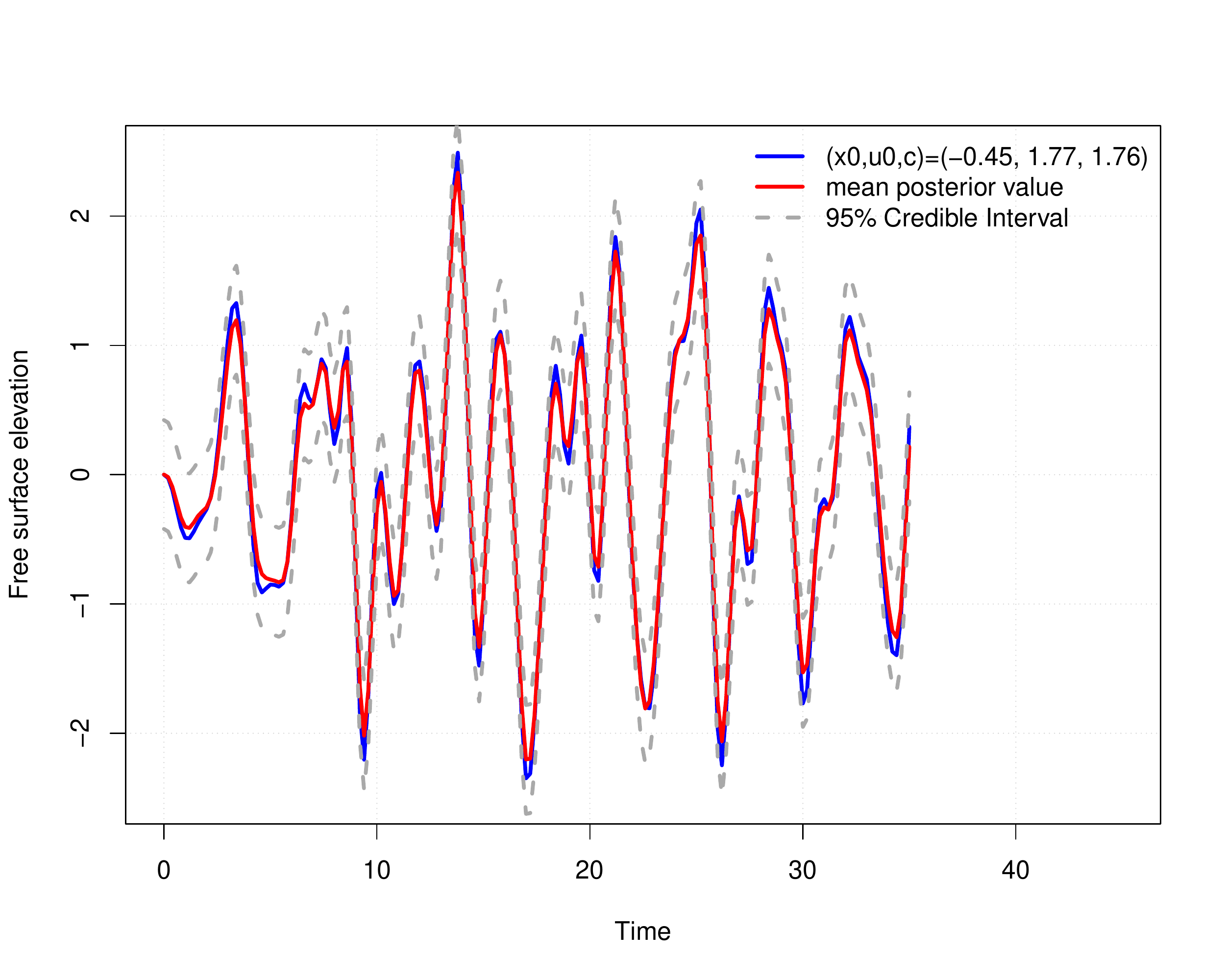}}\\
\subfloat[$(x_0,u_0,c)=(-0.77,1.84,1.49)$ -- point number 24]{\label{LOO24}\includegraphics[trim=0cm 0cm 0cm 2cm, clip=true,height=0.335\textwidth]{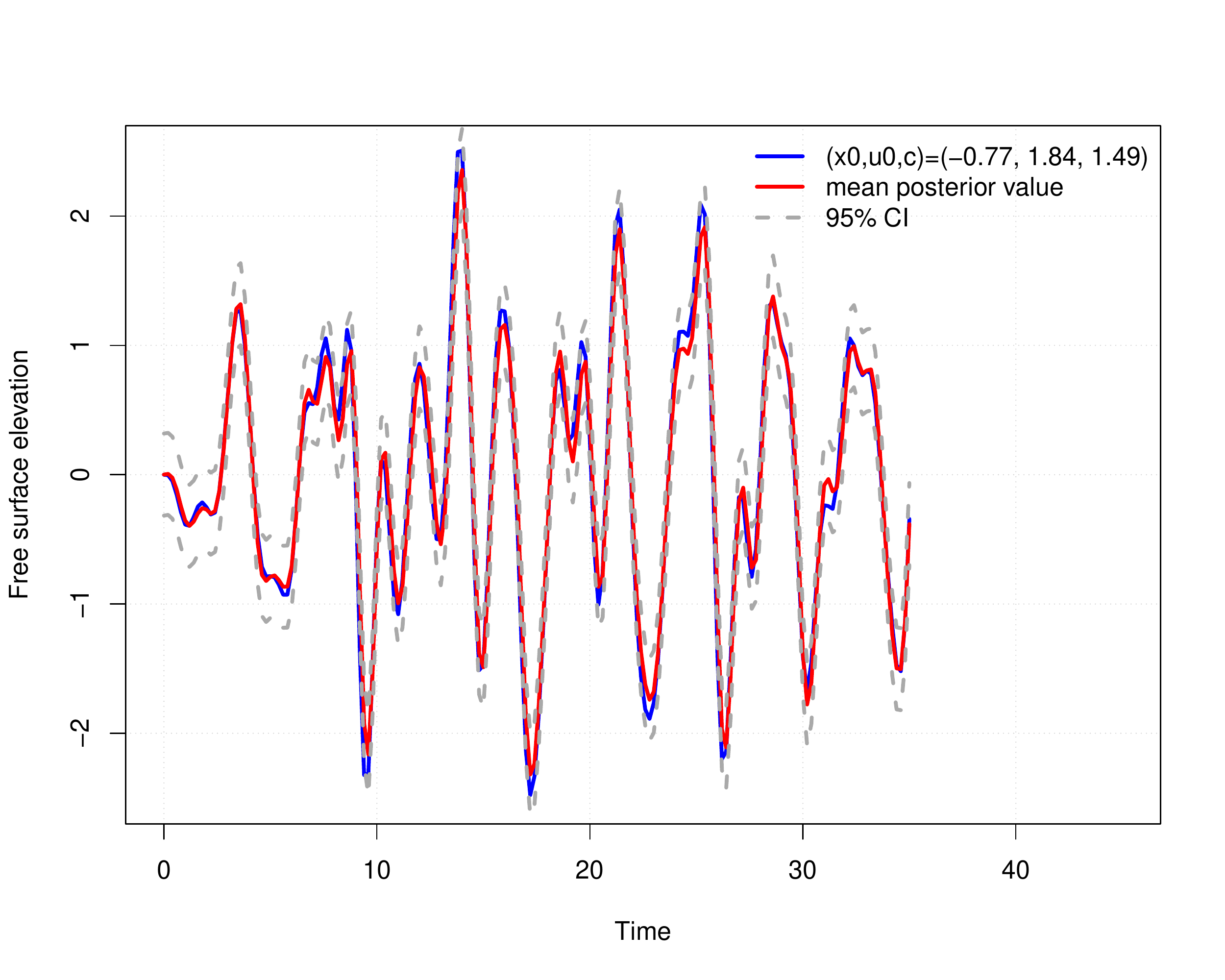}}
\subfloat[$(x_0,u_0,c)=(0.98,1.94,1.06)$ -- point number 25]{\label{LOO25}\includegraphics[trim=0cm 0cm 0cm 2cm, clip=true,height=0.335\textwidth]{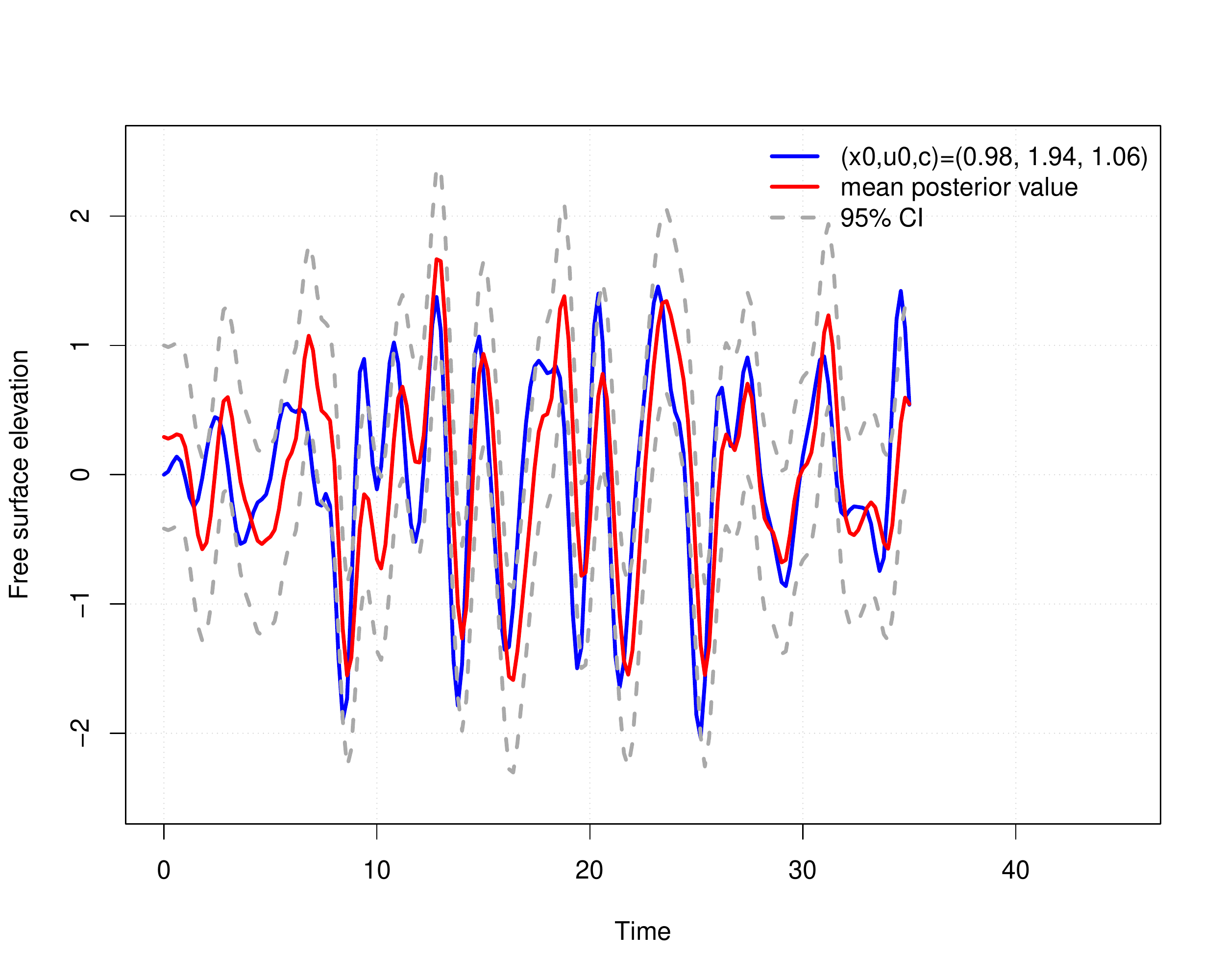}}
\end{center}
\caption{Diagnostic plots for some of the input points looking at $(x,y) = (0,8.38)$. Blue line is the simulator's evaluation, red is the mean value of the posterior distribution and dotted grey is the 95\% credible interval of the posterior distribution.}
\label{diagnostics}
\end{figure*}

\begin{figure}[htb]
\vspace*{2mm}
\ContinuedFloat
\begin{center}
\subfloat[$(x_0,u_0,c)=(-0.94,1.82,1.65)$ -- point number 27]{\label{LOO27}\includegraphics[trim=0cm 0cm 0cm 2cm, clip=true,height=0.335\textwidth]{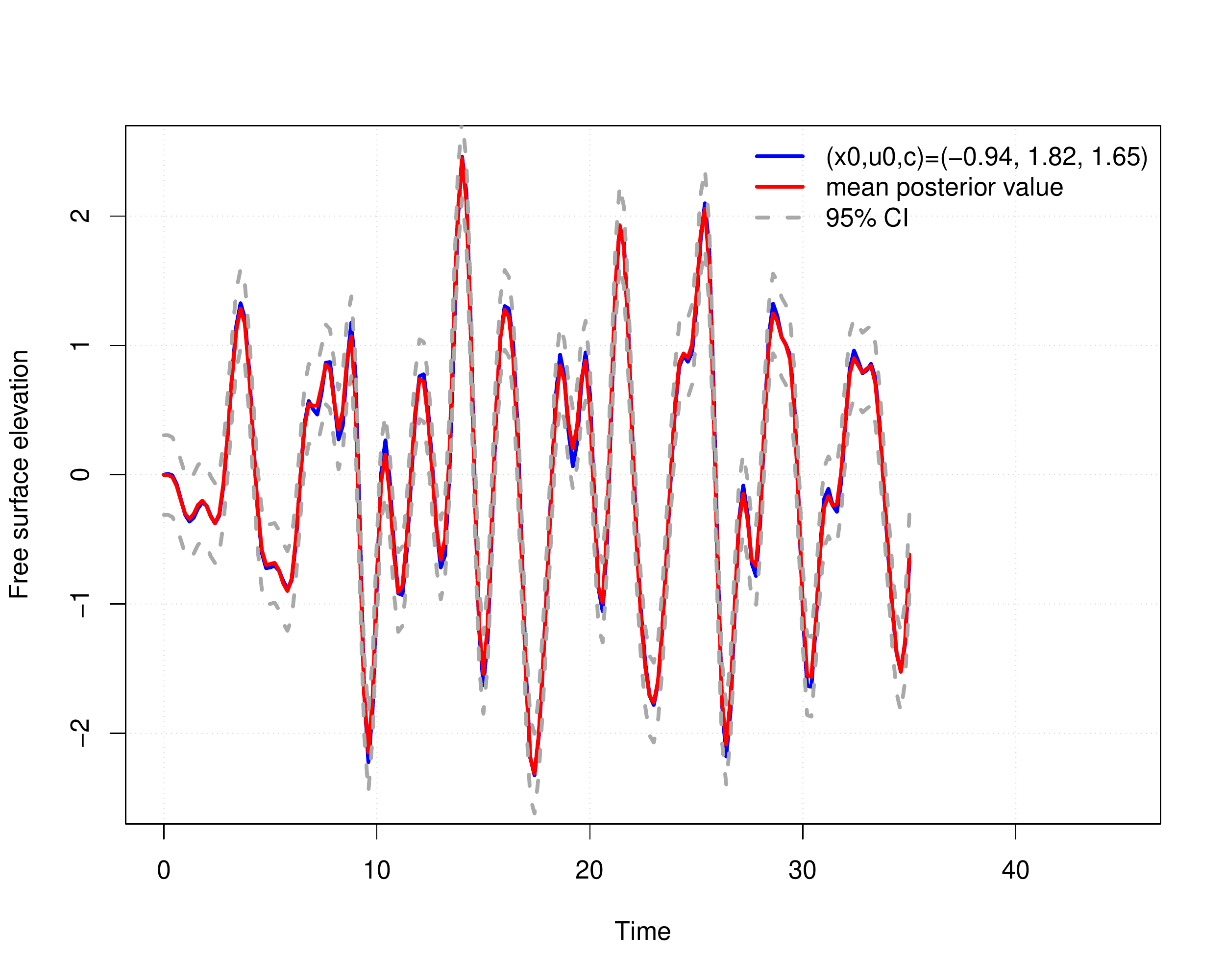}}\\
\subfloat[$(x_0,u_0,c)=(-0.86,1.67,2.15)$ -- point number 36]{\label{LOO36}\includegraphics[trim=0cm 0cm 0cm 2cm, clip=true,height=0.335\textwidth]{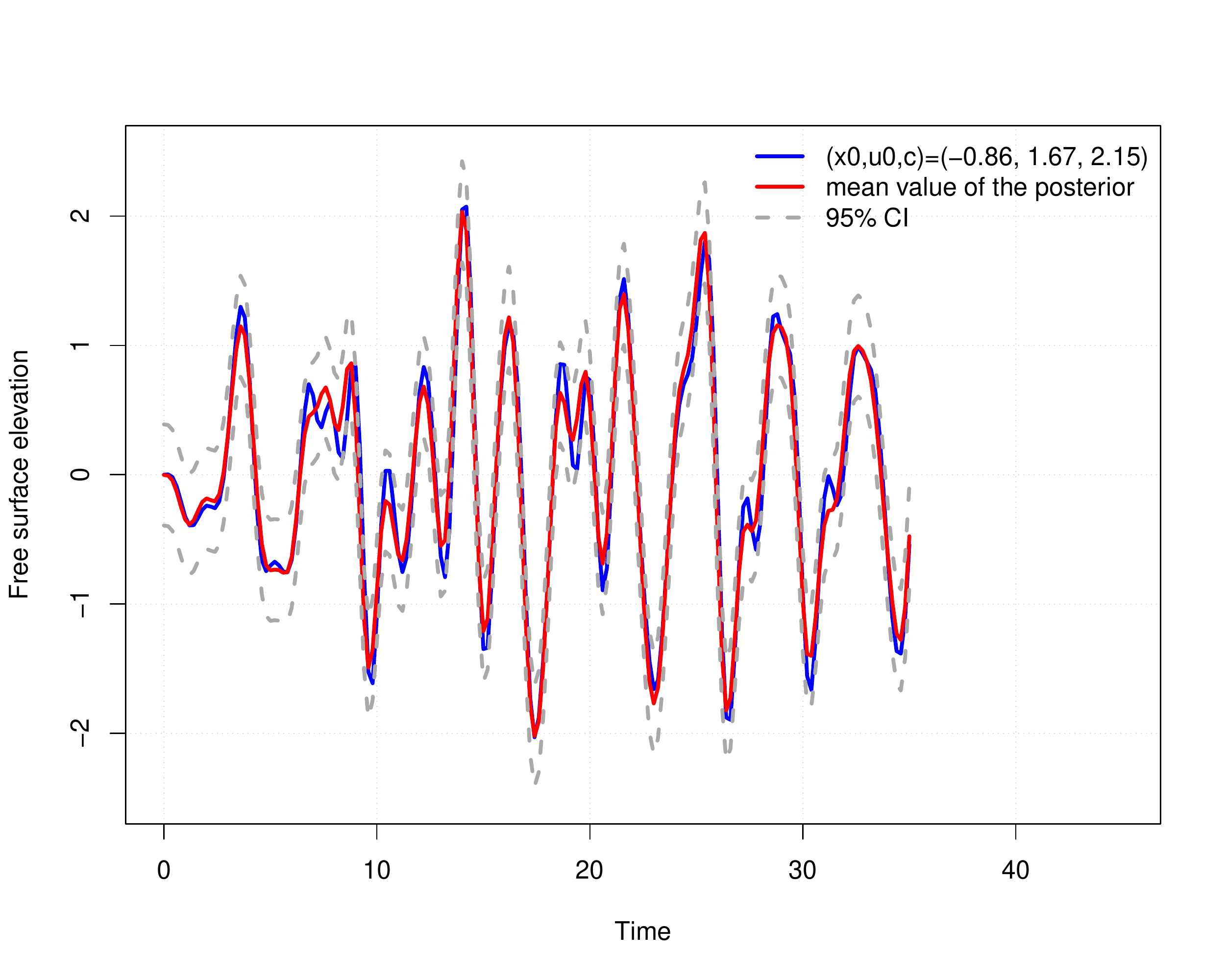}}
\end{center}
\caption{Diagnostic plots for some of the input points looking at $(x,y) = (0,8.38)$. Blue line is the simulator's evaluation, red is the mean value of the posterior distribution and dotted grey is the 95\% credible interval of the posterior distribution.}
\label{diagnostics}
\end{figure}

However, on some of the diagnostic plots, the prediction is not very accurate. One of the fundamental reasons affecting the emulator performance is the position of the point at which we try to predict in the input space. Generally, it is expected to obtain more accurate predictions in the cases where the points at which we try to predict are surrounded closely by other design points, compared to the cases where the points are located in a sparsely covered region, since more information can be obtained by the neighbouring points. The behaviour at each point is significantly linked to the behaviour at the points close to it and this influence decays rapidly with the distance separating the two points. To quantify this, the Euclidean distances in the three-dimensional input space between a point and the other 39 points are obtained. Then the mean values of these distances (MED) for each of the 40 input points are calculated:
\begin{equation}
MED = \frac{ \sum_{i=1}^{39} \sqrt {(x_1-x_2)^2 +(u_1 - u_2)^2 + (c_1 - c_2)^2}}{39}
\end{equation}

\begin{figure}[htb]
\vspace*{2mm}
\begin{center}
\subfloat{\includegraphics[trim=4cm 8cm 2.5cm 9cm, clip=true,height=0.345\textwidth]{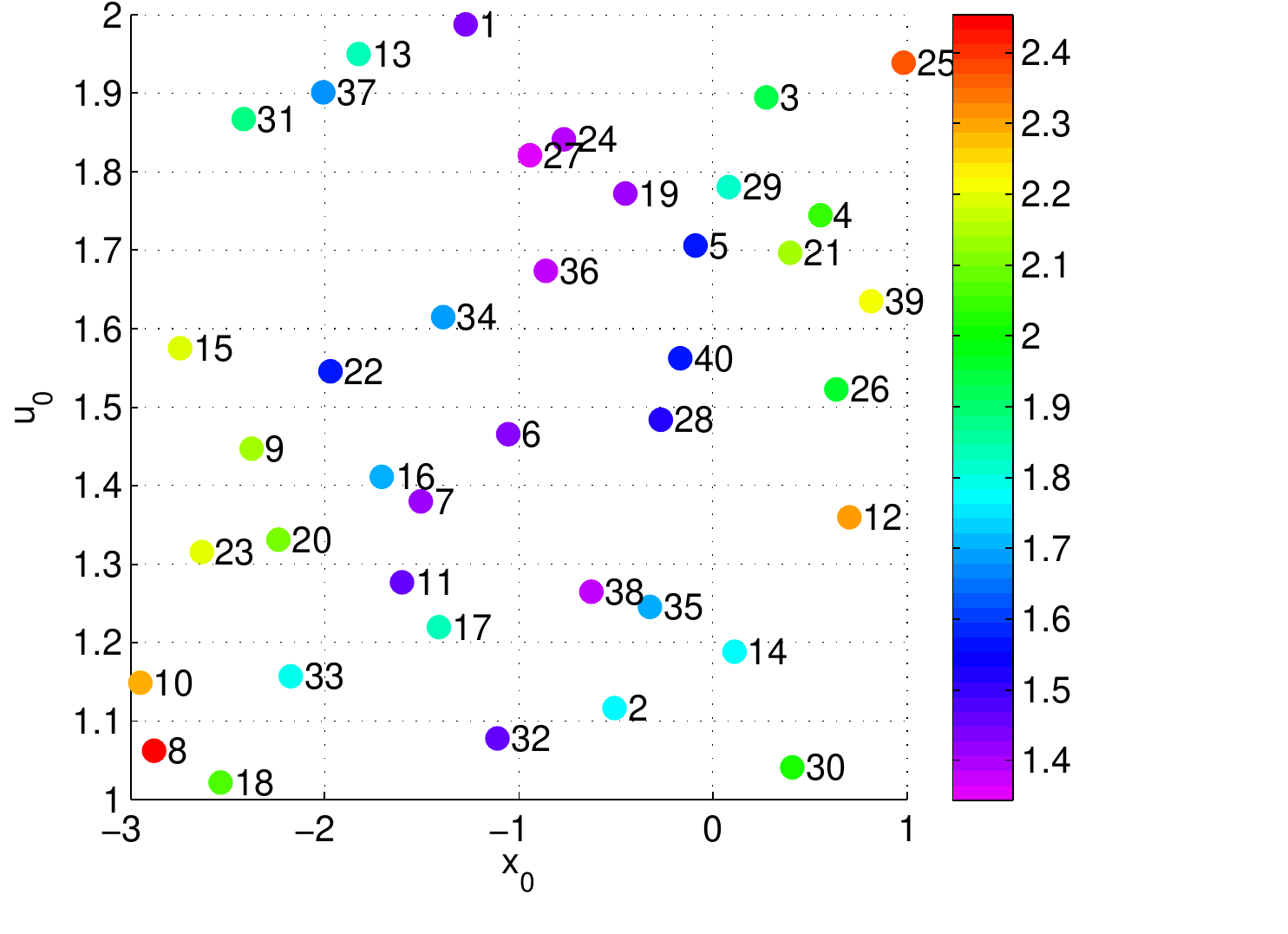}}\\
\subfloat{\includegraphics[trim=4cm 8cm 2.5cm 9cm, clip=true,height=0.345\textwidth]{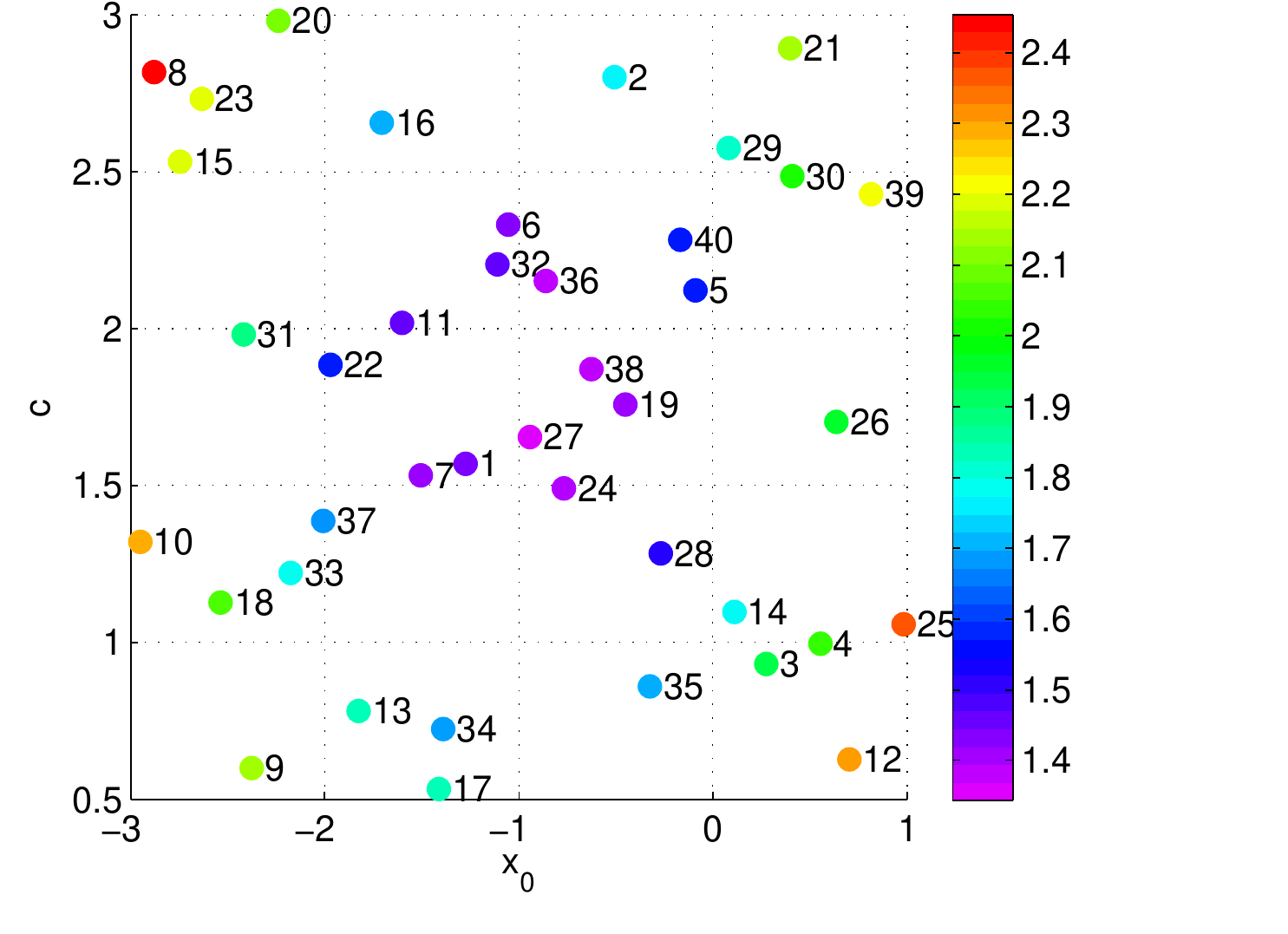}}\\
\subfloat{\includegraphics[trim=4cm 8cm 2.5cm 9cm, clip=true,height=0.345\textwidth]{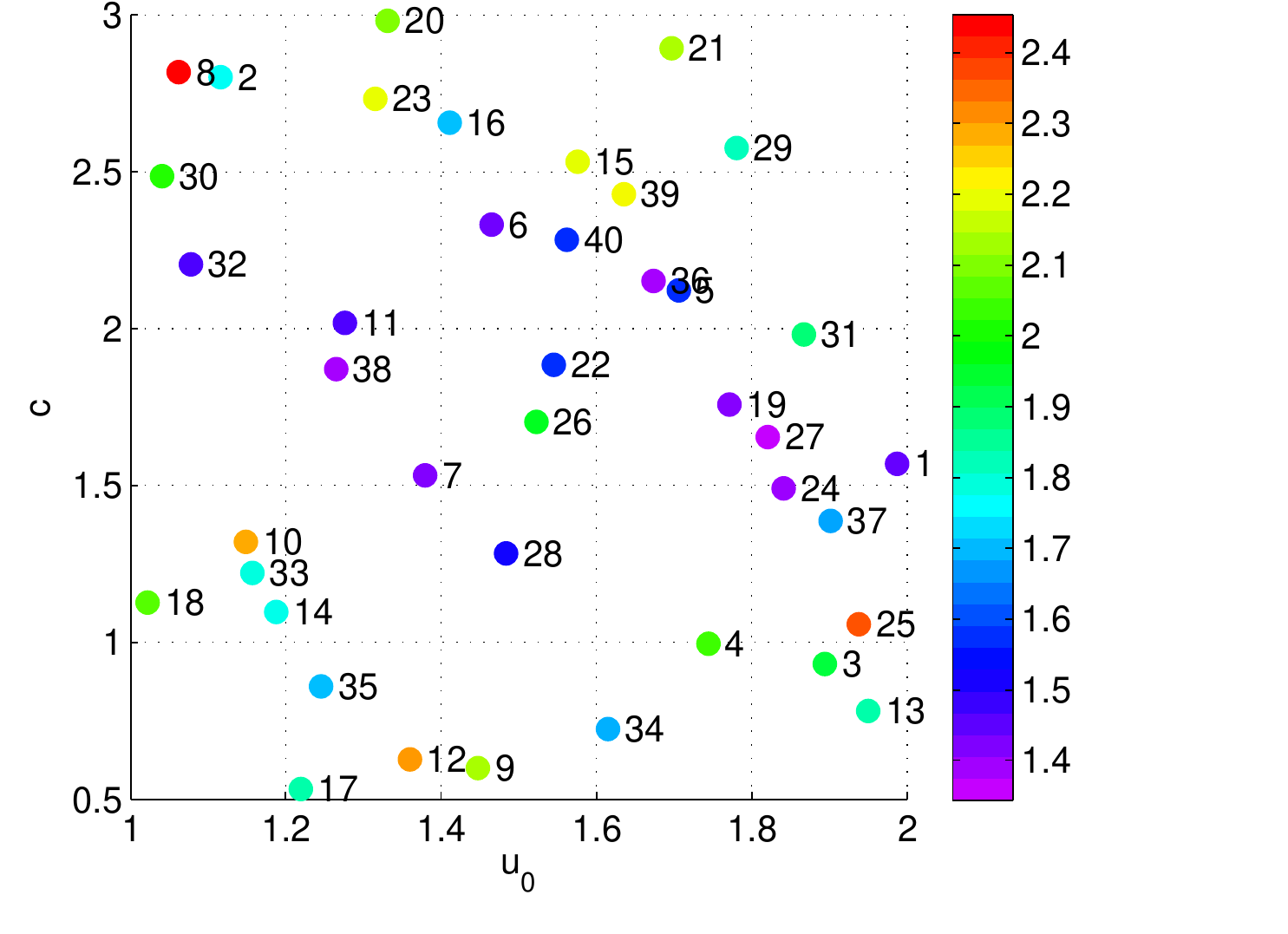}}
\end{center}
\caption{Euclidean distance between each of the points and the other 39 points in the three-dimensional parameter space.}
\label{ed}
\end{figure}

Figure \ref{ed} displays the mean Euclidean distances for all the design input points. We can see that the points 8, 10, 12 and 25 show a large MED from the rest of the 39 points. Looking at the LOO diagnostics of these four points in Figs \ref{LOO8}, \ref{LOO10}, \ref{LOO12}, \ref{LOO25}, we can easily observe that the predictions are not very accurate. However, the maximum wave elevation, which is the most important measurement, is still satisfactory and almost everywhere the simulator evaluation lines are within the 95\% credible intervals. This indicates that, even for the design points that are isolated from the neighbouring points, the emulator predictions are still usable. 

On the other hand, points such as 19, 24, 27 and 36 are affected significantly by the other points, separated by small Euclidean distances from the rest of the 39 points in space. Looking at the diagnostic plots of these points (Figs \ref{LOO19}, \ref{LOO24}, \ref{LOO27}, \ref{LOO36}), it is obvious that the emulator does an excellent job in prediction, since all the features of the wave are predicted accurately by the emulator. 

Two measures that can be used to quantify the emulator's accuracy are the mean credible interval length (MCIL) and the root mean square error (RMSE) between the observed and the predicted evaluations at each of the 40 input points. The RMSE is given by the equation
\begin{equation}
RMSE= \sqrt {\frac{\sum_{i=1}^{n}(\hat{x_{i}}-x_{i})^2}{n}}
\end{equation}
where $x_{i}$ and $\hat{x_{i}}$ are the observed and predicted values at each time step $i$, respectively, and $n$ is the number of time steps.

\begin{figure}[htb]
\vspace*{2mm}
\begin{center}
\includegraphics[trim=0cm 0cm 0cm 2cm, clip=true,height=0.22\textwidth]{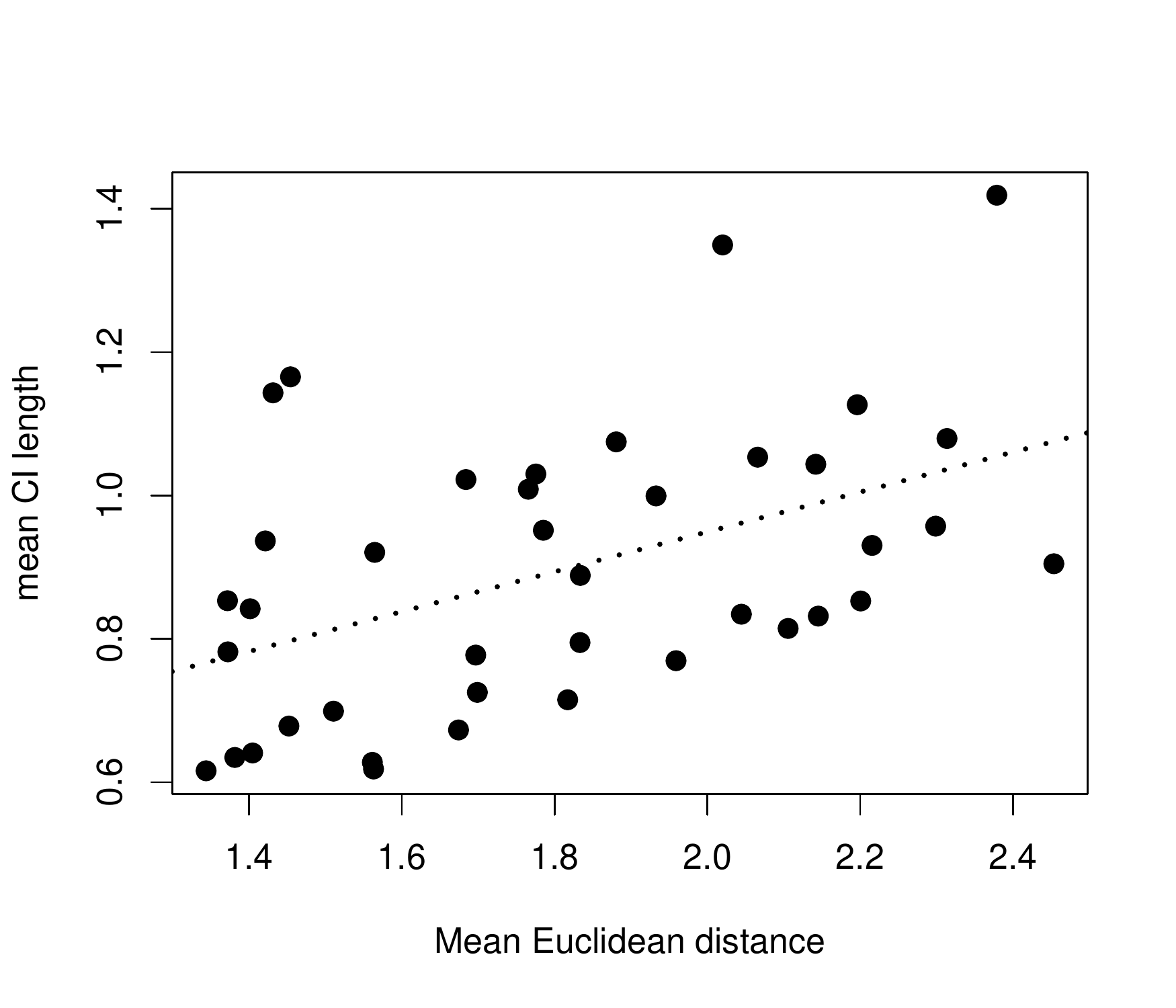}
\end{center}
\caption{Mean Euclidean distance vs. mean 95\% credible interval length for the location $(x,y) = (0,8.38)$, where the dotted line is the linear regression.}
\label{CI}
\end{figure}

\begin{figure}[htb]
\vspace*{2mm}
\begin{center}
\includegraphics[trim=0cm 0cm 0cm 2cm, clip=true,height=0.22\textwidth]{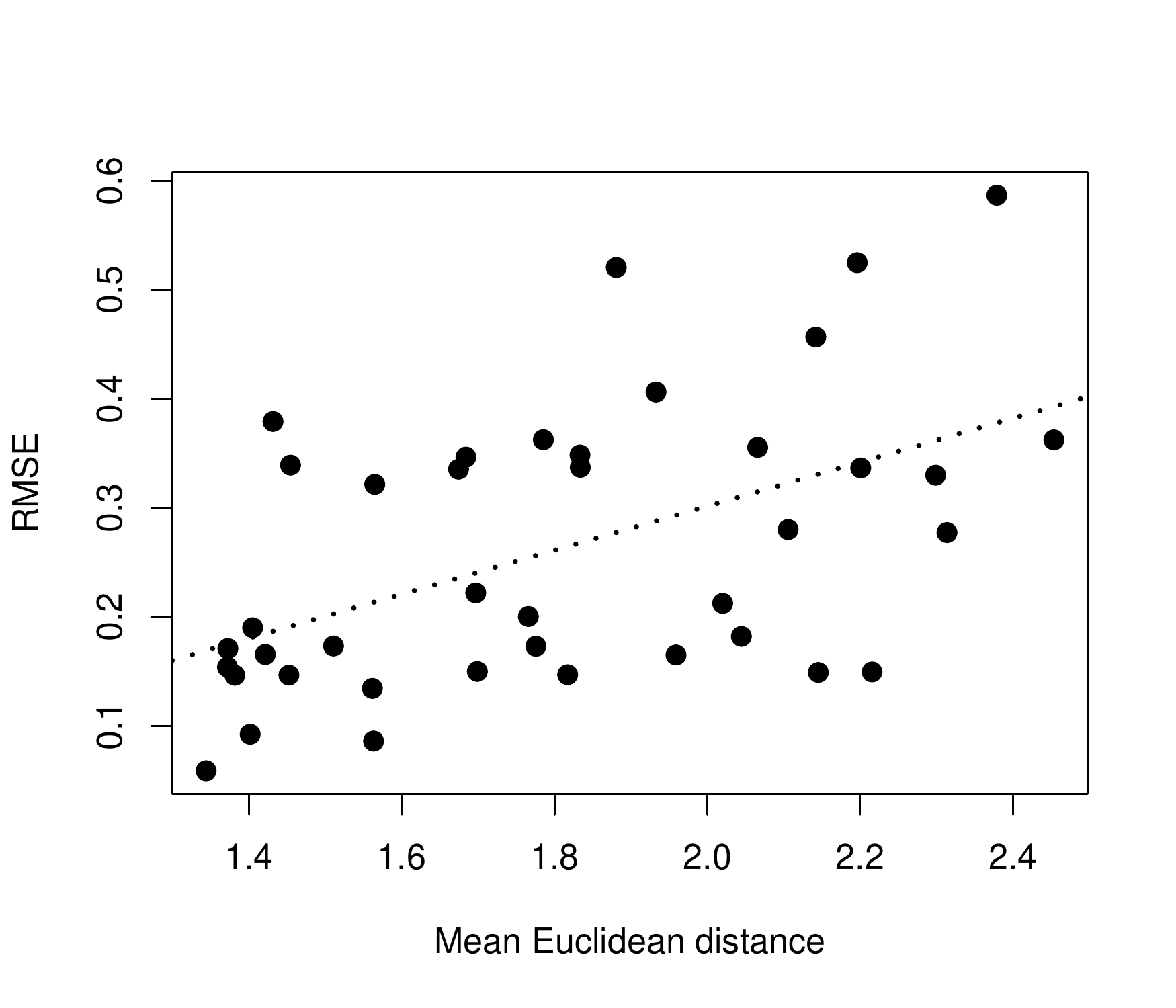}-
\end{center}
\caption{Mean Euclidean distance vs. RMSE for the location $(x,y) = (0,8.38)$, where the dotted line is the linear regression.}
\label{RMSD}
\end{figure}

Figures \ref{CI} and \ref{RMSD} display the MCIL and the RMSE versus MED, respectively, for all the input points, looking at the case of the location $(x,y) = (0,8.38)$. We observe a positive correlation between the MED and both the MCIL and the RMSE. Therefore, this confirms that the distance separating the points in space is a fundamental factor that affects the predictive power of the emulator and hence this highlights the importance of a good experimental design. This positive correlation is also satisfied for the other locations examined. 

\begin{figure}[htb]
\vspace*{2mm}
\begin{center}
\includegraphics[trim=0cm 0cm 1cm 2.5cm, clip=true,height=0.35\textwidth]{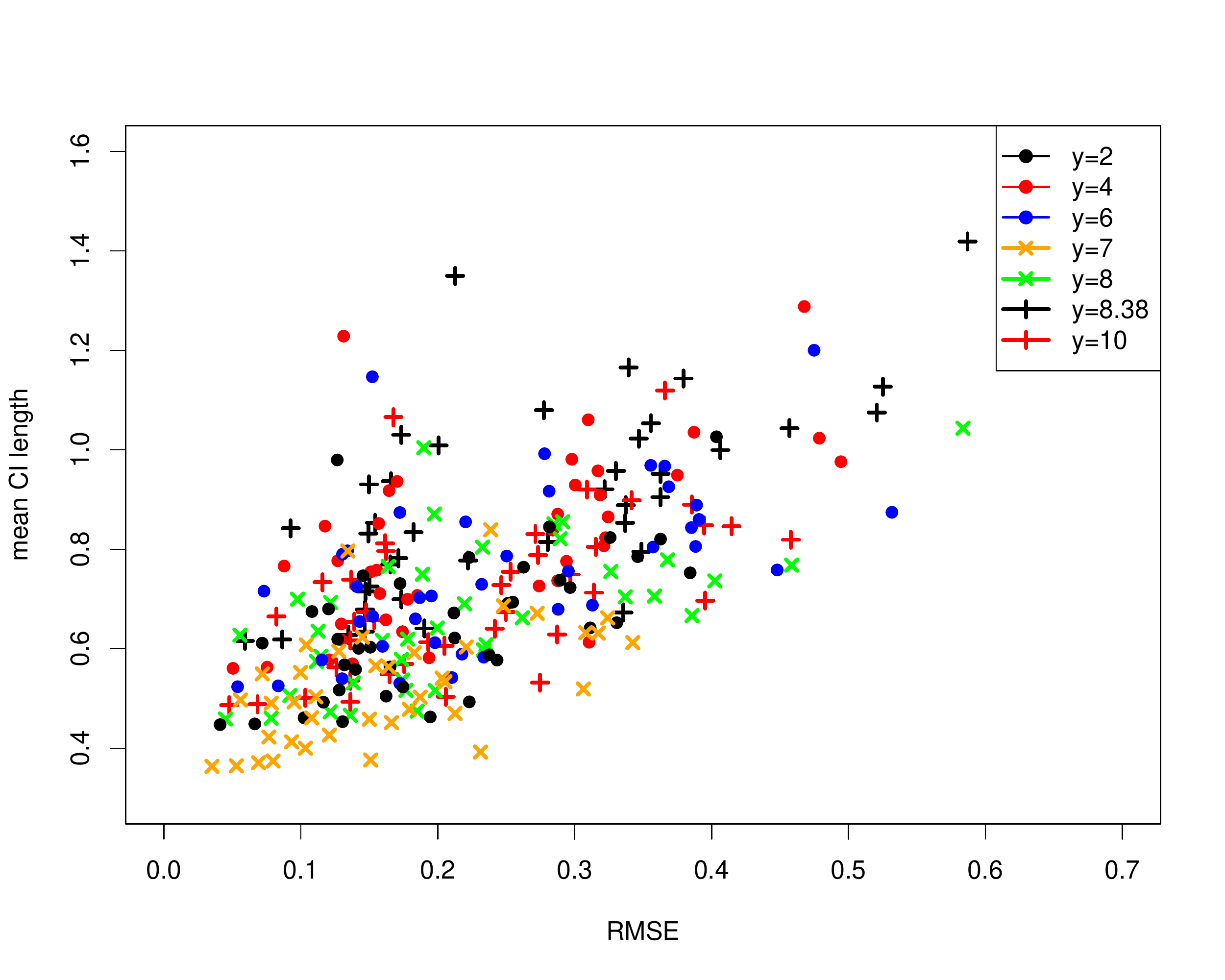}
\end{center}
\caption{Root Mean Square Error vs. mean CI length. Different types and colors represent different locations along the shoreline.}
\label{meanCI_RMSE}
\end{figure}

In Fig. \ref{meanCI_RMSE}, the RMSE with respect to MCIL is presented for all the 40 diagnostics for the seven locations along the shoreline investigated, in order to compare the emulator's performance when applied to different locations. A combination of both small RMSE and MCIL is desirable, indicating both small error and small uncertainty in emulator's predictions. The figure clearly shows that the emulator performs similarly for all the locations investigated. Therefore, the emulator can be applied to different locations along the shoreline, resulting in accurate enough representations of the simulator output. The reasons that we have slightly better predictions at some locations compared to others is an area of further investigation. Nevertheless, the location along the shoreline with $y=8.38$ shows the worst results in this Figure. Therefore, the predictions of the emulator for the other locations are better than the ones given in Fig.  \ref{diagnostics}. This reinforces the confidence we have in our emulator.

\section{Sensitivity and Uncertainty Analyses}

In Section 5 we have presented the process to create a statistical emulator that can predict the simulator's output with sufficient accuracy, for a number of different locations along the shoreline. Therefore, the emulator can be used in place of the expensive-to-run simulator to efficiently perform analyses that require a large number of evaluations, in order to save time without sacrificing accuracy. In this Section, we demonstrate a sensitivity and uncertainty analyses using the emulator. 

\subsection{Sensitivity analysis}
The statistical emulator is used to carry out a sensitivity analysis of the model, where we investigate how sensitive is the maximum wave elevation for $t\leq35$ to changes in inputs. Additionally, we examine whether the individual locations along the shoreline present consistent sensitivity to inputs' variation.

\begin{figure}[htb]
\vspace*{2mm}
\begin{center}
\subfloat[max. elevation w.r.t. initial position for two different speeds and spread ratios]{\label{x0u0}\includegraphics[trim=0cm 0.5cm 1cm 2.5cm, clip=true,height=0.319\textwidth]{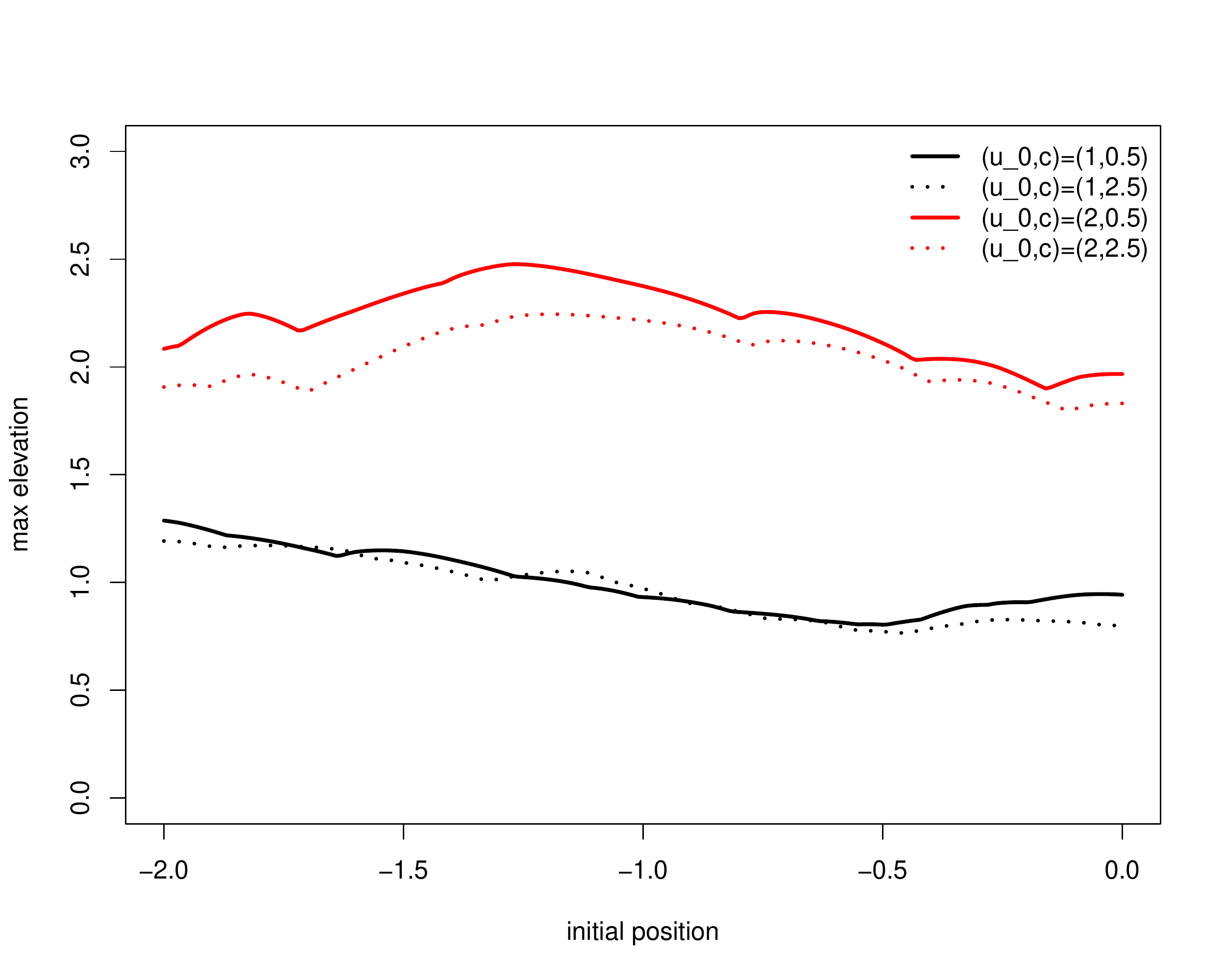}}\\
\subfloat[max. elevation w.r.t. speed for two different initial positions and spread ratios]{\label{x0c}\includegraphics[trim=0cm 0.5cm 1cm 2.5cm, clip=true,height=0.319\textwidth]{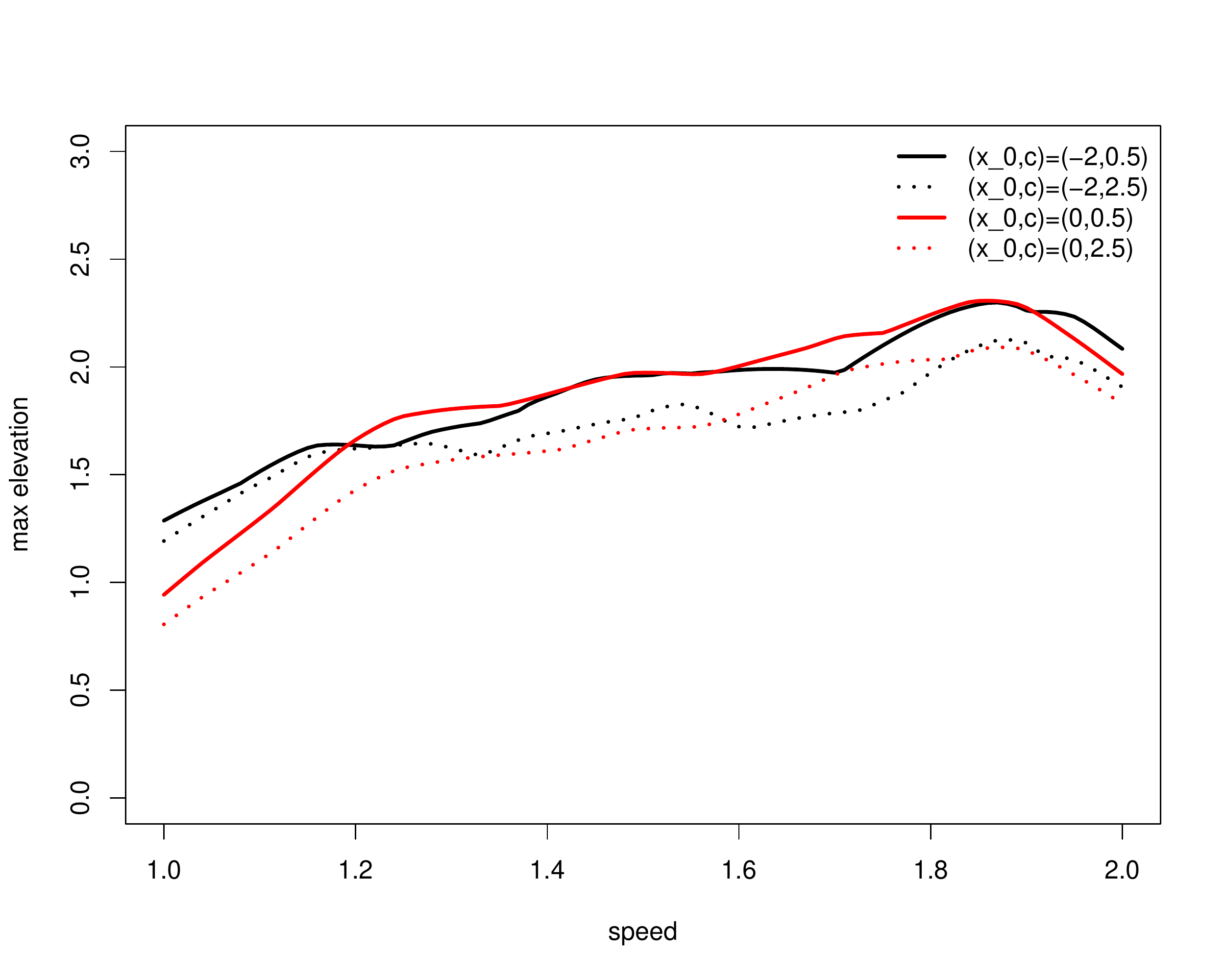}}\\
\subfloat[max. elevation w.r.t. spread ratio for two different initial positions and speeds]{\label{u0c}\includegraphics[trim=0cm 0.5cm 1cm 2.5cm, clip=true,height=0.319\textwidth]{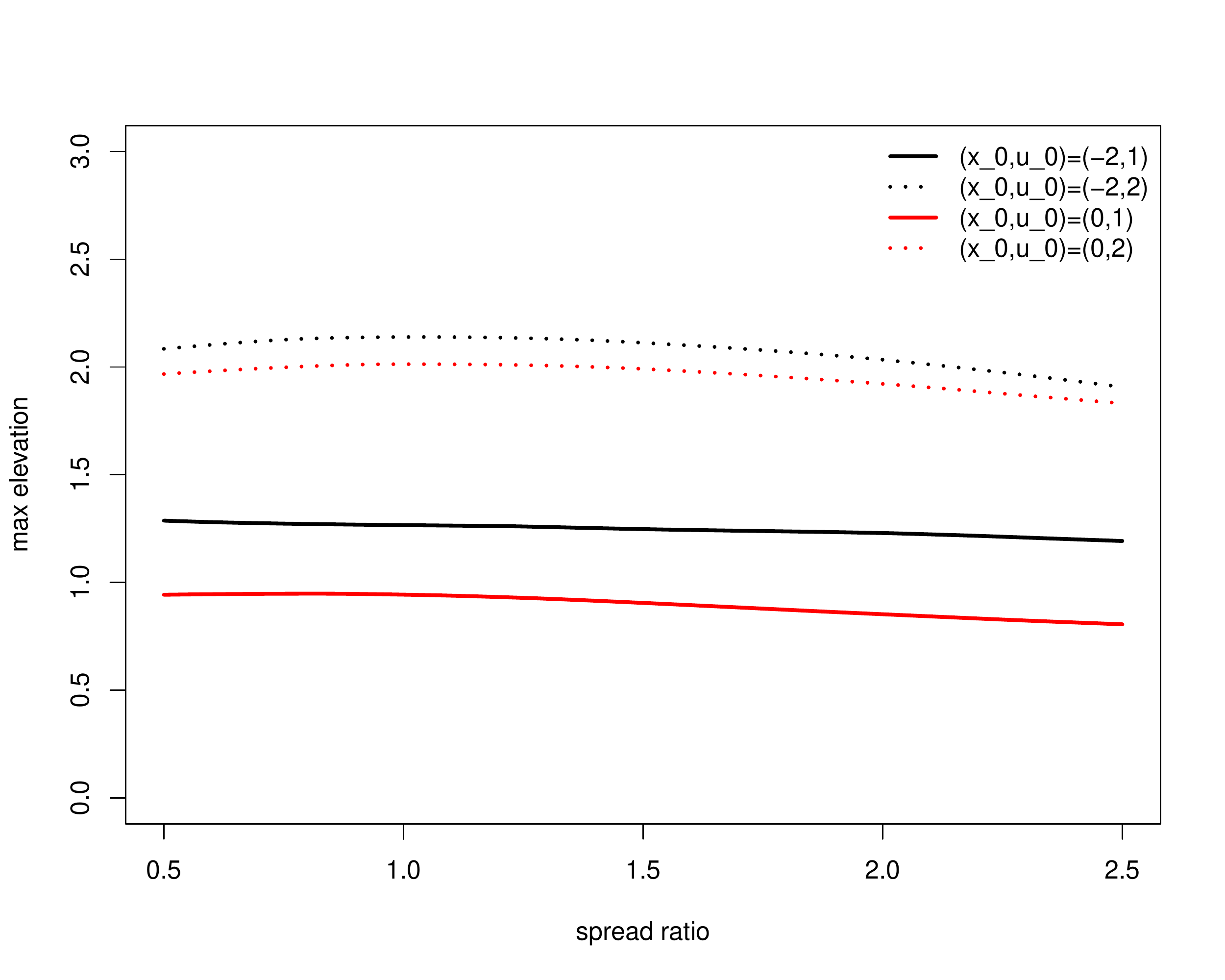}}
\end{center}
\caption{Maximum sea free-surface elevation with respect to (a) initial position, (b) speed and (c) shape, for the time interval $[0,35]$ and position $(x,y) = (0,8.38)$.}
\label{x0u0c}
\end{figure}

Fig. \ref{x0u0c} displays the case for the location $(x,y) = (0,8.38)$. In each of the three plots, the maximum elevation  is plotted against the initial position $x_0$, speed $u_0$ and spread ratio $c$ of the landslide, respectively, with the other two input parameters being kept constant. To ensure maximum emulator's accuracy and keep RMSE to the minimum, the input domain in sensitivity analysis is chosen to be the subset of the whole domain where the mean Euclidean distance between the points are small as presented in Fig. \ref{ed}. Specifically, we consider $x_0\in[-2,0]$, $u_0\in[1,2]$ and $c\in[0.5,2.5]$.

From Fig. \ref{x0u0} we can see an obvious relationship between the landslide's speed and the maximum elevation. Specifically, a landslide with a larger $u_0$ gives larger maximum sea free-surface elevations. No strong dependency of the maximum elevation on initial position and spread ratio can be observed. Figure \ref{x0c} highlights the positive relationship between $u_0$ and the maximum elevation, with the larger the $u_0$, the larger the maximum elevation. Finally, Fig. \ref{u0c} shows that a landslide initiating from a subaerial position shows larger maximum sea free-surface elevations compared to a landslide starting from the origin. So, a relationship between the $x_0$ value and the maximum elevation is indicated. Also, a landslide moving with a larger speed yields larger maximum elevations. Moreover, we cannot say that the spread ratio is a significant factor at the specific range investigated. The same conclusions result by repeating the sensitivity analysis for the other six locations. We could easily perform similar analyses in which the output is another important aspect of the tsunami, different from the maximum elevation.

\begin{figure}[htb]
\vspace*{2mm}
\begin{center}
\subfloat[]{\includegraphics[trim=0.2cm 0.5cm 1cm 2.5cm, clip=true,height=0.1817\textwidth]{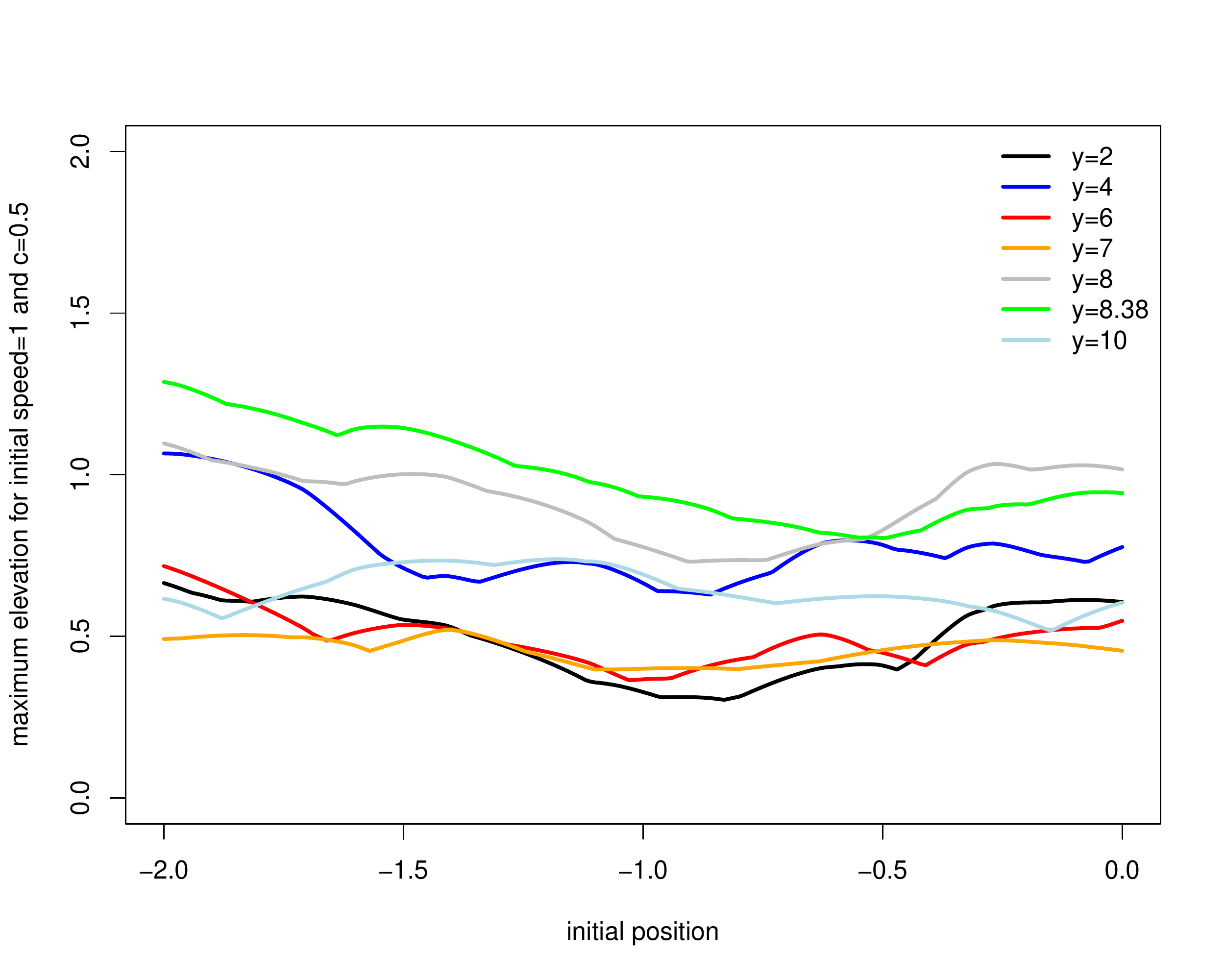}}
\subfloat[]{\includegraphics[trim=0.2cm 0.5cm 1cm 2.5cm, clip=true,height=0.1817\textwidth]{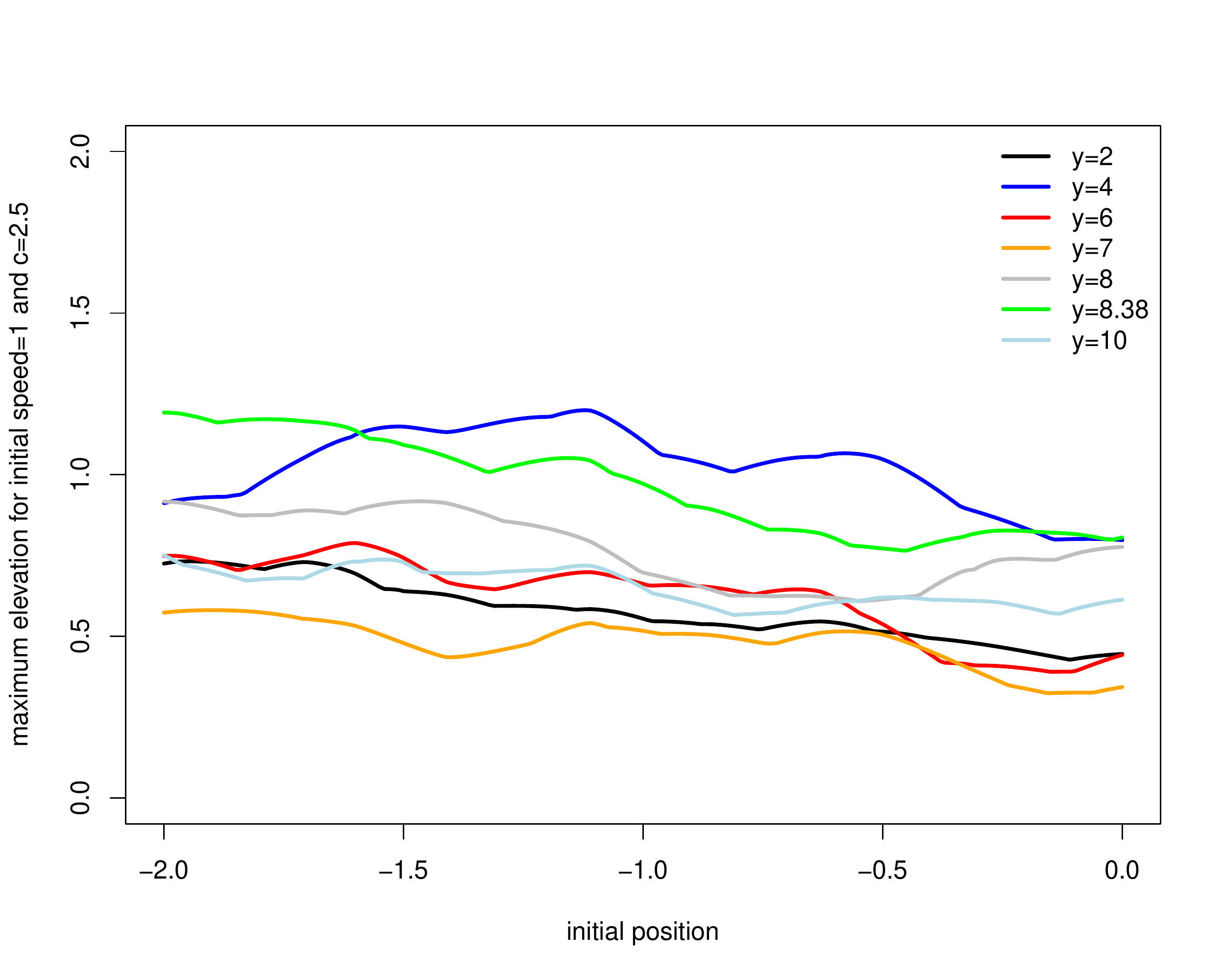}}\\
\subfloat[]{\includegraphics[trim=0.2cm 0.5cm 1cm 2.5cm, clip=true,height=0.1817\textwidth]{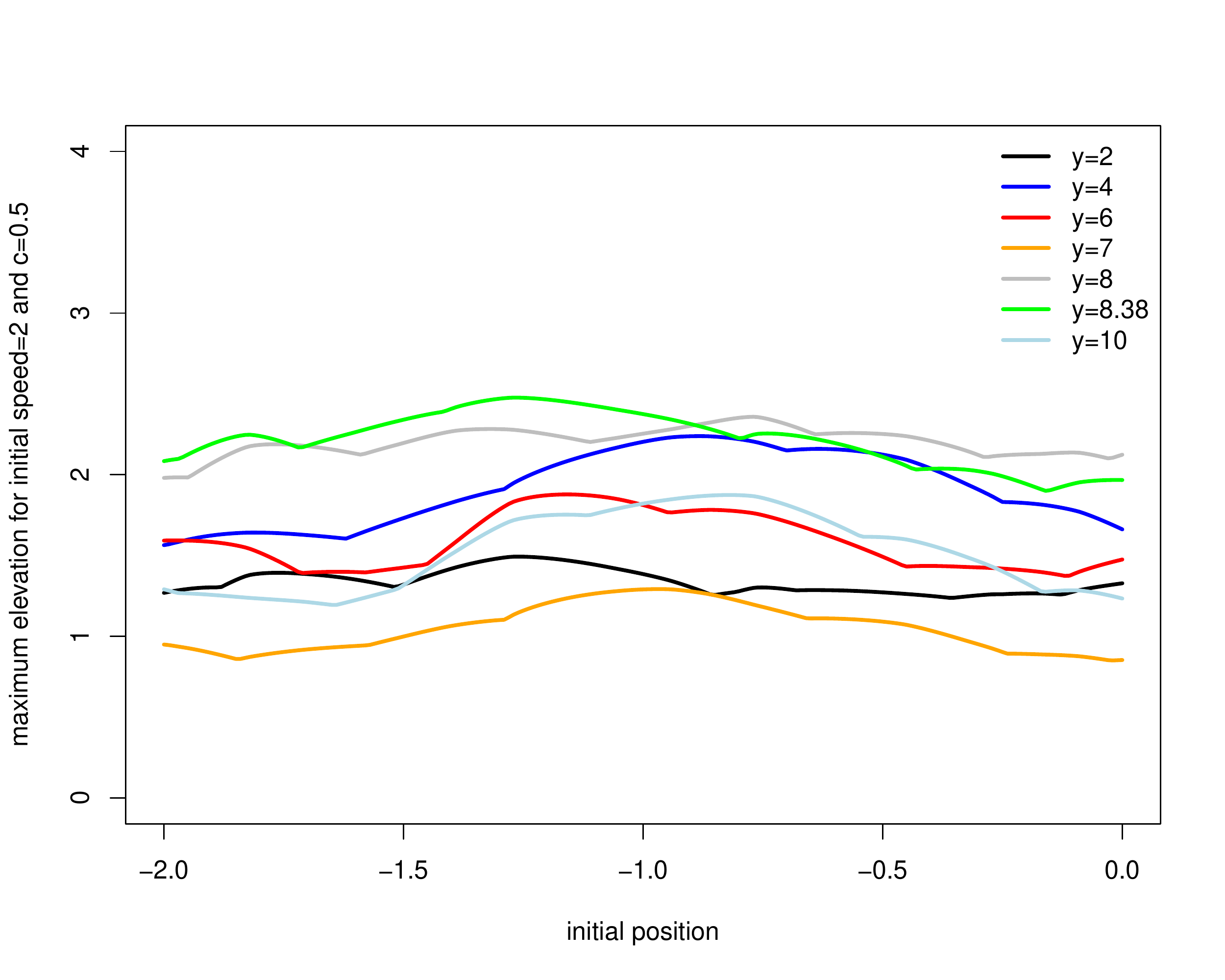}}
\subfloat[]{\includegraphics[trim=0.2cm 0.5cm 1cm 2.5cm, clip=true,height=0.1817\textwidth]{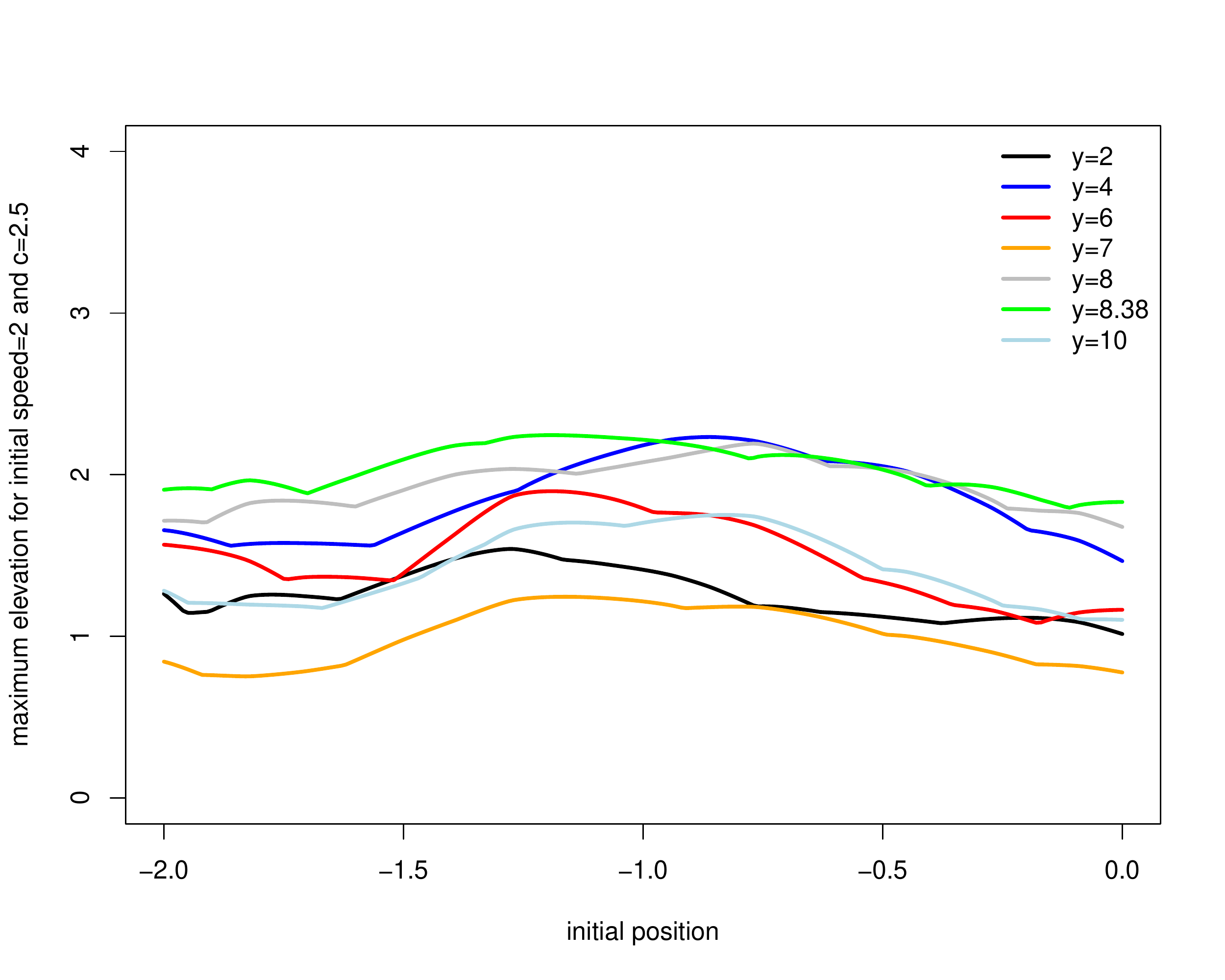}}
\end{center}
\caption{Maximum sea free-surface elevation with respect to initial position for (a) $(u_0,c) = (1,0.5)$, (b) $(u_0,c) = (1,2.5)$, (c) $(u_0,c) = (2,0.5)$ and (d) $(u_0,c) = (2,2.5)$, for the time interval $[0,35]$.}
\label{max_x0_4}
\end{figure}

\begin{figure}[htb]
\vspace*{2mm}
\begin{center}
\subfloat[]{\includegraphics[trim=0.2cm 0.5cm 1cm 2.5cm, clip=true,height=0.1817\textwidth]{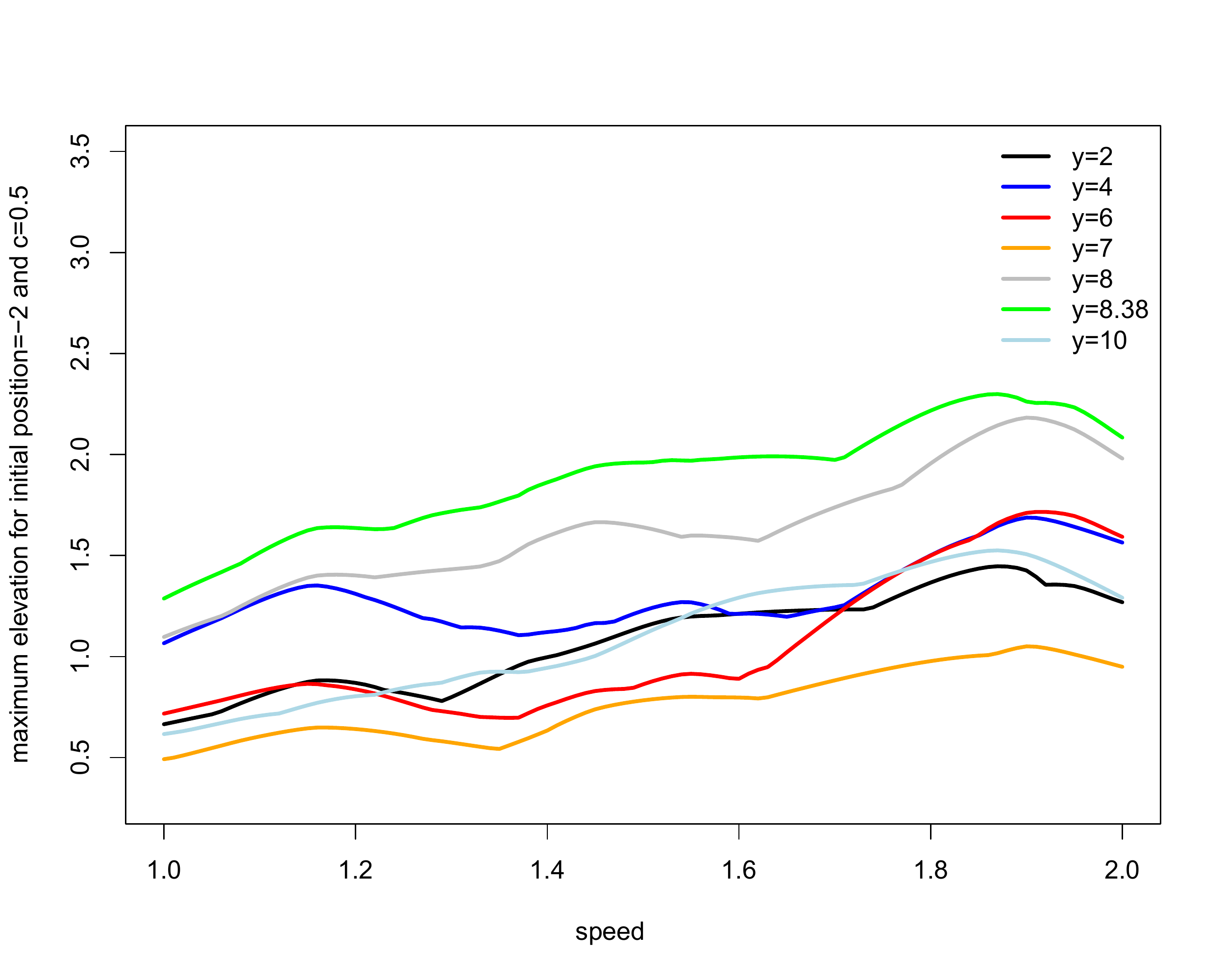}}
\subfloat[]{\includegraphics[trim=0.2cm 0.5cm 1cm 2.5cm, clip=true,height=0.1817\textwidth]{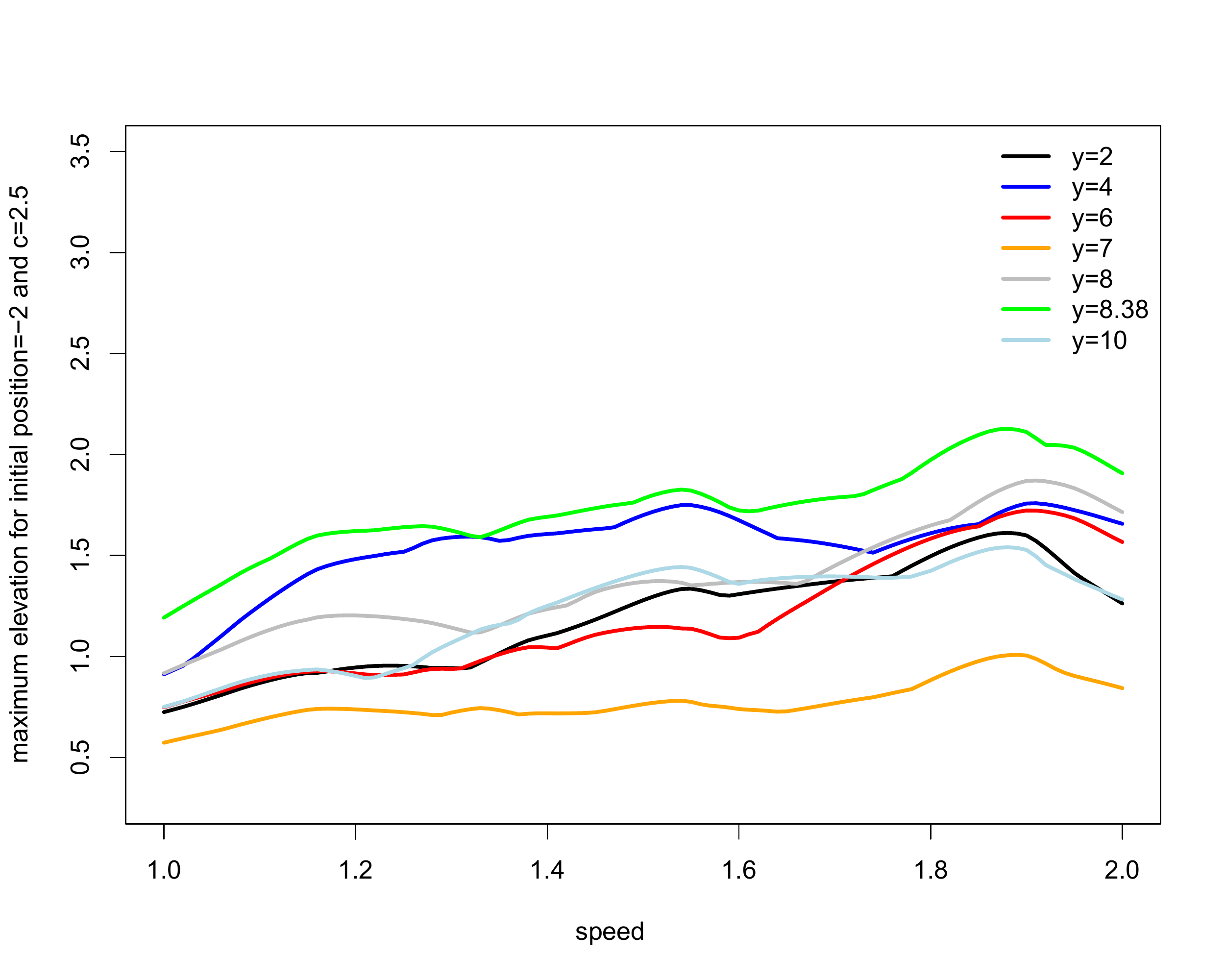}}\\
\subfloat[]{\includegraphics[trim=0.2cm 0.5cm 1cm 2.5cm, clip=true,height=0.1817\textwidth]{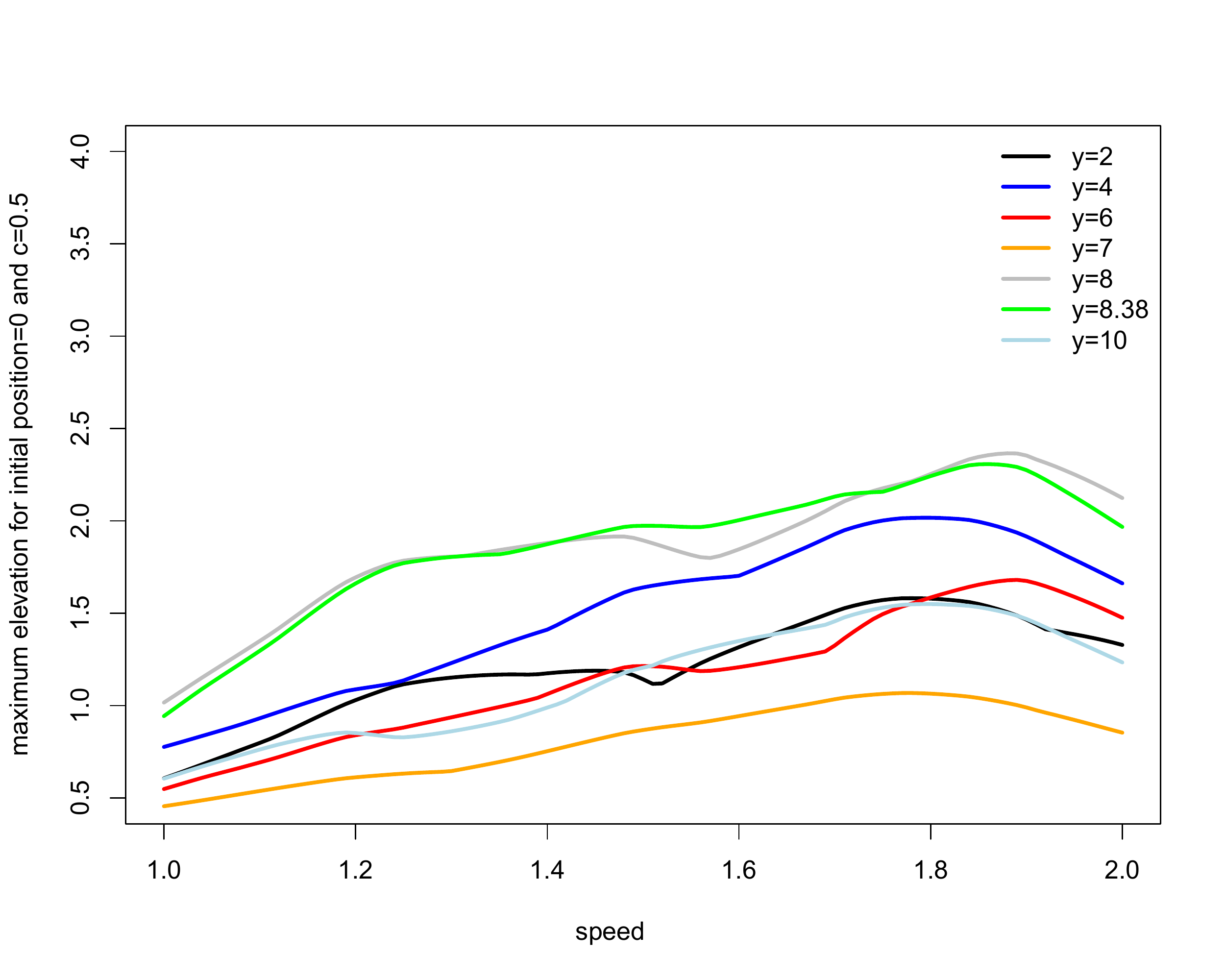}}
\subfloat[]{\includegraphics[trim=0.2cm 0.5cm 1cm 2.5cm, clip=true,height=0.1817\textwidth]{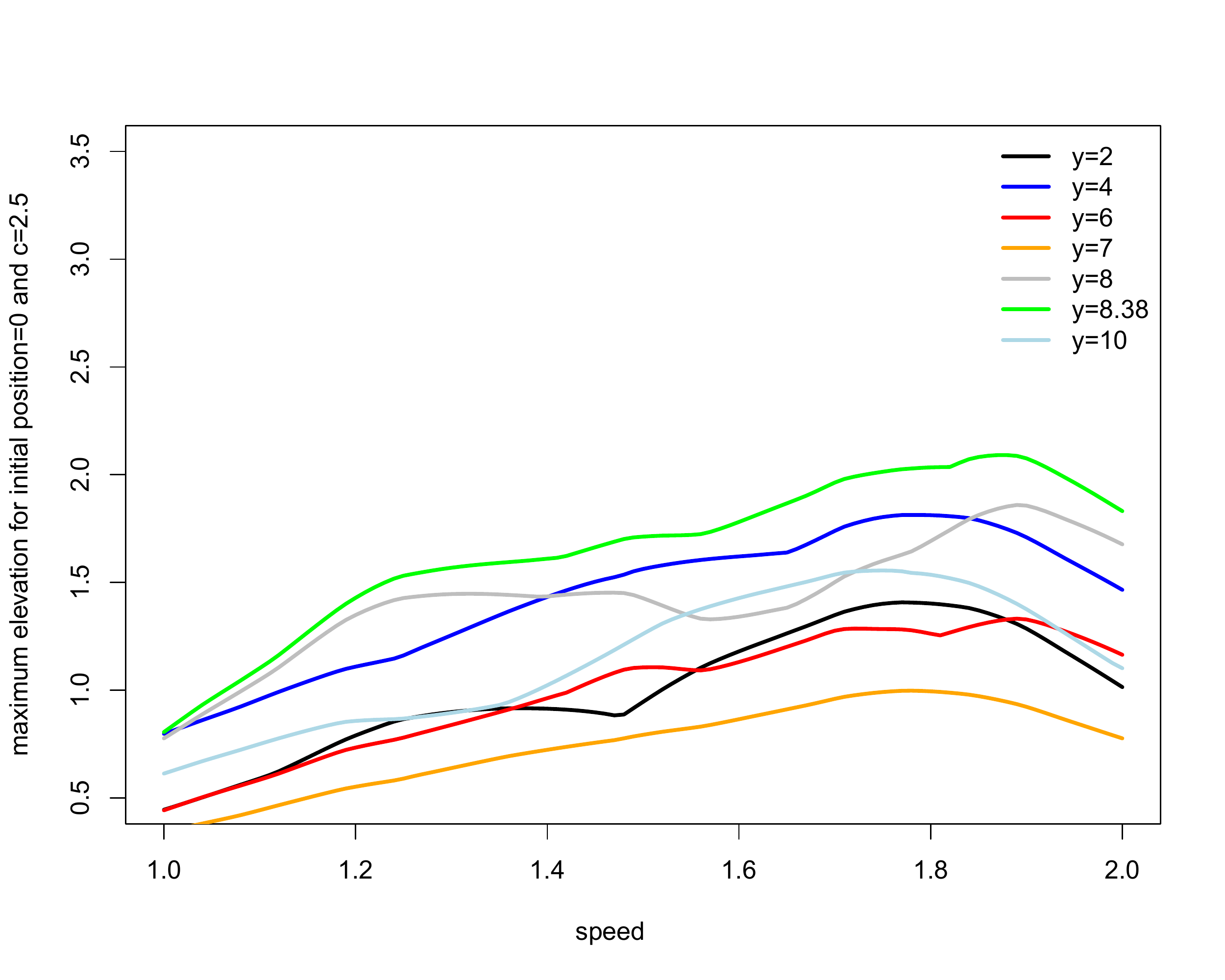}}
\end{center}
\caption{Maximum sea free-surface elevation with respect to landslide's speed for (a) $(x_0,c) = (-2,0.5)$, (b) $(x_0,c) = (-2,2.5)$, (c) $(x_0,c) = (0,0.5)$ and (d) $(x_0,c) = (0,2.5)$, for the time interval $[0,35]$.}
\label{max_u0_4}
\end{figure}

\begin{figure}[htb]
\vspace*{2mm}
\begin{center}
\subfloat[]{\includegraphics[trim=0.2cm 0.5cm 1cm 2.5cm, clip=true,height=0.181\textwidth]{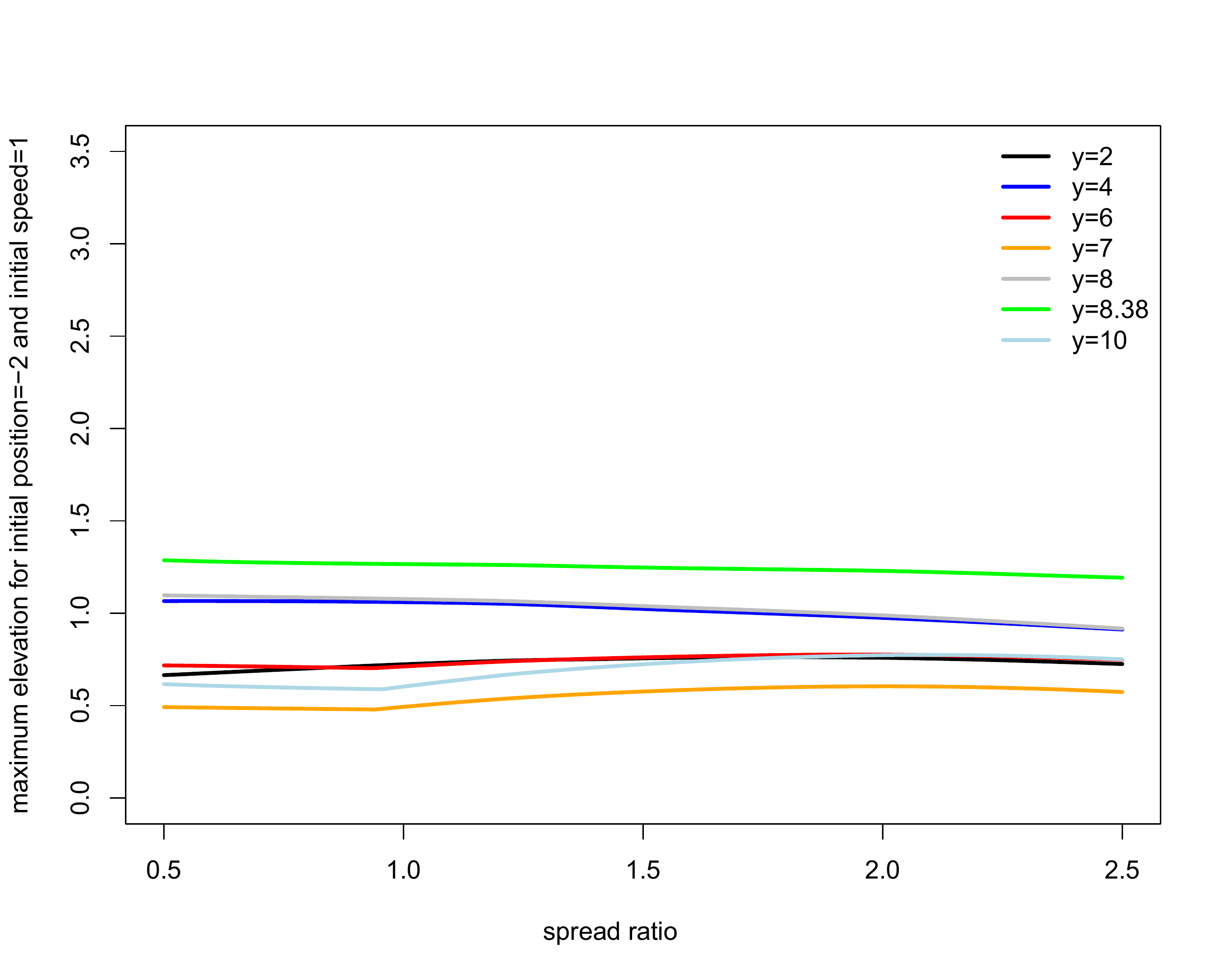}}
\subfloat[]{\includegraphics[trim=0.2cm 0.5cm 1cm 2.5cm, clip=true,height=0.181\textwidth]{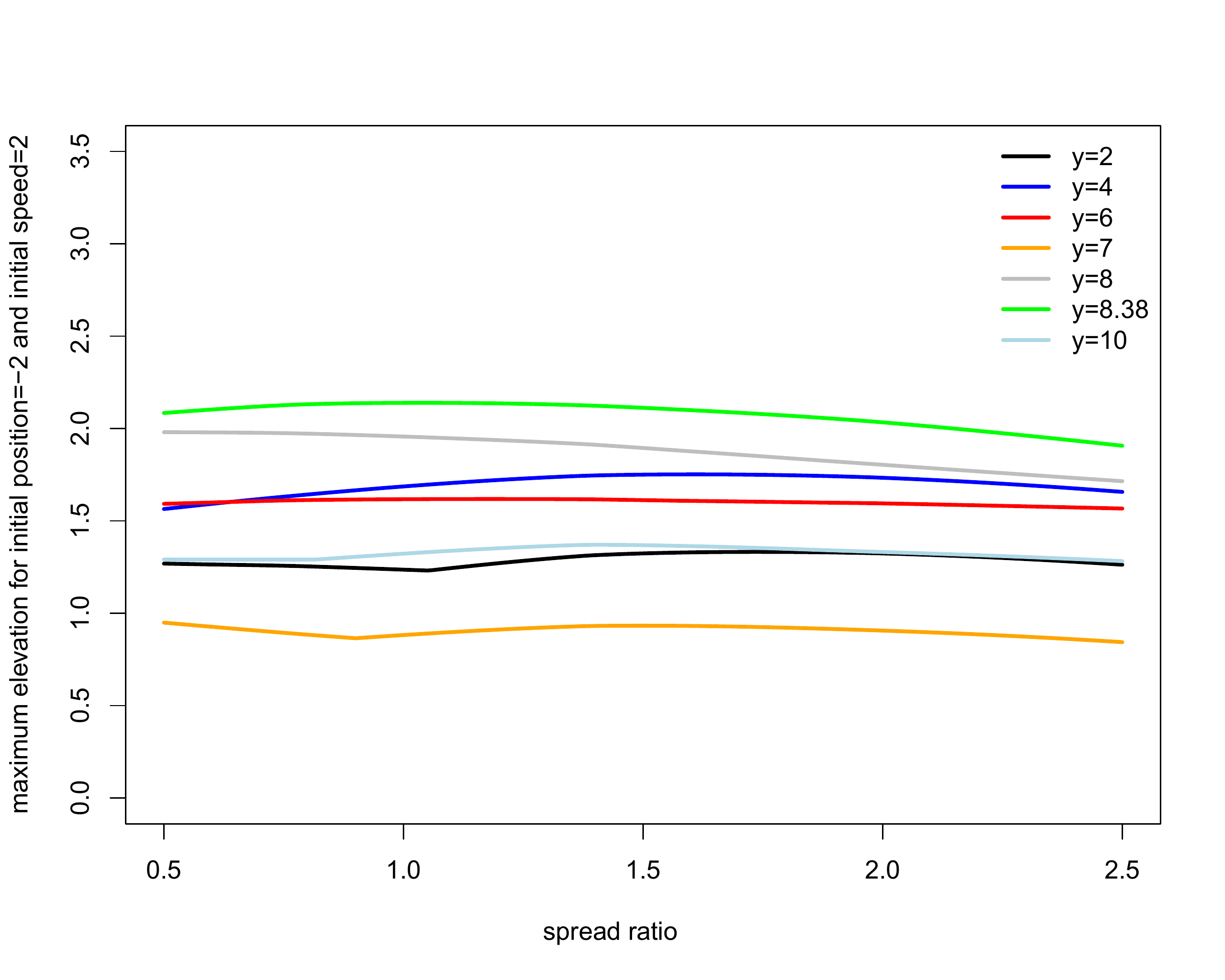}}\\
\subfloat[]{\includegraphics[trim=0.2cm 0.5cm 1cm 2.5cm, clip=true,height=0.181\textwidth]{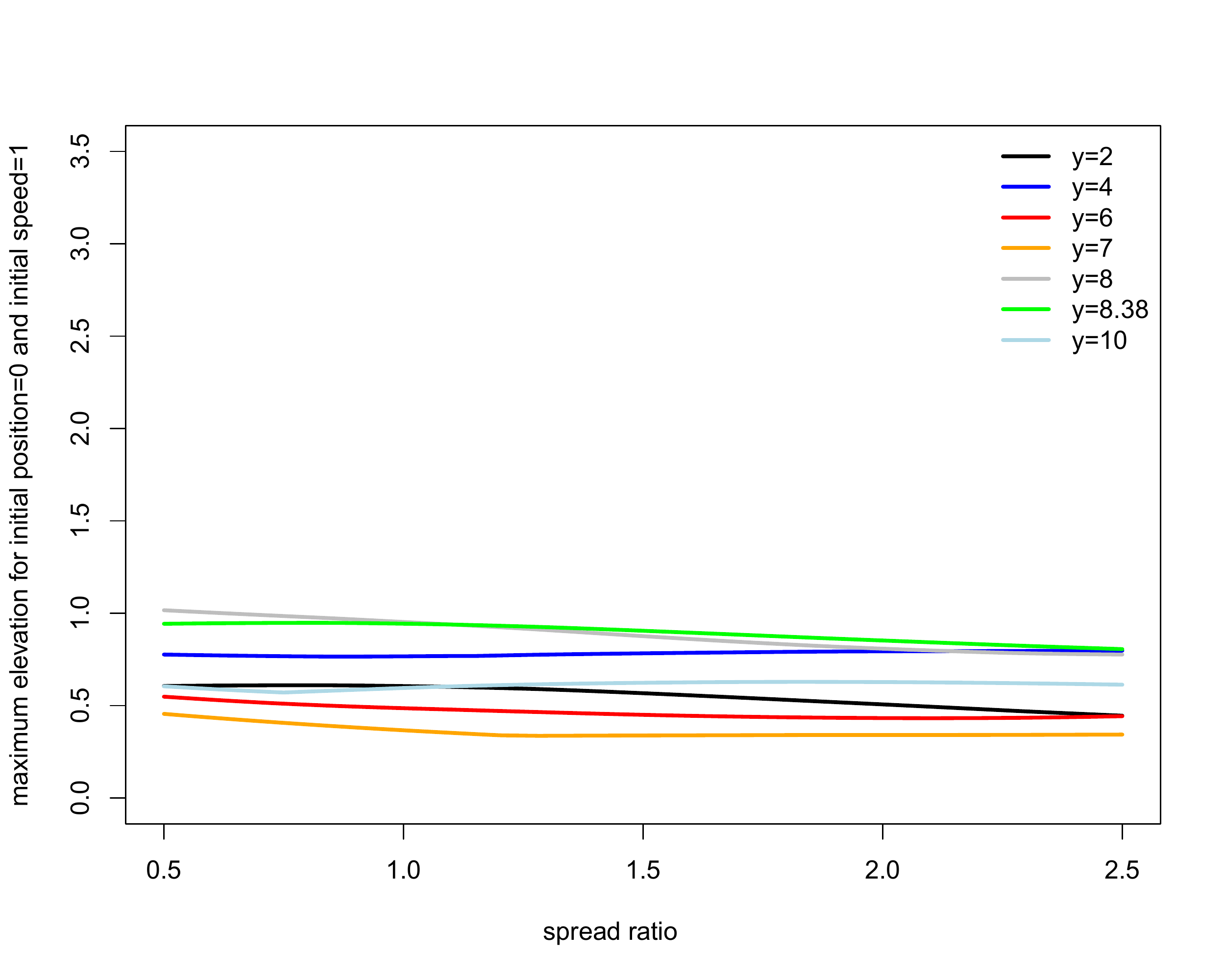}}
\subfloat[]{\includegraphics[trim=0.2cm 0.5cm 1cm 2.5cm, clip=true,height=0.181\textwidth]{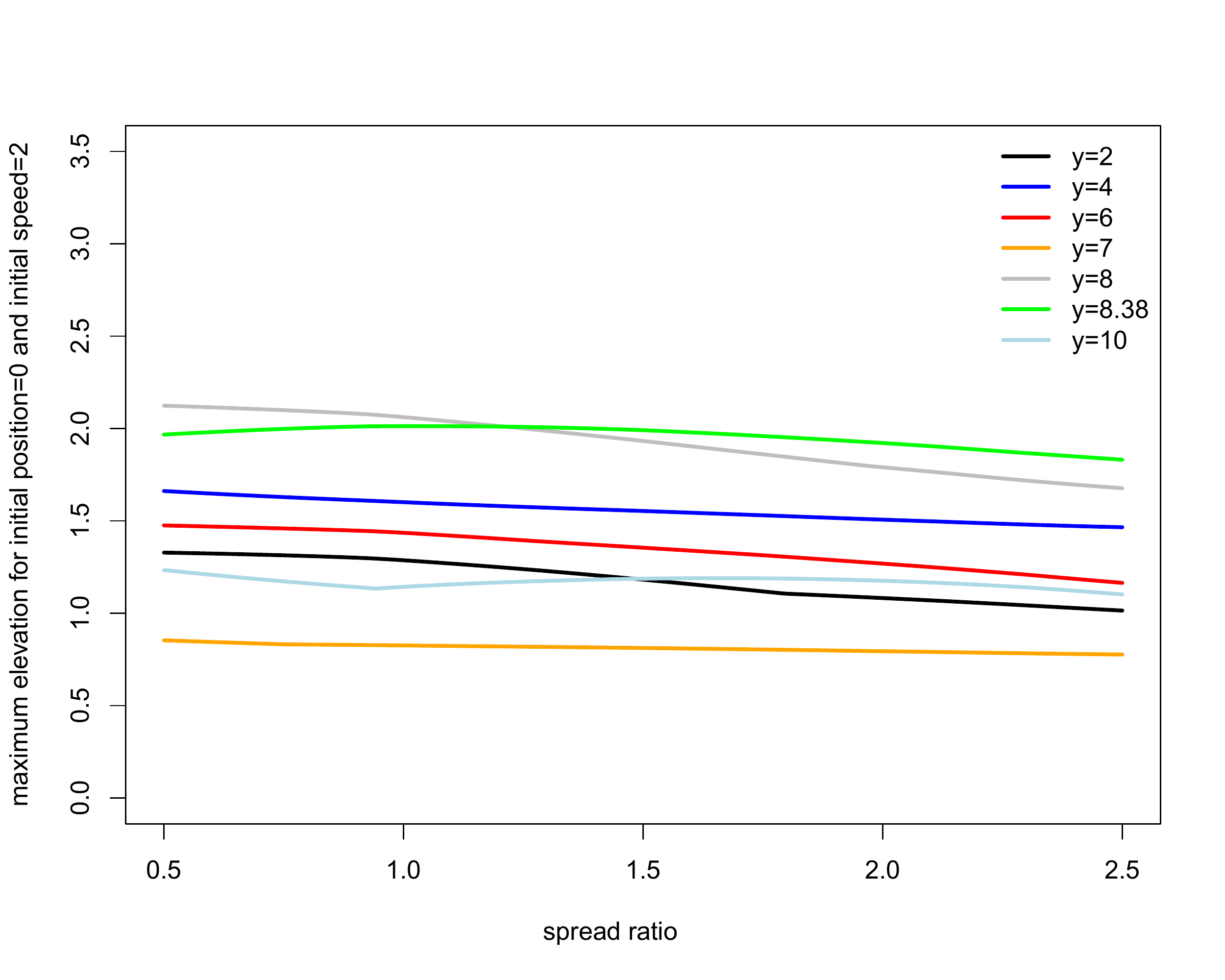}}
\end{center}
\caption{Maximum sea free-surface elevation with respect to landslide's spread ratio for (a) $(x_0,u_0) = (-2,1)$, (b) $(x_0,u_0) = (-2,2)$, (c) $(x_0,u_0) = (0,1)$ and (d) $(x_0,u_0) = (0,2)$, for the time interval $[0,35]$.}
\label{max_c_4}
\end{figure}

A comparison of how sensitive is the maximum wave elevation at different locations to changes in the input parameters is showed in Fig. \ref{max_x0_4}, \ref{max_u0_4} and \ref{max_c_4}. Each of the figures illustrate the change in maximum sea free-surface elevation with respect to variations in one of the input parameters, keeping the other two constant. We look at four different combinations of the constant parameters. We conclude that the sensitivity of maximum elevation is very similar for all the investigated locations along the shoreline.

Overall, the conclusions reached by using the emulator are the same as those obtained using the simulator as shown in Fig. \ref{1}. However, the emulator has the fundamental advantage that it is much faster compared to the simulator. Therefore, it can be evaluated at a much larger number of inputs, leading to higher resolution and smoother plots. Figure \ref{x0u0c} plots required a large number of emulator evaluations, specifically 2012. Importantly, the required emulator running time is very short. A total time for this entire analysis for a specific location was around 186.6 seconds on a Dual Core 3.06GHz computer. Using a simulator to perform the same analysis would take much longer, as a single run to reconstruct the sea free-surface elevation time series up to time 35 with the SR analytical model takes about 30 minutes.

\subsection{Uncertainty Analysis}
Usually the largest amount of uncertainty induced in simulator evaluations comes from the high uncertainty of tsunami trigger features. It is impossible to know exactly the initial position, speed and spread ratio of the landslide that cause the tsunami. Since, as we have shown, the emulator can provide accurate enough predictions of the simulator's outputs, an uncertainty analysis is performed by employing the emulator in the place of the simulator. The uncertainty analysis will give us the amount of uncertainty in the predictions that is due to the uncertain inputs, as well as from the use of emulator in place of the simulator. Usually experts have some knowledge about the most likely distribution of the inputs. Using these distributions, one can draw a number of random input samples, that can be given to the emulator in order to estimate the posterior distribution of key tsunamis features (e.g. maximum elevation). 

We assume that some collection of emergency management experts (in landslides or in real-time remote sensing) come to the conclusion that the inputs follow a beta distribution with some skewness and that the input domain is the same as with the sensitivity analysis. The beta distribution is a flexible distribution over a finite interval that can enable experts to express their believes. The distributions of input parameters are given by
\begin{equation}
\label{x0eq}
x_0 \sim Be(5,2)\quad \text{for} \quad x_0\in[-2,0]
\end{equation}
\begin{equation}
\label{u0eq}
u_0 \sim Be(2,5)\quad \text{for} \quad u_0\in[1,2]
\end{equation}
\begin{equation}
\label{ceq}
c \sim Be(2,5)\quad \text{for} \quad c\in[0.5,2.5]
\end{equation}
More specifically, the initial position of the landslide follows a distribution that indicates that a starting position near the origin is more likely. Both the speed and spread ratio distributions are skewed to the left, in order to highlight landslide's speeds most likely close to one and characteristic length and width of the landslide to be most likely of similar dimensions.

\begin{figure}[htb]
\vspace*{2mm}
\begin{center}
\subfloat[]{\label{x0P}\includegraphics[trim=1cm 0.5cm 0cm 2cm, clip=true,height=0.3\textwidth]{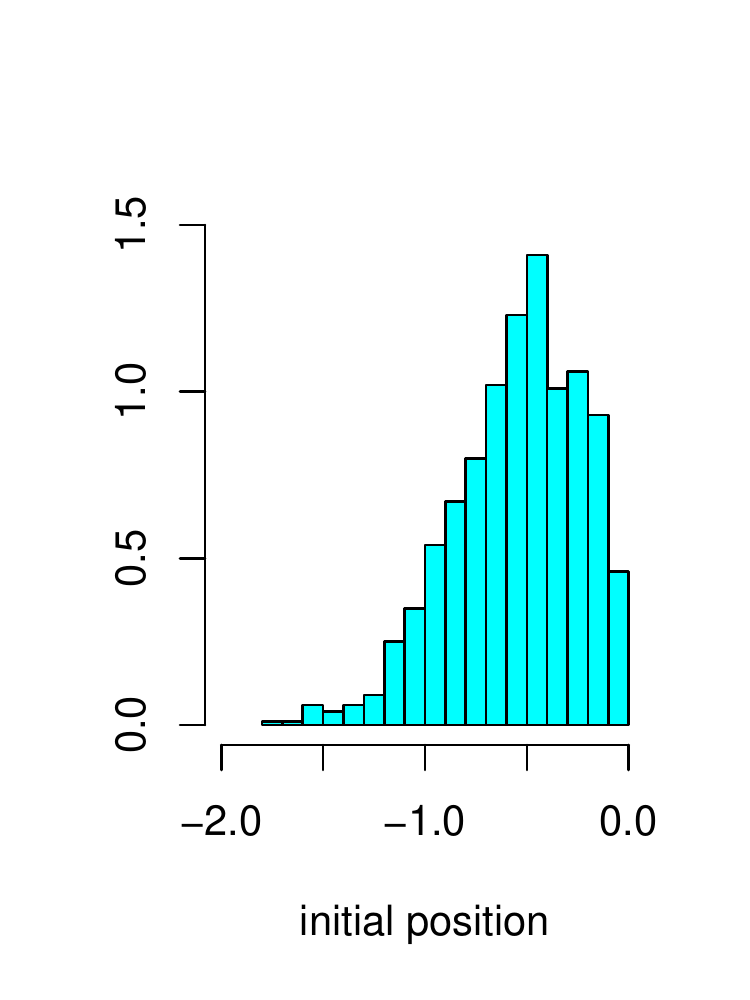}}\\
\subfloat[]{\label{u0P}\includegraphics[trim=1cm 0.5cm 1cm 2cm, clip=true,height=0.3\textwidth]{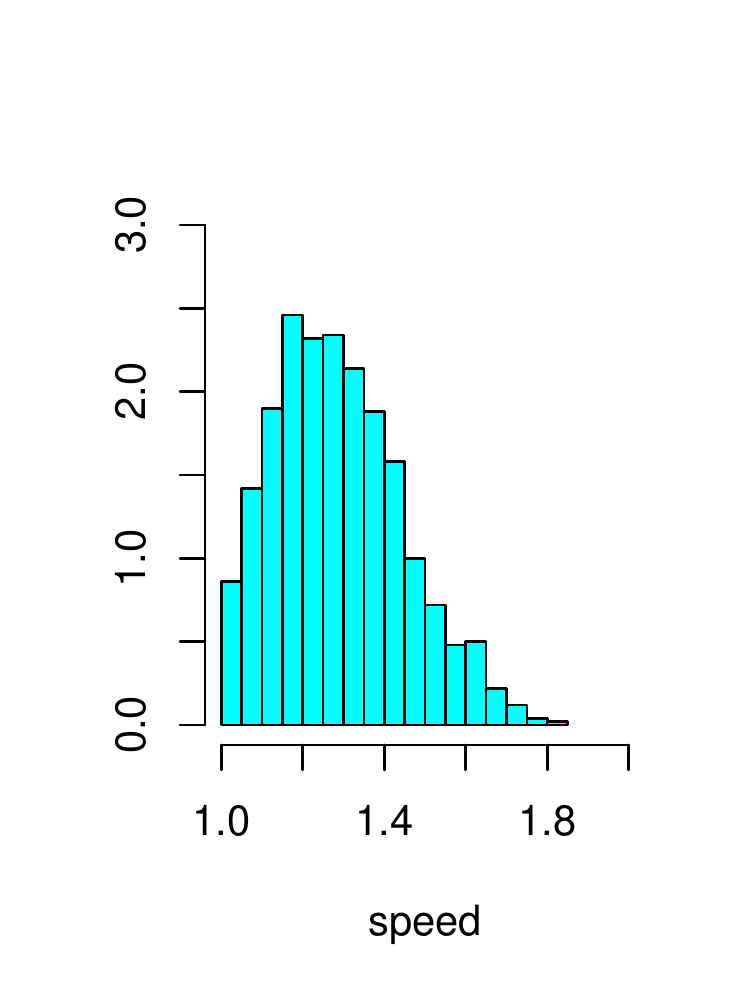}}
\subfloat[]{\label{cP}\includegraphics[trim=1cm 0.5cm 0.5cm 2cm, clip=true,height=0.3\textwidth]{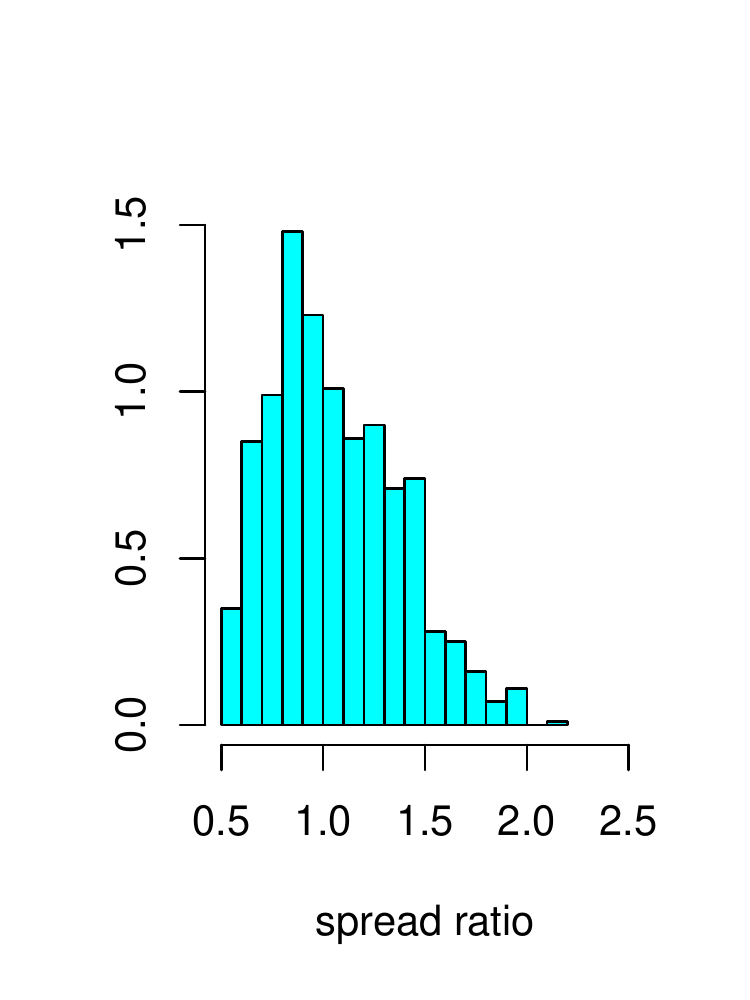}}
\end{center}
\caption{Histograms showing the prior knowledge about the distribution of input points.}
\label{hist_inputs}
\end{figure}

For this analysis we draw one thousand random samples for the inputs from the distributions given in (\ref{x0eq}), (\ref{u0eq}), (\ref{ceq}), resulting in the prior input distributions shown as histograms in Fig. \ref{hist_inputs}.

\begin{figure}[htb]
\vspace*{2mm}
\begin{center}
\includegraphics[trim=1cm 0.5cm 0cm 2cm, clip=true,height=0.45\textwidth]{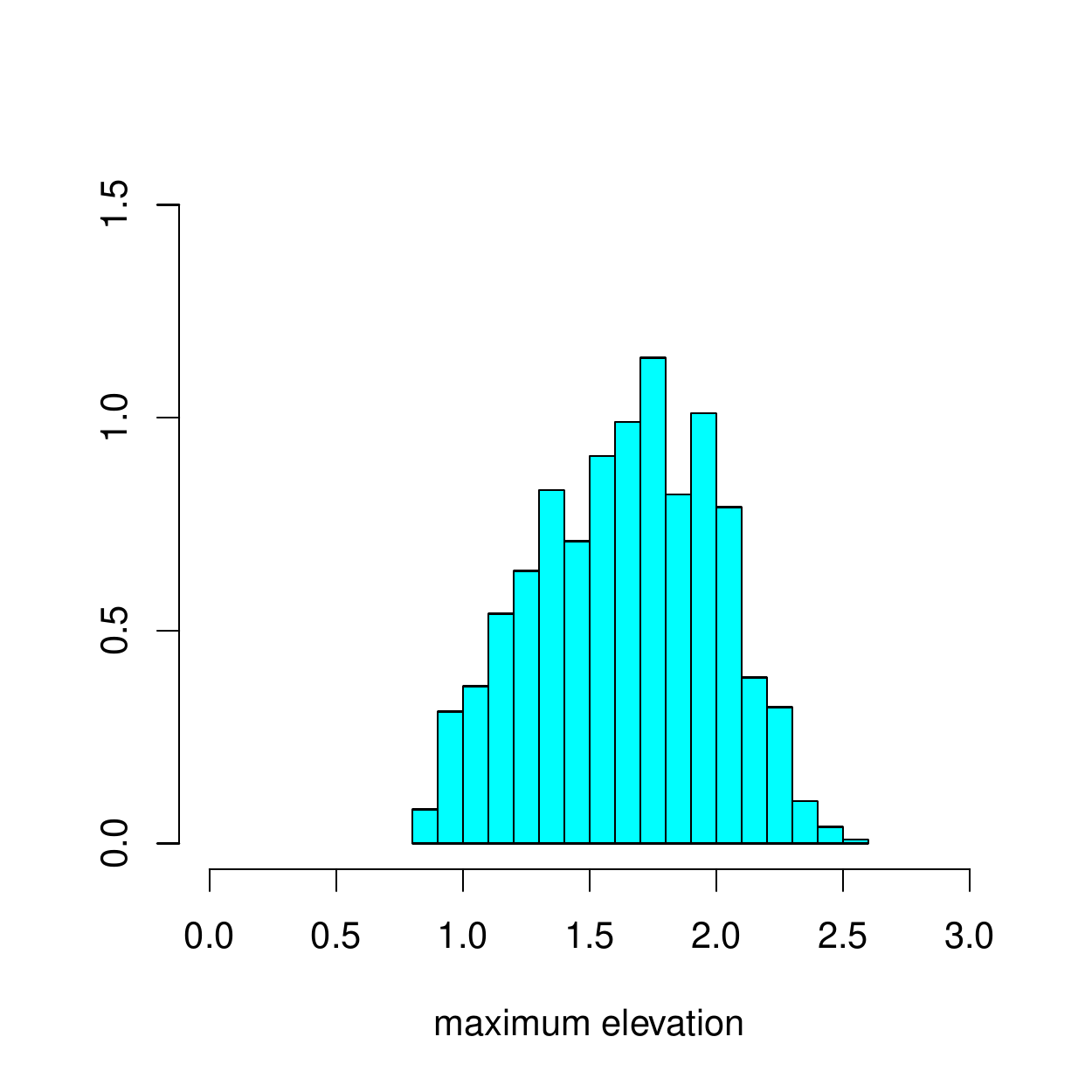}
\end{center}
\caption{Output distribution for maximum wave elevation at the location $(x,y) = (0,8.38)$.}
\label{hist_outputs_P}
\end{figure}

We run the emulator using the selected inputs and therefore, we get one thousand predictions for the wave elevation at a fixed position along the shoreline for times up to 35 at 0.2 intervals. From each of these time series, the maximum elevation and the mean CI length are estimated, resulting in one thousand estimates for each one. The variation among the thousand values are quantified using quantiles. The same process is repeated for all the examined locations along the shoreline. The quantiles for the case of $(x,y) = (0,8.38)$ are summarized in Table \ref{tab:percentiles}. The posterior distribution of the maximum elevation is plotted in Fig. \ref{hist_outputs_P}. This information summarizes the expected tsunami wave elevation and the associated uncertainty in prediction.

\begin{table}[ht]
\begin{center}
\begin{tabular}{|c|ccccc|}
  \hline
  & 1\% & 5\%  & 50\% &  95\% & 99\% \\ 
  \hline
maximum elevation & 0.92 & 1.03 & 1.66  & 2.18 & 2.35  \\ 
 mean CI length  & 0.28 & 0.40  & 0.66 & 0.90 & 1.03  \\ 
   \hline
\end{tabular}
\end{center}
\caption{Maximum elevation and mean CI length percentiles for the position $(x,y) = (0,8.38)$.}
\label{tab:percentiles}
\end{table}

Therefore, for a tsunami wave caused by the postulated landslide features, we are 95\% confident that the resulting tsunami wave will have maximum elevation less than 2.18, and 99\% confident that it will be less than 2.35, looking at a location along the shoreline and far away from the source ($y = 8.38$). The same analysis can be performed similarly for other locations along the shoreline. Again the ability of the emulator to make predictions almost immediately is highlighted in this case, since the total running time was just 83.9 seconds for 1000 runs at each of the locations compared to 30 minutes on the same computer for a single run of the SR tsunami model.

\conclusions
A statistical emulator of the analytical landslide-generated tsunami model developed by \cite{renzi2008} has been obtained using the Outer Product Emulator. This surrogate model is built using a combination of prior knowledge about the simulator, appropriate choices of functions and parameters and a limited number of simulator evaluations. The simulator is computationally expensive to evaluate, while the emulator produces estimates almost instantaneously. However, since the emulator is an approximation of the simulator an additional error is induced in predictions. But this amount of error can be estimated, since the predictions of the emulator are given as statistical distributions, not just values. Moreover, an accurate enough emulator represents the actual model with an almost negligible error. 

The emulator can be used for sensitivity and uncertainty analysis of the simulator, since these analyses are almost impossible to perform using the simulator. We have demonstrated these two analyses and the potential for reducing significantly the computational time. Where the emulator requires 83.9 seconds to get a thousand evaluations, the simulator requires 30 minutes for a single evaluation. Therefore, in critical situations where early warnings are necessary, an emulator can be a life saver by providing accurate prediction in a very short time. 

There are several possible avenues for extensions of this work. First, in this paper we only examined the wave motion at specific positions in space. To describe the space-time variations of the tsunami wave using an emulator, one needs to choose an enhanced formulation that includes spatial correlations of the outputs. This is a logical step but requires statistical expertise. Secondly, the source (landslide here) is still not realistic and prior expert knowledge could be included in a more factual way on a case study. Finally, more detailed simulations using more advanced physical-based models with a complex bathymetry need to be carried out to provide better quantifications of the subsequent sea free-surface elevations as well as more accurate run-ups on the shore with the help of a detailed orography.

\begin{acknowledgements}
The authors would like to acknowledge Dr. Jonathan Rougier for the constructive dialogue on emulation and Ms. Laura O'Brien for kindly providing the MATLAB code of the Sammarco and Renzi model. Also, they thank the UCL Institute of Risk and Disaster Reduction (IRDR) for providing funding for this research project. Additionally, the authors appreciate the help of Dr. Maurizio Filippone and Dr. Emiliano Renzi. Last but not least, they are grateful to the significant contribution of the referees to the improvement of the paper.
\end{acknowledgements}

\bibliographystyle{copernicus}
\bibliography{refSg+FD}

\begin{thebibliography}{16}
\providecommand{\natexlab}[1]{#1}
\providecommand{\url}[1]{{\tt #1}}
\providecommand{\urlprefix}{URL }
\expandafter\ifx\csname urlstyle\endcsname\relax
  \providecommand{\doi}[1]{doi:\discretionary{}{}{}#1}\else
  \providecommand{\doi}{doi:\discretionary{}{}{}\begingroup
  \urlstyle{rm}\Url}\fi

\bibitem[{Bardet et~al.(2003)Bardet, Synolakis, Davies, Imamura, and
  Okal}]{bardetetal2003}
Bardet, J.-P., Synolakis, C.~E., Davies, H.~L., Imamura, F., and Okal, E.~A.:
  Landslide tsunamis: recent findings and research directions, Pure appl.
  geophys., 160, 1793--1809, 2003.

\bibitem[{Liu et~al.(2005)Liu, Wu, Raichlen, Synolakis, and
  Borrero}]{liuetal2005}
Liu, P. L.~F., Wu, T.~R., Raichlen, F., Synolakis, C.~E., and Borrero, J.~C.:
  Runup and rundown generated by three-dimensional sliding masses, J. Fluid.
  Mech., 536, 107--144, 2005.

\bibitem[{Lynett and Liu(2005)}]{liu_lynett}
Lynett, P. and Liu, P. L.~F.: A numerical study of the run-up generated by
  three-dimensional landslides, J. Geophys. Res., 110, C03\,006,
  \doi{10.1029/2004JC002443}, 2005.

\bibitem[{Oakley and O'Hagan(2002)}]{OO_unc}
Oakley, J. and O'Hagan, A.: Bayesian inference for the uncertainty distribution
  of computer model outputs, Biometrika, 89, 769--784,
  \doi{10.1093/biomet/89.4.769}, 2002.

\bibitem[{Oakley and O'Hagan(2004)}]{Oakley02probabilisticsensitivity}
Oakley, J. and O'Hagan, A.: Probabilistic sensitivity analysis of complex
  models: a Bayesian approach, J. Roy. Stat. Soc. B, 66, 751--769,
  \doi{10.1111/j.1467-9868.2004.05304.x}, 2004.

\bibitem[{O'Hagan(2006)}]{OHag2006}
O'Hagan, A.: Bayesian analysis of computer code outputs: A tutorial, Reliab.
  Eng. Syst. Safe., 91, 1290--1300, \doi{10.1016/j.ress.2005.11.025}, 2006.

\bibitem[{Panizzo et~al.(2005)Panizzo, De~Girolamo, and
  Petaccia}]{forecasting_impulse_waves}
Panizzo, A., De~Girolamo, P., and Petaccia, A.: Forecasting impulse waves
  generated by subaerial landslides, J. Geophys. Res., 110, C12\,025,
  \doi{10.1029/2004JC002778}, 2005.

\bibitem[{Rasmussen and Williams(2006)}]{GPbook}
Rasmussen, C.~E. and Williams, C., K.: Gaussian Processes for Machine Learning,
  MIT Press, 2006.

\bibitem[{Renzi and Sammarco(2012)}]{renzi2012}
Renzi, E. and Sammarco, P.: The influence of landslide shape and continental
  shelf on landslide generated tsunamis along a plane beach (in press), Nat.
  Hazards Earth Syst. Sci., 2012.

\bibitem[{Rougier(2008)}]{Rougier2008}
Rougier, J.: Efficient Emulators for Multivariate Deterministic Functions, J.
  Comput. Graph. Stat., 17, 827--843, \doi{10.1198/106186008X384032}, 2008.

\bibitem[{Rougier et~al.(2009)Rougier, Guillas, Maute, and Richmond}]{rohtua}
Rougier, J., Guillas, S., Maute, A., and Richmond, A.~D.: Expert Knowledge and
  Multivariate Emulation: The Thermosphere – Ionosphere Electrodynamics General
  Circulation Model (TIE-GCM), Technometrics, 51, 414--424,
  \doi{10.1198/TECH.2009.07123}, 2009.

\bibitem[{Sammarco and Renzi(2008)}]{renzi2008}
Sammarco, P. and Renzi, E.: Landslide tsunamis propagating along a plane beach,
  J. Fluid Mech., 598, 107--119, \doi{10.1017/S0022112007009731}, 2008.

\bibitem[{Synolakis et~al.(2002)Synolakis, Bardet, Borrero, Davies, Okal,
  Silver, Sweet, and Tappin}]{synolakisetal2002}
Synolakis, C.~E., Bardet, J.-P., Borrero, J.~C., Davies, H.~L., Okal, E.~A.,
  Silver, E.~A., Sweet, S., and Tappin, D.~R.: The slump origin of the 1998
  {P}apua {N}ew {G}uinea {T}sunami, Proc. R. Soc. Lond. A, 458, 763--789, 2002.

\bibitem[{Tinti et~al.(2008)Tinti, Zaniboni, Pagnoni, and
  Manucci}]{tintietal2008}
Tinti, S., Zaniboni, F., Pagnoni, G., and Manucci, A.: Stromboli Island
  (Italy): Scenarios of tsunamis generated by submarine landslides, Pure appl.
  geophys., 165, 2143--2167, 2008.

\bibitem[{Urban and Fricker(2010)}]{comparisonLH_grid}
Urban, N., M. and Fricker, T., E.: A comparison of Latin hypercube and grid
  ensemble designs for the multivariate emulation of an Earth system model,
  Comput. Geosci., 36, 746--755, \doi{10.1016/j.cageo.2009.11.004}, 2010.

\bibitem[{Wiegel(1955)}]{oldpaper}
Wiegel, R.~L.: Laboratory studies of gravity waves generated by the movement of
  a submerged body, Trans. AGU, 36, 759--774, 1955.

\end{thebibliography}
\end{document}